\documentclass[journal]{IEEEtran}
\usepackage[utf8]{inputenc}
\usepackage{amssymb,amsmath,amsthm, mathtools}
\usepackage{graphicx}
\usepackage{cite}
\usepackage{latexsym}
\usepackage{verbatim}
\usepackage{amsmath}
\usepackage{bm}
\usepackage{amssymb}
\usepackage{subfigure}
\usepackage{url}
\usepackage{epstopdf}
\usepackage{algorithm}
\usepackage{times}
\usepackage{stfloats}
\usepackage{multirow}
\usepackage{enumerate}
\usepackage{xspace}
\usepackage{longtable}
\usepackage{mathrsfs}
\usepackage{mathtools}
\usepackage{algcompatible}
\usepackage[noend]{algpseudocode}
\usepackage{amssymb}
\usepackage{pifont}
\usepackage{multicol}
\usepackage{caption}
\usepackage{color}
\usepackage{booktabs}
\usepackage{tabularx}
\usepackage[dvipsnames]{xcolor}

 \makeatletter

\renewcommand\subsubsection{\@startsection{subsubsection}{3}{\parindent}{0ex plus 0.1ex minus 0.1ex}%
{0.5ex}{\normalfont\normalsize\itshape}}%

\renewcommand\paragraph{\@startsection{paragraph}{4}{2\parindent}{0ex plus 0.1ex minus 0.1ex}%
{0.3ex}{\normalfont\normalsize\itshape}}%

\renewcommand{\figurename}{Figure}

\makeatother

\newcommand{\ccomment}[1]{ }
\newcommand{\fc}{\ensuremath {\mathit{FC}}{\xspace}}
\newcommand{\cogr}{\ensuremath {\mathit{CR}}{\xspace}}
\newcommand{\db}{\ensuremath {\mathit{DB}}{\xspace}}
\newcommand{\su}{\ensuremath {\mathit{SU}}{\xspace}}
\newcommand{\pu}{\ensuremath {\mathit{PU}}{\xspace}}
\newcommand{\fcc}{\ensuremath {\mathit{FCC}}{\xspace}}
\newcommand{\crn}{\ensuremath {\mathit{CRN}}{\xspace}}
\newcommand{\dsa}{\ensuremath {\mathit{DSA}}{\xspace}}
\newcommand{\bs}{\ensuremath {\mathit{BS}}{\xspace}}
\newcommand{\aoa}{\ensuremath {\mathit{AoA}}{\xspace}}
\newcommand{\toa}{\ensuremath {\mathit{ToA}}{\xspace}}
\newcommand{\tdoa}{\ensuremath {\mathit{TDoA}}{\xspace}}

\newcommand{\rss}{\ensuremath {\mathit{RSS}}{\xspace}}
\newcommand{\rem}{\ensuremath {\mathit{REM}}{\xspace}}
\newcommand{\tm}{\ensuremath {\mathit{t}}{\xspace}}
\newcommand{\sse}{\ensuremath {\mathit{SSE}}{\xspace}}
\newcommand{\snr}{\ensuremath {\mathit{SNR}}{\xspace}}
\newcommand{\sinr}{\ensuremath {\mathit{SINR}}{\xspace}}

\newcommand{\ch}{\ensuremath {\mathit{chn}}{\xspace}}
\newcommand{\qos}{\ensuremath {\mathit{QoS}}{\xspace}}
\newcommand{\pir}{\ensuremath {\mathit{PIR}}{\xspace}}
\newcommand{\oram}{\ensuremath {\mathit{ORAM}}{\xspace}}
\newcommand{\ot}{\ensuremath {\mathit{OT}}{\xspace}}
\newcommand{\mpc}{\ensuremath {\mathit{MPC}}{\xspace}}
\newcommand{\nbr}{\ensuremath {\mathit{n}}{\xspace}}
\newcommand{\mac}{\ensuremath {\mathit{MAC}}{\xspace}}
\newcommand{\ope}{\ensuremath {\mathit{OPE}}{\xspace}}
\newcommand{\gw}{\ensuremath {\mathit{GW}}{\xspace}}
\newcommand{\spr}{\ensuremath {\mathit{SP}}{\xspace}}

{\bfseries}{\rmfamily}
\usepackage{xcolor,cite,etoolbox}
\makeatletter
\pretocmd\@bibitem{\color{black}\csname keycolor#1\endcsname}{}{\fail}
\newcommand\citecolor[1]{\@namedef{keycolor#1}{\color{red}}}
\makeatother

\begin{document}

 \title{Location Privacy in Cognitive Radio Networks: A Survey
}

\author{Mohamed~Grissa,
		Bechir~Hamdaoui
        and~Attila~A.~Yavuz\\

        \small Oregon State University, Corvallis, OR 97331, grissam,hamdaoui,attila.yavuz@oregonstate.edu
\thanks{This work was supported in part by the US National Science Foundation under NSF award CNS-1162296. Mohamed Grissa, Bechir~Hamdaoui and~Attila~A.~Yavuz are with the Electrical Engineering and Computer Science (EECS) Department, Oregon State University, Corvallis, OR 97331-5501, USA (e-mail: grissam,hamdaoui,attila.yavuz@oregonstate.edu).}
\thanks{\copyright~2017 IEEE. Personal use of this material is permitted. Permission from IEEE must be obtained for all other uses, in any current or future media, including reprinting/republishing this material for advertising or promotional purposes, creating new collective works, for resale or redistribution to servers or lists, or reuse of any copyrighted component of this work in other works.}

}
\maketitle
{\let\thefootnote\relax\footnote{{\\Digital Object Identifier 10.1109/COMST.2017.2693965}}}
\begin{abstract}
Cognitive radio networks (\crn s) have emerged as an essential technology to enable dynamic and opportunistic spectrum access which aims to exploit underutilized licensed channels to solve the spectrum scarcity problem. Despite the great benefits that \crn s offer in terms of their ability to improve spectrum utilization efficiency, they suffer from user location privacy issues. Knowing that their whereabouts may be exposed can discourage users from joining and participating in the \crn s, thereby potentially hindering the adoption and deployment of this technology in future generation networks. The location information leakage issue in the \crn~context has recently started to gain attention from the research community due to its importance, and several research efforts have been made to tackle it. However, to the best of our knowledge, none of these works have tried to identify the vulnerabilities that are behind this issue or discuss the approaches that could be deployed to prevent it. In this paper, we try to fill this gap by providing a comprehensive survey that investigates the various location privacy risks and threats that may arise from the different components of this \crn~technology, and explores the different privacy attacks and countermeasure solutions that have been proposed in the literature to cope with this location privacy issue. We also discuss some open research problems, related to this issue, that need to be overcome by the research community to take advantage of the benefits of this key \crn~technology without having to sacrifice the users' privacy.
\end{abstract}

\begin{keywords}
Location privacy, cognitive radio networks, dynamic spectrum access, privacy preserving protocols.
\end{keywords}


\section{Introduction}
\label{introduction}
{\em Cognitive radio networks} (\crn s) have been widely adopted as an efficient way to improve the spectrum utilization efficiency and alleviate the spectrum scarcity crisis caused by the huge demand on radio frequency resources. This technology has several applications and is considered as one of the main enablers for 5G wireless networks to deal with its stringent spectrum requirement.
This paradigm, first coined by Mitola~\cite{mitola1999cognitive}, could be thought of as an intelligent wireless communication system that is aware of its surrounding and that can adapt dynamically to the changes in the RF environment. It enables {\em dynamic spectrum access} (\dsa) and improves the spectrum utilization efficiency by allowing unlicensed/secondary users (\su s) to exploit unused spectrum bands of licensed/primary users (\pu s). That is, \su s can opportunistically use unused spectrum bands (aka spectrum holes or white spaces), which are defined by FCC as the channels that are unused at a specific location and time~\cite{akhtar2016white},
so long as doing so does not cause harmful interference to \pu s.

\subsection{The \crn~location privacy problem}
Despite its great potential for improving spectrum utilization efficiency, the \crn~technology suffers from serious privacy and security risks.
Although the survey covers location privacy issues arising at the various \crn~components, for motivation purposes, we focus in this section on the {\em spectrum discovery component} only, in which white spaces are identified using either the {\em cooperative spectrum sensing} or the {\em database-driven spectrum access} functions.
\subsubsection{Cooperative spectrum sensing} In cooperative sensing, a central entity called Fusion Center (FC) orchestrates the sensing operations as follows: It selects one channel for sensing and, through a control channel, requests that each \su~perform local sensing on that channel to detect the presence of \pu~signals and send its sensing report back to it. Fusion Center then combines the received sensing reports, makes a decision about the channel availability, and diffuses the decision back to the \su s.
Here, a sensing report is essentially a sensed/measured quantity characterising some \pu~signal strength the \su~observed on some \pu~channel, and what quantity the \su~measures depends on the spectrum sensing method it uses (e.g., waveform~\cite{tian2006wavelet}, energy detection~\cite{poor2013introduction}, cyclostationarity~\cite{ghozzi2006cyclostatilonarilty}, etc.; see Section~\ref{section2-sensing} for more details). For example, when using the energy detection method, the sensed quantity is the energy strength of the sensed \pu~signal, often referred to as {\bf received signal strength} (RSS)~\cite{fatemieh2011using}.

In cooperative sensing, communications between \su s and Fusion Center could be done via one of the following: $(i)$ direct, single-hop wireless links; $(ii)$ multi-hop links (with first link being wireless); $(iii)$ wired links (whether single or multiple hops).
%
In the first and second types, location privacy information can be easily leaked by observing the wireless radio signals sent by \su s to Fusion Center. In this case, existing (mostly mature) location privacy preservation technologies (e.g., see~\cite{conti2013providing,xi2006preserving} for sensor, ~\cite{jiang2007preserving} for WiFi and~\cite{gorlatova2011managing} for cellular) can be applied here to protect the location privacy of \su s during cooperative sensing.
%
In the third communication type when \su s communicate with Fusion Center via wired links, wireless signal-based localization techniques can no longer be used here to locate \su s.

\begin{figure}[h!]
\vspace{-10pt}
  \centering
  \subfigure[\small { Cooperative sensing}]{\includegraphics[width=0.24\textwidth]{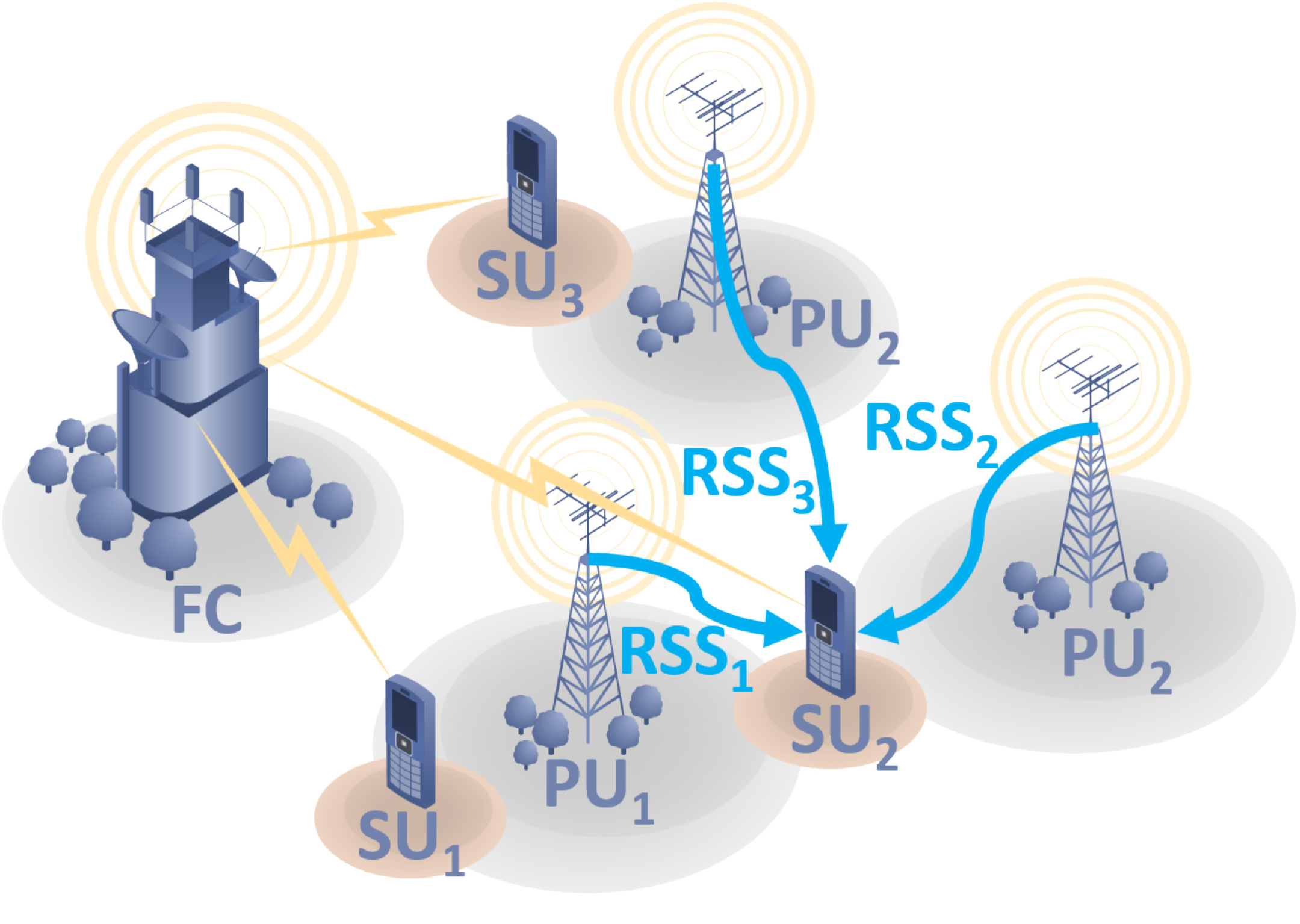}\label{coopLocation}}
  \subfigure[\small { \db-driven access}]{\includegraphics[width=0.24\textwidth]{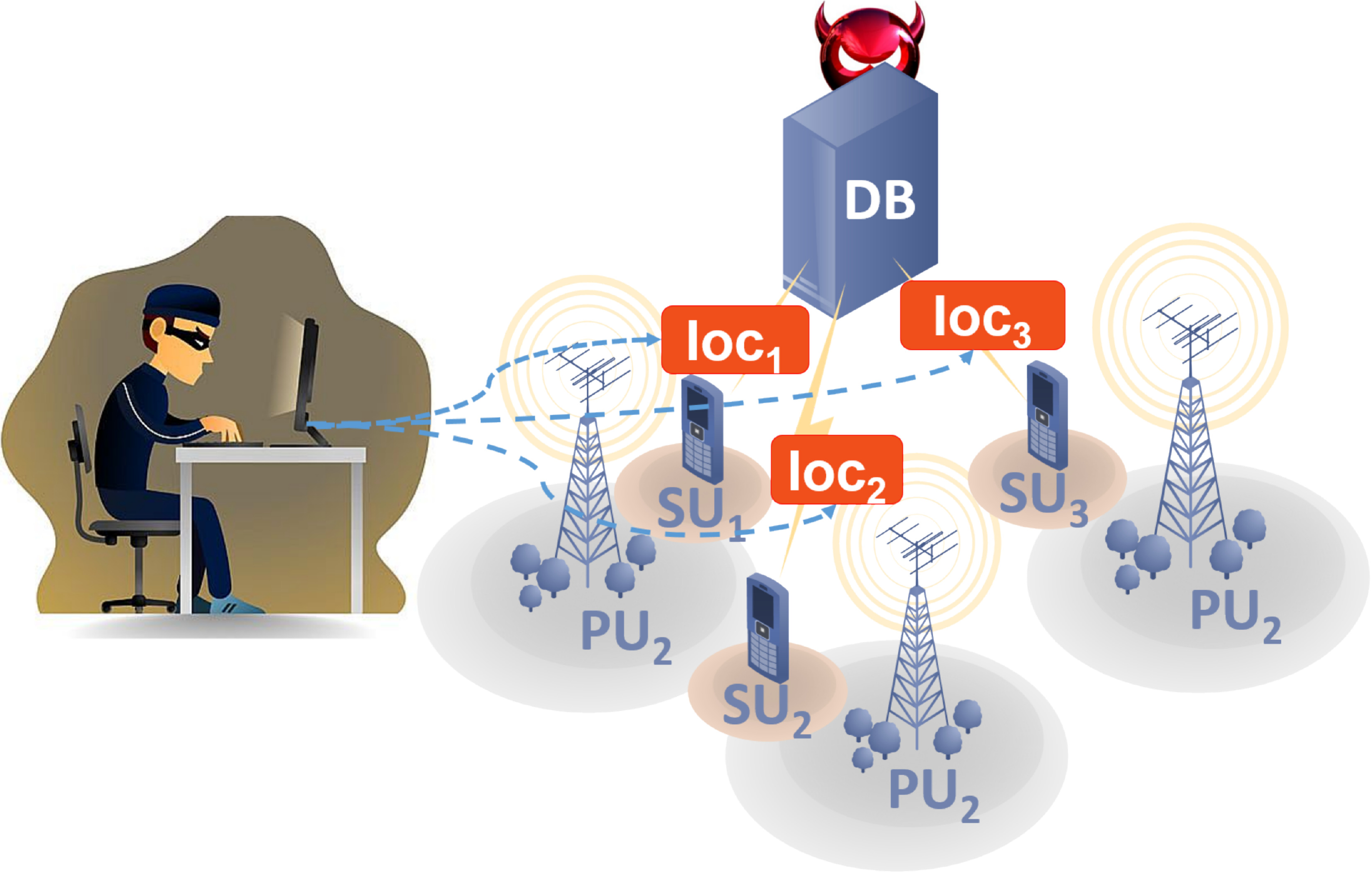}\label{DBLocation}}
  \vspace{-5pt}
  \caption{\small {Location privacy issues during spectrum discovery}} \label{}

\end{figure}

However, unlike traditional wireless networks, in the case of cooperative sensing, preventing leakage of location information from wireless signals (e.g., by communicating via wired links) does not guarantee the preservation of \su s' location privacy. This is because location information can also be leaked from the sensing reports sent by \su s to the Fusion Center during cooperative sensing\cite{li2012location}.
Recall that a sensing report is essentially the {received signal strength} (RSS) value of some \pu~signal that the \su~observed on a specific \pu~channel.
And it has been shown that these values are highly correlated to the physical location of the reporting \su~\cite{li2012location}. Now Fusion Center may know the actual physical locations of few \pu s as well as the channels these \pu s communicate on, and thus, by knowing the RSS values measured by an \su~on each of these \pu~channels, Fusion Center can easily locate the \su. Some illustrative scenarios, showing how Fusion Center can easily infer the physical locations of \su s by simply looking at few sensing reports on different \pu~channels, are given in~\cite{li2012location}. This is also illustrated in Figure~\ref{coopLocation}.

\subsubsection{ Database-driven access} In database-driven spectrum access, spectrum availability information is provided to \su s by querying a spectrum database, often maintained and controlled by a third
party (e.g., Google, Spectrum Bridge, RadioSoft, etc.). Here, although \su~queries' final destination is the database, which is often located far away from the \su s, location information can also be leaked from wireless radio signals if the \su s' first hop is a wireless link; e.g., a cellular base station or a WiFi access point. In this case, the aforementioned, existing location privacy preservation techniques that overcome wireless signal-based leakage can also be applied to protect \su s' location privacy.
However, as illustrated in Figure~\ref{DBLocation}, there is a more straightforward location privacy threat specific to the database-driven access method: In order for an \su~to acquire spectrum availability information, it is required to query the database with its physical location, so that the database can inform it about spectrum availability in its vicinity. This explicit exposure of \su s' location information to third (commercial) parties raises serious privacy concerns and has some undesired consequences, as discussed next.

\subsection{Why worry about location information privacy?}
Most users will not be okay with having their whereabouts and private location information made public, especially in the presence of malicious entities that may be eager to exploit this information for malicious purposes and to gain more knowledge about other sensitive and private information.
A survey conducted in 2015 by Pew Research found that ''{\it Most Americans hold strong views about the importance of privacy in their everyday lives}", and that ''{\it These feelings also extend to their wishes that they be able to maintain privacy in their homes, at work, during social gatherings, at times when they want to be alone and when they are moving around in public}"(Madden et al.~\cite{privacy-report}). This same survey also reports that ''{\it 90\% say that controlling what information is collected about them is important}'' and
''{\it 88\% say it is important that they not have someone watch or listen to them without their permission}''.
For instance, with an operation as simple as a succession of database accesses, a database can easily monitor and track the \su's daily life activities and communications, allowing the database to learn various behavioral information about the user; e.g., where he/she goes for shopping, which social circles he/she attends, and where and when he/she eats.
As spectrum databases are being managed by business entities, such a private information is at the risk of being sold and shared with other commercial entities.
Indeed, a \su's fine-grained location information, when combined with other publicly available information, could easily be exploited to infer other personal information about an individual including his/her behavior, preferences, personal habits or even beliefs. For instance, an adversary can learn an individual's religious belief by observing that a he/she frequently visits places with religious affiliations.
Location traces could also reveal some information about the health condition of a user if the adversary observes that the user regularly goes to a hospital for example. The frequency and duration of these visits can even reveal the seriousness of a user illness and even the type of illness if the location corresponds to that of a specialty clinic. The adversary could sell this health information to pharmaceutical advertisers without the user's consent. Moreover, malicious adversaries with criminal intent could use the location information to pose a threat to individuals' security and privacy; for instance, they can commit crimes of theft and burglary when users are absent.

We envision that the public's acceptance of the dynamic and opportunistic
spectrum sharing paradigm will greatly depend on the robustness and trustworthiness of \crn s vis-a-vis of their ability to address these privacy concerns. It is, therefore, imperative that techniques and tools to be developed by the research community for \crn s be enabled with privacy preserving capabilities that protect the location privacy of \su s while allowing them to harness the benefits of the \crn~paradigm without disrupting the functionalities that these techniques are designed for to promote {\em dynamic spectrum access}.


\subsection{Location privacy protection: pros and cons}

Ensuring that the location privacy information of \su s is protected has great benefits.
First and most importantly, it promotes dynamic and opportunistic sharing of spectrum resources, thereby increasing spectrum utilization efficiency. Knowing that their location privacy is protected so that they do not have to worry about their whereabouts being tracked and their privacy being compromised, \su s will be encouraged to participate in the cooperative spectrum sensing process, and to query spectrum databases for spectrum availability.
Ensuring location privacy protection can also be beneficial to \pu s. For example, being concerned that their location privacy information may be leaked to spectrum databases, \su s may attempt to use \pu~channels without registering and querying spectrum databases for spectrum availability, thereby causing harmful interference to \pu s.

Providing location privacy preservation guarantees cannot, however, be done without a cost. It does introduce additional communication, computation and storage overheads, which may, in turn, also introduce additional delay when it comes to learning about the availability status of some channel, and can, in the extreme case, make the spectrum availability information outdated, thus possibly resulting in using a channel that is not vacant.
The pros and cons of providing location privacy protection are summarized in Table~\ref{advDis}.

\begin{table*}[th!]
\vspace{-5pt}
\centering
\caption{\small Pros and cons of preserving the location privacy of \su s}
\label{advDis}
\resizebox{\textwidth}{!}{%
\renewcommand{\arraystretch}{1.25}{
\begin{tabular}{@{}lp{8.5cm}p{8.5cm}@{}}
\toprule[1.5pt]
                 & Pros                    & Cons                                                                             \\ \midrule

{$\boldsymbol{\su}$}    &  \begin{tabular}[c]{@{}p{8.5cm}@{}} {- Encourages \su s to participate in the cooperative spectrum sensing process, and hence in accurately locating spectrum opportunities. }  \\{ - Discourages \su s from using spectrum opportunities without checking for availability first, either through spectrum databases or cooperative sensing, and thus prevents \su s from violating secondary spectrum access policies. } \\{- Promotes dynamic spectrum sharing, and thus increases spectrum utilization efficiency and helps address the spectrum supply shortage problem.} \end{tabular} &   \begin{tabular}[c]{@{}p{8.5cm}@{}} {- Incurs additional \su s' communication, computation, and storage overheads; this can be problematic when \su s are resource-limited devices (e.g., IoT devices, sensors, etc.).}  \\ {- Introduces delay in the process of querying spectrum databases for spectrum availability information in the case of database-driven \crn~approach.}  \\ {- Introduces delay when locating and deciding about spectrum availability through the cooperative spectrum sensing approach. } \end{tabular} \\ \addlinespace[5pt] \hline \addlinespace[5pt]

{$\boldsymbol{\pu}$  }    &  \begin{tabular}[c]{@{}p{8.5cm}@{}}{- Protects \pu s from harmful interference that might come from \su s not willing to check for spectrum availability (either via the cooperative spectrum sensing approach or database-driven access approach) before using \pu~channels.} \end{tabular} & \begin{tabular}[c]{@{}p{8.5cm}@{}} {- Outdated spectrum availability information due to the delays incurred as a result of protecting the location privacy may lead to the use of occupied \pu~channels by \su s, thereby causing interference to \pu s.} \end{tabular} \\ \addlinespace[5pt]
 \bottomrule[1.5pt]
\end{tabular}}}
\end{table*}

\begin{figure*}[th!]
\center
    \includegraphics[width=0.8\textwidth]{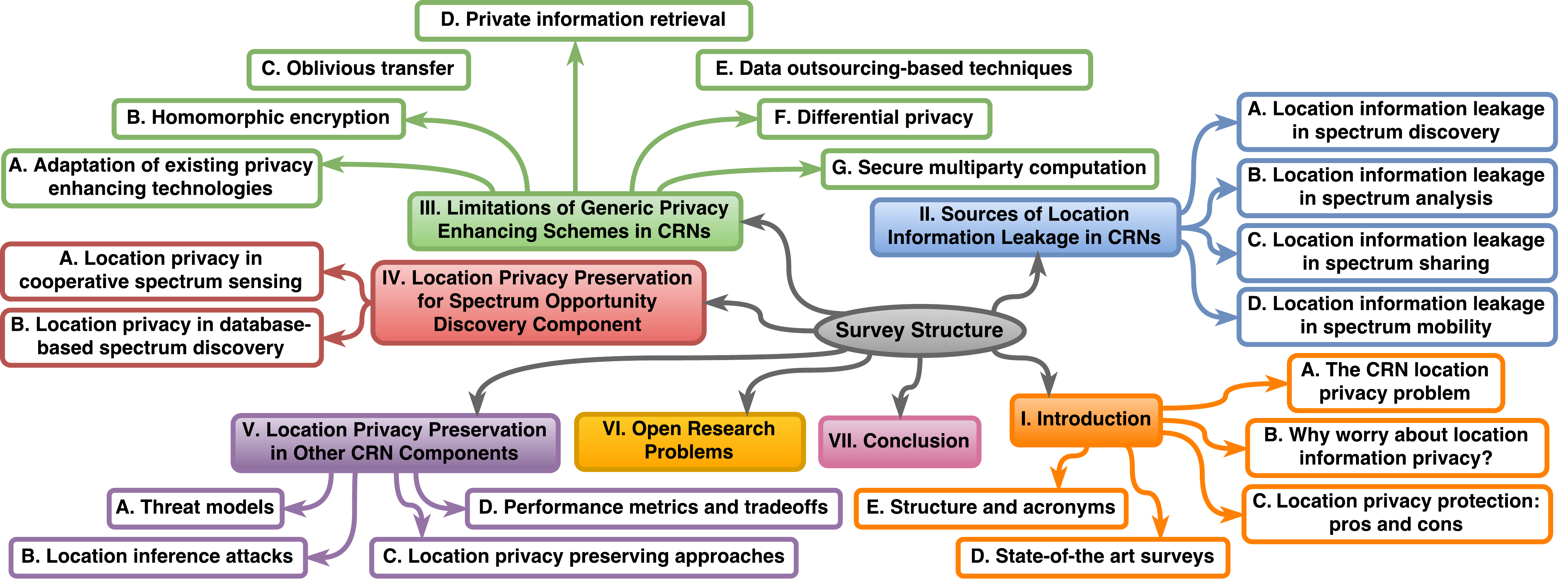}
\caption{\small Survey structure}
\label{fig:structure}
\end{figure*}

\subsection{State-of-the art surveys}

There have been several existing works that investigate and address the various \crn~vulnerabilities and security issues~\cite{araujo2012security,fragkiadakis2013survey,ling2015application,el2011survey,sharma2015advances}. However, most of these survey works focus on security and privacy issues in general with little or no attention to the location privacy issue that we address in this survey. For instance, Mee et al.~\cite{ling2015application} present an extensive review on the use of reinforcement learning to achieve security enhancement in the context of \crn s while dealing with jamming and byzantine attacks. El-Hajj et al.~\cite{el2011survey} provide a per-layer classification of attacks targeting \crn s, and discuss existing countermeasure solutions that address these attacks. Sharma et al.~\cite{sharma2015advances} discuss security threats, attacks, and countermeasures in \crn s for both \pu s and \su s with focus on the physical layer. There have also been few surveys that aimed at exploring location privacy issues, but they are generally not focusing on \crn s. For instance, Zhang et al.~\cite{zhang2015location} present a high-level overview of fundamental approaches for user localization and privacy preservation but mainly in the context of location-based services (LBS). They also discuss this issue, but only briefly, in the context of indoor environments, wireless sensor networks, and cognitive radio networks. To the best of our knowledge, this is the first comprehensive survey that digs into the different privacy threats and attacks that target the location information of \su s at the different \crn~components, along with the different techniques that have been proposed in the literature to mitigate and address these threats.
\label{related}

\subsection{Structure and acronyms}

This paper provides a comprehensive survey of the location privacy threats and vulnerabilities arising at the various components of \crn s, as well as the different techniques proposed in the literature to overcome these privacy issues. The general survey structure is depicted in Figure~\ref{fig:structure}, and is as follows:
\begin{itemize}
\item Section~\ref{sec:sources} investigates the vulnerabilities and sources of location information leakage in \crn s, and provides insights on how these vulnerabilities could become potential threats to \su s' location privacy.
\item Section~\ref{limitGeneric} explores the privacy enhancing technologies (PETs) that are most relevant to \crn s. The goal is to show how these techniques, that are widely adopted in other contexts, could not be applied off-the-shelf as they are in the context of \crn s unless judiciously adapted to the unique requirements of \crn s.

\item Sections~\ref{lpsd} and~\ref{lpoc} discuss threats and attacks that have been identified in the literature with respect to the spectrum opportunity discovery component (Section~\ref{lpsd}), as well as other \crn~components (Section~\ref{lpoc}). They also discuss their impacts on \su s' privacy, and  investigate countermeasure solutions that have been proposed in the literature to deal with these attacks. These two sections also explore and present the different metrics used to assess and evaluate both the achievable performance and the privacy level of these proposed solutions.
\item Section~\ref{openproblems} discusses unsolved research challenges pertaining to the location privacy in \crn s, with a special focus on the \cogr~components that have received the least attention from the research community. It also discusses open research problems arising from alternative \crn~architectures and from emerging \cogr-based technologies.
\item Section~\ref{con} concludes the survey.

\end{itemize}
For convenience, we summarize the used acronyms in Table~\ref{t:notations}.

\begin{table}[h!]
\vspace{-7pt}
\caption{\small Acronyms}
\centering
\resizebox{0.4\textwidth}{!}{
\label{t:notations}
\begin{tabular}{l l}

\hline
\noalign{\medskip }
$\aoa$ & Angle of arrival \\
$\bs$ & Base station \\
$\cogr$ & Cognitive radio \\
$\crn$ & Cognitive radio network \\
$\db$ & Spectrum database \\
$\dsa$ & Dynamic spectrum access \\
$\fc$ & Fusion center\\
$\fcc$ & Federal Communications Commission \\
$\gw$ & Gateway\\
$\mac$ & Medium Access Control \\
$\mpc$ & Secure multiparty computation \\
$MTP$ & Maximum transmit power \\
$\ope$ & Order preserving encryption \\
$\oram$ & Oblivious random access memory \\
$\ot$ & Oblivious transfer \\
PET & Privacy enhancing technology\\
$\pir$ & Private information retrieval \\
$\qos$ & Quality of service \\
$\rem$ & Radio environment map \\
$\rss$ & Received signal strength\\
$\sinr$ & Signal to interference-plus-noise ratio \\
$\snr$ & Signal to noise ratio \\
$\spr$ & Service provider \\
$\sse$ & Searchable symmetric encryption \\
$\su$ & Secondary user \\
$\pu$ & Primary user \\
$\toa$ & Time of arrival \\
$\tdoa$ & Time difference of arrival \\
$WSN$ & Wireless sensor network \\
\noalign{\smallskip} \hline \noalign{\smallskip}
\end{tabular}
}
\end{table}

\section{Sources of location information leakage in \crn s}
\label{sec:sources}
\label{sec:sourcesof}
\begin{figure}[h!]
\center
\includegraphics[width=0.36\textwidth]{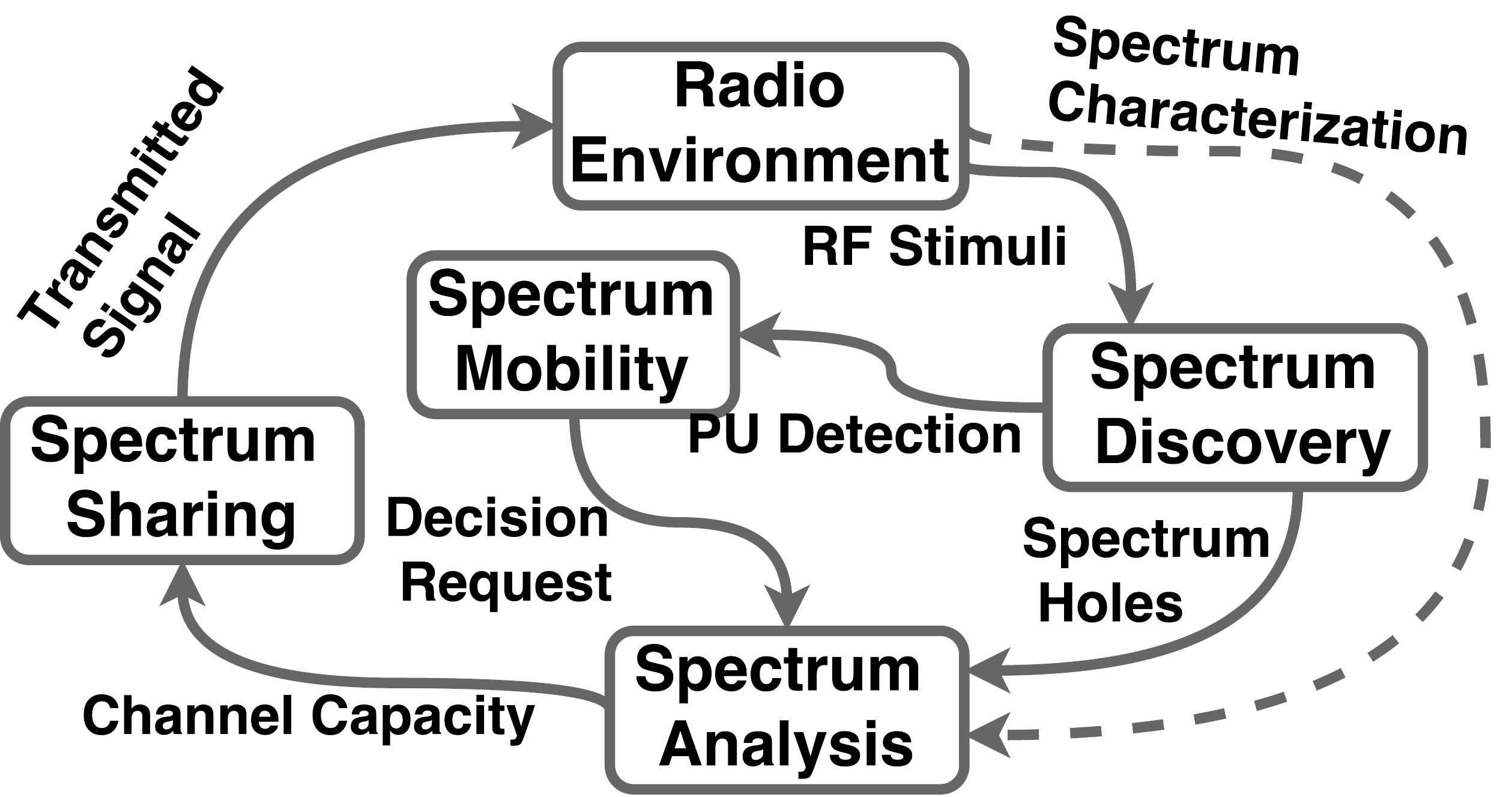}
\caption{\small Cognitive radio cycle~\cite{akyildiz2009crahns}.}
\label{cogCycle}
\vspace{-10pt}
\end{figure}

\crn s need to perform a number of spectrum-aware operations to adapt to the dynamic spectrum environment. These operations form what is called a {\em cognition cycle} \cite{mitola1999cognitive,haykin2005cognitive,akyildiz2009crahns,hossain2009dynamic}, which mainly consists of four spectrum functions as shown in Figure~\ref{cogCycle}: Spectrum opportunity discovery, spectrum analysis, spectrum sharing and spectrum mobility. Despite their merit in enhancing the spectrum utilization, \crn s may present some privacy risks to \su s especially in terms of their location privacy. In this section, we investigate the different aspects of the cognitive spectrum functions and we discuss the different threats that can compromise the location privacy of \su s in \crn s during the execution of these functions.

\subsection{Location information leakage in spectrum discovery}
\label{specDisc}
This is considered to be one of the most important components of the cognition cycle, as it provides information about spectrum holes and \pu s' presence. Mainly, there are two approaches to obtain this information: spectrum sensing, to be performed by \su s \cite{yucek2009survey}, and geolocation database. We first describe these two approaches, and then investigate the sources of location information leakage that they may have.

\subsubsection{Spectrum sensing}
\label{section2-sensing}
Spectrum sensing enables \su s to be aware of their surroundings and to be able to identify spectrum holes in their vicinity so that they can exploit them opportunistically. It basically requires \su s to sense and detect primary signals without interfering with \pu's transmissions~\cite{ghasemi2008spectrum,axell2012spectrum}. Spectrum sensing could be divided into two main functionalities, \pu~detection~and cooperation, which are detailed next.

\paragraph{\pu~detection}
\label{pudetection}
The first step towards discovering spectrum opportunities is to detect \pu s' signals. To do so, each \su~needs to sense its local radio environment, as it is generally assumed not to have any prior knowledge about \pu s' activities.
%
%
We now present existing techniques that have been proposed in the literature to detect primary signals.

%
\begin{itemize}
\item {\em Energy detection~\cite{poor2013introduction}}: This is the simplest and the most popular approach for signal detection~\cite{letaief2009cooperative}. It is also considered as the optimal sensing approach when no information about the primary signal is available~\cite{hoven2005some}.
    The presence or absence of a \pu~is decided by measuring \pu~signal's energy (aka the received signal strength (\rss)) on a target channel and comparing it against a detection energy threshold $\lambda$~\cite{fatemieh2011using,ma2009signal}.




\item {\em Matched filter detection~\cite{proakis2001intersymbol}}: It is considered as the optimal signal detection method~\cite{cabric2004implementation,letaief2009cooperative} as it maximizes the signal to noise ratio. It requires a full knowledge of \pu's signal features such as modulation format, data rate, etc. It compares the known signal (aka template) with the input signal to detect the presence of the template signal in the unknown signal. The output of the matched filter is then compared to a predetermined threshold to decide on \pu's presence or absence. 


\item {\em Cyclostationary detection \cite{lunden2009collaborative,letaief2009cooperative}}:
\pu s' transmitted signals are usually cyclostationary, i.e. their statistics exhibit periodicity~\cite{ma2009signal}. Such a periodicity is usually introduced to the primary signals so that receivers can use it for timing and channel estimation purposes. But it can also be exploited for detecting \pu s~\cite{hossain2009dynamic}. \su s can detect this periodicity in the modulated signals by analyzing a spectral correlation function. This spectrum sensing technique is appealing because of its capability of differentiating the primary signal from noise and interference even in very low \snr~environments~\cite{ma2009signal}. 


  \item {\em Wavelet detection \cite{tian2006wavelet,ma2009signal}}:
  This method uses wavelet transform, an attractive mathematical tool used to investigate signal local regularity to analyze singularities and irregular structures in the power spectrum density caused by spectrum usage~\cite{hossain2009dynamic}. Wavelets are used for detecting edges, which are boundaries between spectrum holes and occupied bands, in the power spectral density (PSD) of a wideband channel. 

%
%
\end{itemize}

Most of the above mentioned techniques are based on a set of measurements
sampled at the Nyquist rate and can sense only one band at a time because of the hardware limitations~\cite{salahdine2016survey}. In addition, sensing a wideband spectrum requires dividing it into narrow bands and making \su~sense each band using multiple RF frontends simultaneously~\cite{salahdine2016survey}. This may result in a very high processing time and hardware cost which makes these approaches not suitable for wideband sensing. Compressive sensing~\cite{donoho2006compressed} is proposed to overcome these issues. In compressive sensing theory, a sparse signal can be acquired and compressed simultaneously in the same process with
only the essential information at rates significantly lower than Nyquist rate. This means that the signal can be recovered from far fewer measurements and at a lower rate (below Nyquist rate) compared to that of traditional methods~\cite{salahdine2016survey,sharma2016application}.
As the wideband spectrum is inherently sparse due to its low spectrum utilization, compressive sensing becomes a promising approach to realize wideband spectrum sensing.

\paragraph{Cooperation}
\label{coop}

One widely adopted approach for improving spectrum sensing accuracy is cooperation, where \su s share their local sensing observations and collaboratively make spectrum availability decisions.
These observations can be made using one of \pu~detection techniques discussed in Section~\ref{pudetection}.

The idea behind cooperation is to exploit spatial diversity of spatially distributed \su s to cope with problems, like shadowing, multipath fading, and receiver uncertainty, that may impact individual local observations of \su s~\cite{yucek2009survey}. There have been many cooperative approaches proposed in the literature~\cite{cheng2012energy,ganesan2005cooperative,ganesan2007cooperative,letaief2009cooperative,althunibat2013optimizing}, and cooperative spectrum sensing has been widely adopted in many cognitive radio standards, e.g. WhiteFi~\cite{bahl2009white}, IEEE 802.22 WRAN~\cite{ieee802.22} and IEEE 802.11af~\cite{ieee802.11af}. The collaboration between \su s is usually performed through a control channel~\cite{cabric2004implementation} and could be realized in two major ways: centralized and distributed~\cite{akyildiz2011cooperative}.

\begin{figure}[h!]
  \vspace{-5pt}
  \centering
  \subfigure[\small Centralized]{\includegraphics[width=0.24\textwidth]{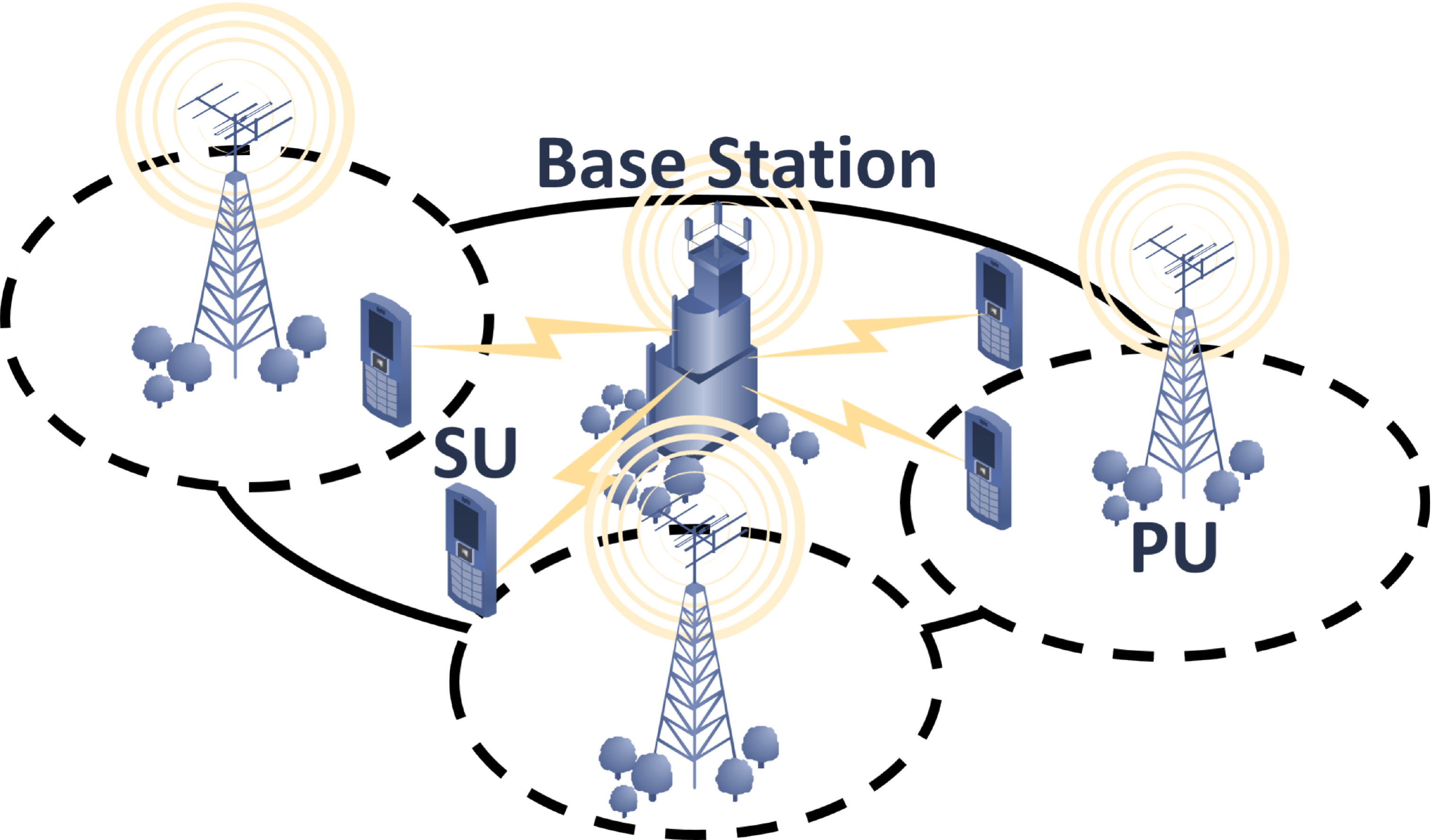}\label{centralized}}
  \subfigure[ \small Distributed]{\includegraphics[width=0.24\textwidth]{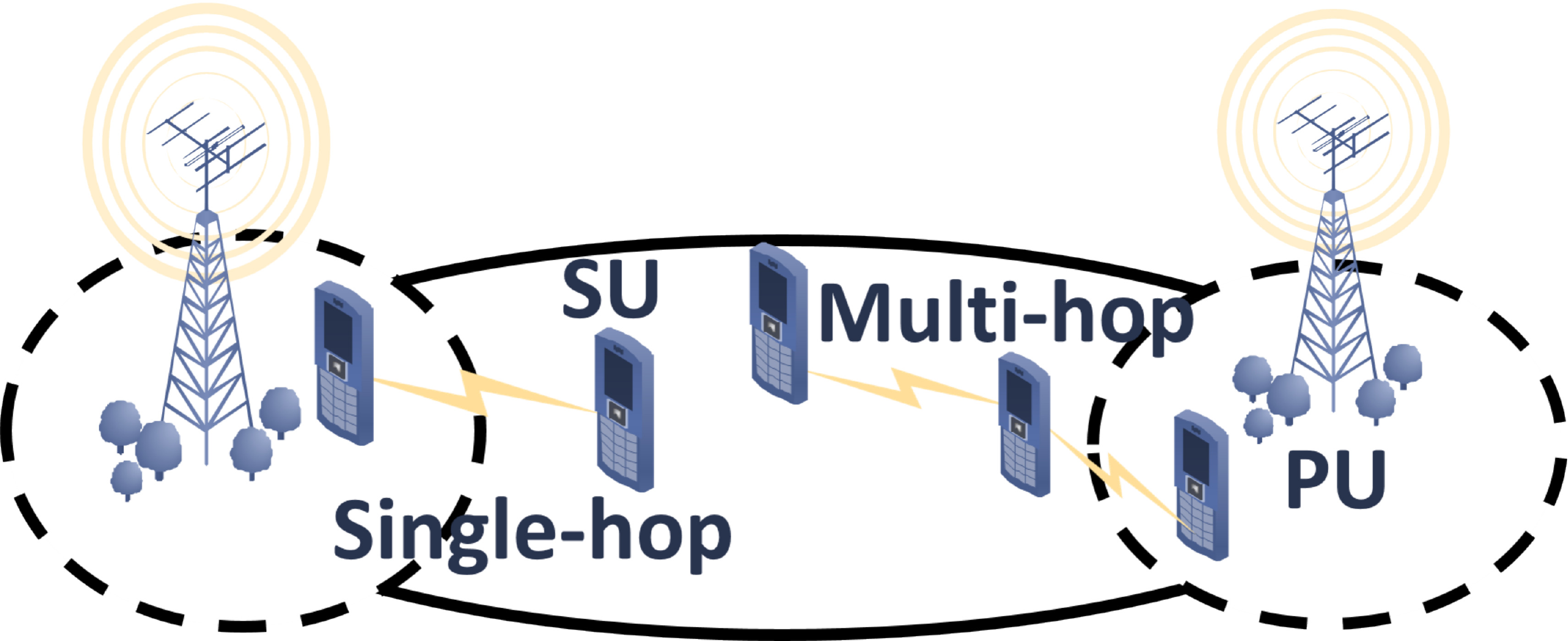}\label{distributed}}
  \vspace{-5pt}
  \caption{Cooperative spectrum sensing} \label{}
\end{figure}

In the centralized approach a central entity, referred to as {\em fusion center} (\fc), orchestrates the cooperative sensing task among \su s through a control channel as shown in Figure~\ref{centralized}. \fc~collects the local observations from \su s and combines them to determine \pu's presence on a specific channel.
In the distributed approach, \su s do not rely on \fc~for making channel availability decisions. They instead exchange sensing information among one another to come to a unified decision~\cite{grissa2017preserving,akyildiz2011cooperative} as shown in Figure~\ref{distributed}

%

Another promising approach for enabling effective cooperative spectrum sensing over a large geographic area is to exploit the emerging {\em crowdsourcing}
paradigm, in which spectrum service providers outsource spectrum sensing tasks to distributed mobile users~\cite{fatemieh2010secure,fatemieh2011using,zhang2013secure,feng2014trac,jin2016privacy}.
Crowdsourcing is formally defined as the act of taking a job traditionally performed by a designated agent and outsourcing it to an undefined, generally large group of people in the form of an open call. This concept has been adopted in many contexts~\cite{sun2016securefind}, and has been first applied to \crn s by Fatemiah et al.~\cite{fatemieh2010secure}.

The use of crowdsourcing for enabling spectrum sensing is motivated by several facts and trends. First, according to a recent Cisco report~\cite{cisco2016}, the number of mobile-connected devices is expected to hit $11.6$ billion. This huge number guarantees sufficient geographic coverage, especially in highly populated regions such as metropolitan areas~\cite{zhang2013secure} where {\em dynamic spectrum access (\dsa)} systems are expected to play important roles~\cite{jin2016privacy}. Moreover, future mobile devices are widely expected to be able to perform spectrum sensing tasks given the expected pervasiveness of \dsa~future wireless systems~\cite{nika2014towards,zhang2013secure}. Finally, mobile devices are
increasingly equipped with more powerful communication and
computation resources, and are enabled with self-localization capabilities, making mobile crowdsourcing even more appealing and attractive~\cite{jin2016privacy}.

The cooperative spectrum sensing process is usually performed by a specified set of nodes that are considered to be trustworthy~\cite{fatemieh2010secure}. Crowdsourcing-based cooperative spectrum sensing, on the other hand, is to be realized by gathering and combining sensing reports from a large group of nodes that could be unreliable, untrustworthy, or even malicious~\cite{fatemieh2010secure}, thereby giving rise to new challenges.

Another important challenge that arises from \su s' cooperation nature is how to combine the various \su s' sensing results or observations for hypothesis testing to decide on the presence of primary signals in an accurate manner. This process consists of sending the sensing results to \fc~or to the neighboring \su s, depending on the topology, to make spectrum availability decisions. It is referred to as data fusion and can be done in one of two ways: soft combining and hard combining~\cite{teguig2012data}. In soft combining, local sensing reports, measured locally by \su s from their received signals, are combined together to compute some statistics using combining rules such as square law combining (SLC), maximal ratio combining (MRC) and selection combining (SC)~\cite{teguig2012data}. In hard combining, \su s make decisions about the availability of the spectrum locally, and share their one-bit decision (i.e., available or not available) outputs to make a voting decision about spectrum availability~\cite{teguig2012data}.

\subsubsection{Geolocation database}
\label{db}
\begin{figure}[h!]
  \centering
  \subfigure[without \bs]{\includegraphics[width=0.24\textwidth]{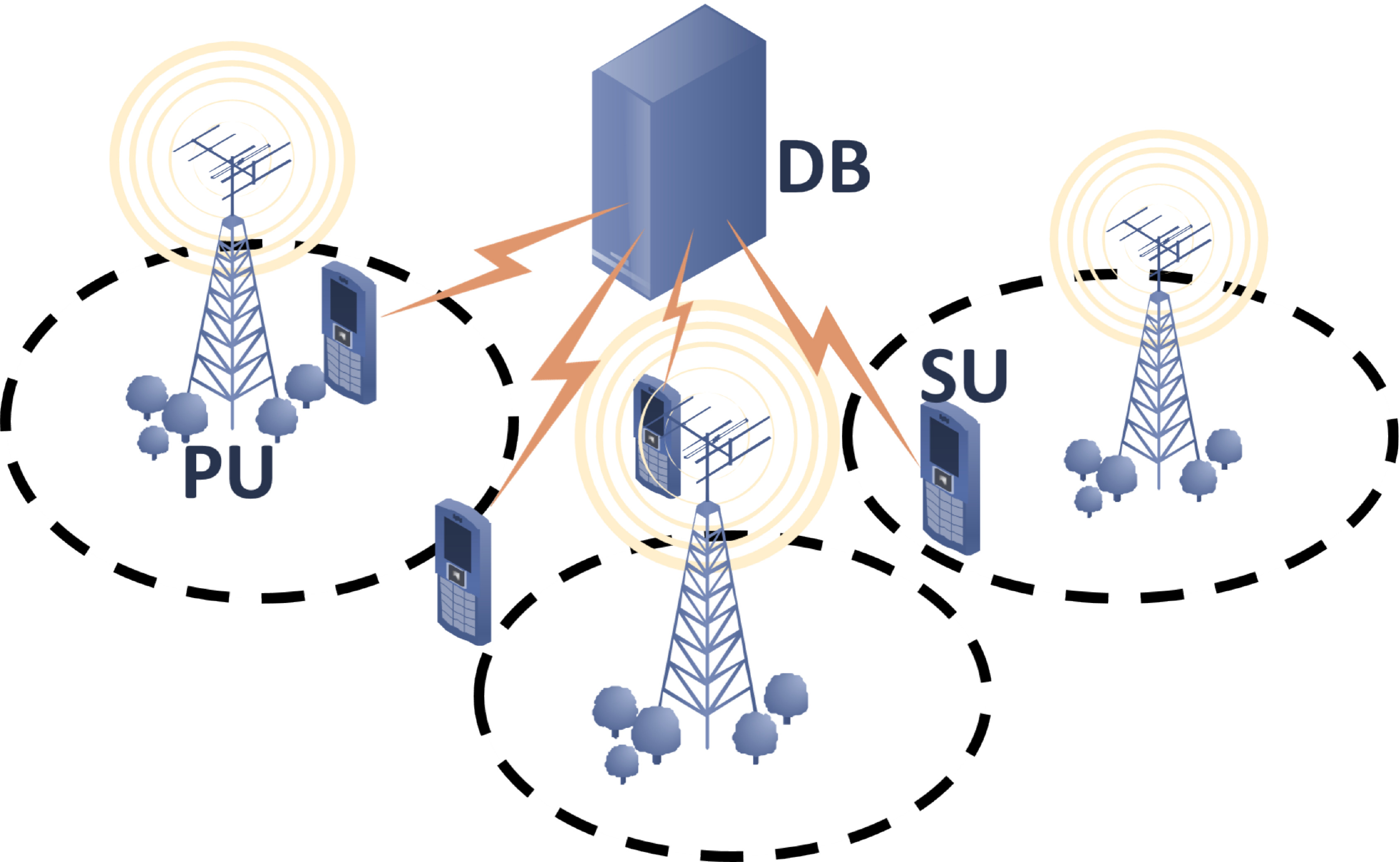}\label{databaseDirect}}
  \subfigure[with \bs]{\includegraphics[width=0.24\textwidth]{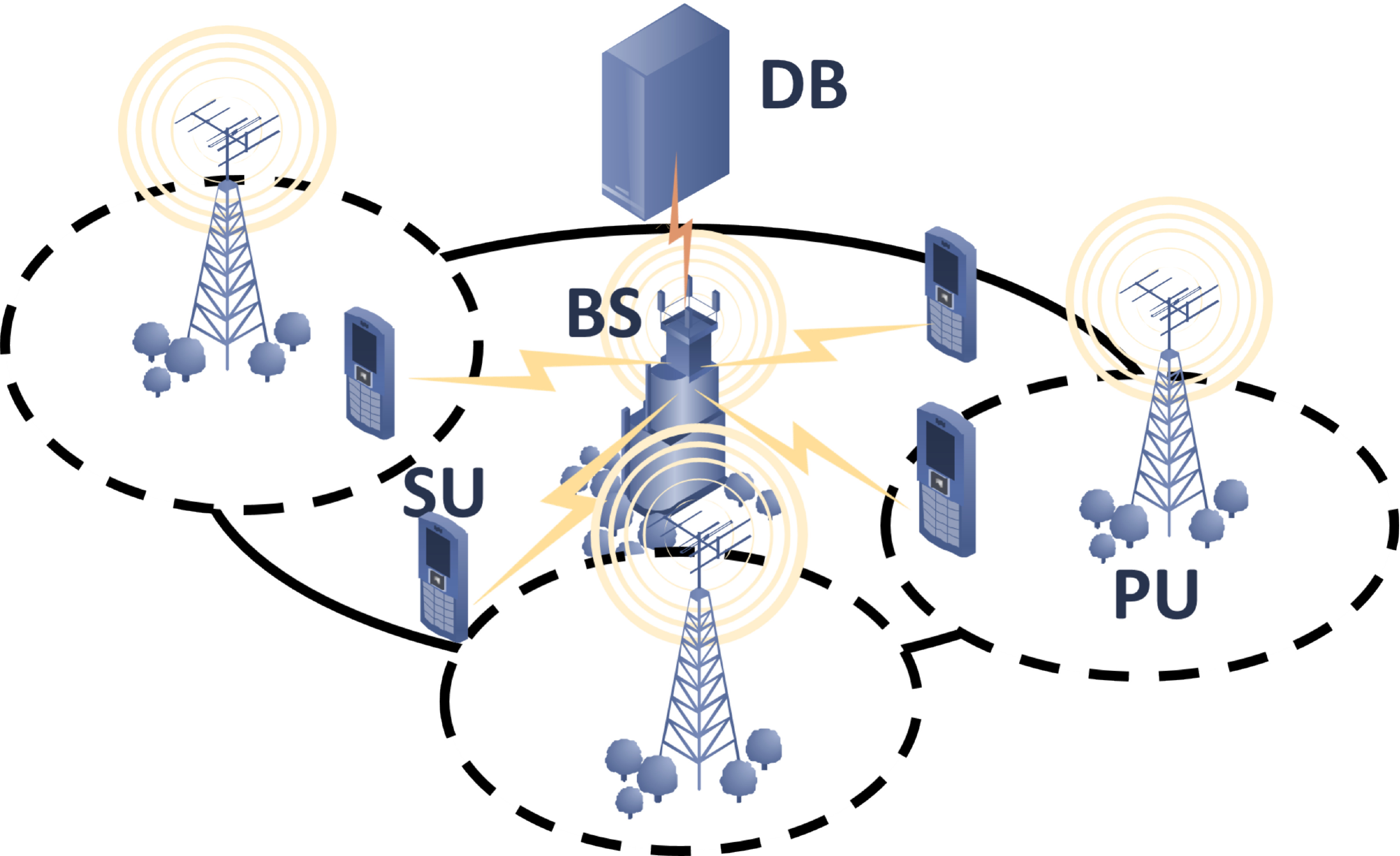}\label{databaseBS}}
  \vspace{-5pt}
  \caption{\small Spectrum database-based \crn~topologies}
\end{figure}

This is another approach for spectrum opportunity discovery that was recommended recently by FCC~\cite{federal2012third}.
A typical database-driven \crn~\cite{murty2012senseless,grissa2015cuckoo} consists of a geolocation database (\db) containing spectrum availability information, a set of \su s and a set of \pu s as shown in Figure~\ref{databaseDirect}. To learn about spectrum opportunities in its vicinity, a \su~is not required to detect the primary signal by itself anymore. Instead, it needs to query \db~by including its exact location in the query. \db~then replies with a set of available channels in \su's location and with the appropriate transmission parameters (e.g. transmit power) for each channel to avoid interfering with the incumbents. Afterwards, depending on the situation, \su~may optionally inform \db~of its choice and registers the channel it is planning to operate on during what is referred to as notification or commitment phase~\cite{gao2013location,zhang2015optimal}. \db~keeps track of this information to have more visibility over the \crn~and make its decision adaptively which allows it to reduce interference among \su s. \su s may be able to communicate directly with \db~as in Figure~\ref{databaseDirect} or via a fixed base station that relays their queries to \db~as in Figure~\ref{databaseBS}.

%

\subsubsection{Sources of location information leakage}

In this Section, we investigate the different vulnerabilities in the spectrum opportunities discovery phase and the potential threats that could exploit them in order to localize \su s. We first begin by exploring the sources of leakage in the cooperative spectrum sensing approach, and then we explore those in the database-based approach.

\paragraph{Cooperative spectrum sensing} \label{coopSources}
In the cooperative spectrum sensing approach, \su s need to communicate with other entities in the \crn~to exchange and share their observations of the spectrum. This collaboration may lead to a significant leakage of information regarding the location of the collaborating \su s. In the following, we investigate and discuss the different vulnerabilities that arise from the cooperation process.


\textbf{Wireless signal:}
This is the most obvious and direct source of leakage in wireless networks in general and in \crn s in particular. The wireless medium and its inherent open and broadcast nature in \crn s makes it much easier for an attacker to compromise a \su's privacy and more specifically its location~\cite{zekavat2011handbook,sithamparanathan2012cognitive}. Despite the many efforts to protect the private location information at the system level, mainly using encrypted signal transmissions, the signal itself can still be used to potentially localize a \su. Classic approaches for localization are usually based on a small set of measurements on the wireless signals, that include time-based ranging, received signal strength (\rss) and angle of arrival (\aoa)~\cite{zekavat2011handbook}.

\begin{itemize}
\item Time-based ranging: This is used to estimate the distance between two communicating nodes by measuring the signal propagation delay, known also as time-of-flight, $\tau_F = d/c$, where $d$ is the actual distance between the nodes and $c$ is the propagation speed ($c\simeq 3.10^8 m/s$)~\cite{sithamparanathan2012cognitive}. This can be accomplished using either time-of-arrival (\toa) or time difference-of-arrival (\tdoa). If at time $\tm_1$ the victim node sends a packet that contains the timestamp $\tm_1$ to a semi-honest or malicious node that receives it at time $\tm_2$, then the latter node can estimate the distance that separates it from the victim node based on $\tau_F = \tm_1 - \tm_2$. This technique is known as \toa~ranging~\cite{sithamparanathan2012cognitive,zekavat2011handbook}.  \toa~needs at least three measurements of distance to localize the target via triangulation~\cite{vossiek2003wireless}, which means that a malicious entity cannot localize precisely a target \su~unless it is mobile or it collaborates with two other malicious entities. \tdoa, on the other hand, does not rely on the absolute distance between a pair of nodes but rather on the measurement of the difference in time between signals arriving at two base nodes.

\item Received signal strength (\rss)-based ranging: In theory, the energy of a radio signal decreases with the square of the distance from the signal’s source. As a result, a node listening to a radio transmission should be able to use the strength of the received signal to estimate its distance from the transmitter~\cite{bachrach2005localization}. More details about the practicality of \rss-based ranging technique and its feasibility given various factors could be found in~\cite{whitehouse2007practical}.

\item Angle of arrival (\aoa)-based ranging: \aoa~could be defined as the angle between the propagation direction of an incident wave and some reference direction known as orientation~\cite{peng2006angle}. The estimation of \aoa~could be done using directive antennas or using an array of uniformly separated receivers~\cite{boukerche2007localization}. The relative angle could then be used to derive the distance between the two communicating nodes~\cite{bachrach2005localization}.

\end{itemize}

\textbf{Observations:}
The spectrum sensing measurements and observations that \su s share to identify spectrum holes may be another source of location information leakage in \crn s. In the case of soft combining rule where \su s have to share their raw measurements, \su s may see their location information exposed. Indeed, it has been shown in~\cite{fatemieh2011using,li2012location} that the sensing reports containing \rss~measurements on \pu s' signal, are highly correlated to \su s' physical location. The \rss~measurements could be used to localize \su s with respect to \pu s whose channels are sensed through these measurements. Note that this \rss~is different from the \rss~that we discussed previously for wireless signal which are obtained through a direct communication through the wireless medium between the adversary and the target victim. If the \crn~uses a hard combining rule for the cooperative sensing, \su s need just to share their binary decision values. This can still leak some information about \su s' location as it can tell whether a \su~belongs to the coverage area of a \pu~especially if the activity of \pu~is known by the attacker.

\textbf{Identity:}
One cannot talk about a location information leakage if the identity of the target victim is not revealed. Therefore, the identity of the user could also be considered as a source of location information leakage in the way that an attacker can match this identity to a specific location. In other words, if an attacker learns that a \su~is located at a specific location but at the same time fails to identify who it is, the location privacy of \su~cannot be considered as compromised. So, as long as a \su~is anonymous, its location privacy is preserved. In some cases, identity could also give an idea about the location of a \su~if combined with publicly known information of this specific \su. Take the example of a user whose identity is revealed. Based on this information, an adversary can learn the profession of this user, for instance a doctor that works at a specific hospital. This allows an attacker to estimate the position of the target \su~with high probability especially during regular working hours. This shows that the identity could be associated with a specific location of a \su.

\textbf{Radio hop count:}
The sensing information needs to be delivered to the appropriate nodes for the final decision, especially in multi-hop \crn s which requires deploying efficient routing protocols. These routing protocols usually rely on hop count information~\cite{chowdhury2009search,youssef2014routing}, and such information turns out to be another potential source of location information leakage~\cite{bachrach2005localization}. Many approaches are proposed in the literature, especially in the context of wireless sensor networks, to estimate node position based on the number of hops between pairs of nodes~\cite{niculescu2001ad,rabaey2002robust}.

\textbf{Clustered network:}
\su s may need to form or join different clusters during the spectrum sensing phase in order to improve the overall sensing performance and overhead. Different approaches are proposed in the literature for cluster formation in \crn s based on several criteria and metrics including geographical location, channel availability, signal strength and channel quality~\cite{yau2014clustering}. This clustering could leak information about \su s' location especially if the clustering criteria is based on the positions of \su s or on some information correlated to this position. Chang et al.~\cite{chan2005using} show that the clustering information along with some knowledge of the position of some anchor nodes in wireless sensor networks can lead to localizing the remaining nodes in the network. The same idea could be exploited in the context of \crn s in case, for example, that some \su s are compromised and their location is known to the adversary. In that case, the adversary can localize the remaining \su s. Moreover, if the clusters are also overlapping, this could further facilitate localization as shown by Youssef et al. in~\cite{youssef2006wsn16}.

\textbf{Signal-to-noise ratio ($\boldsymbol{\snr}$):}
\su s may need to share their measured \snr s, with respect to the channels of interest, with other \su s to cooperate in forming coalitions and selecting the decision making nodes in ad hoc \crn s~\cite{hao2011coalition}. The average \snr~of \pu's received signal measured at \su~$i$ is given by:
\begin{equation}
\overline{\snr}_{i,\pu} = \frac{P_\pu \cdot\kappa}{d_{\pu,i}^\alpha\cdot \sigma^2}
\label{snr}
\end{equation}
with $P_\pu$ is the transmission power of \pu, $\sigma^2$ denotes the Gaussian noise variance, $\kappa$ is a path loss constant, $\alpha$ is the path loss exponent and $d_{\pu,i}$ is the distance between \pu~and \su~$i$~\cite{saad2009coalitional}. As Equation~\eqref{snr} shows, the \snr~value measured by a \su~depends on the distance that separates it from the corresponding \pu. This could present a source of location information leakage as this information could be exploited to localize \su~especially when it has to share its \snr~with other \su s in the same coalition~\cite{kasiri2015privacy}.

The vulnerabilities and sources of leakage that we have raised previously could lead to serious location privacy risks for \su s if exploited by malicious entities in the \crn. This leakage could happen in the following scenarios:

\begin{itemize}

 \item \textbf{Cooperation and sharing observations:} In order to participate in the cooperation for spectrum sensing, \su s need to share their observations of the spectrum either with other \su s or with a central entity. Despite the fact that sharing this information considerably improves the spectrum sensing performance, it exposes, however, individual \su s observations to other entities in the network. This becomes problematic if, during the sharing process, an external or internal malicious entity to the network gains access to these observations. This is due to the fact that these observations could be exploited to compromise \su s' location privacy as discussed earlier.

 \item \textbf{Dynamism:} Due to the dynamic nature of \crn s, \su s  can leave or join the collaborative spectrum sensing task at anytime, making privacy-preserving aggregation
techniques designed for static networks to hide individual observations of \su s unsuitable for \crn s. Indeed, this might allow a malicious entity that is collecting aggregated observations from \su s to estimate individual observations of leaving or joining \su s which, as discussed previously, is a source of location information leakage.

 \item \textbf{Node failure:} The location privacy issue here is very similar to the situation of network dynamism. Indeed, if, for some reason, some \su s cannot sense the spectrum or fail to share their sensing reports during the cooperation process, the individual observations of these \su s could be estimated. Again, these individual observations could be exploited by an adversary for localization purposes.

 \item \textbf{User selection:}
User selection is an important step in cooperative spectrum sensing, which aims to optimally select the
cooperating \su s that lead to the maximization of the cooperative gain and the minimization of the cooperation overhead. \su s are selected such that all the sensing reports are informative and not correlated while saving energy by avoiding unnecessary sensing operations~\cite{akyildiz2011cooperative}. This selection could be done through a clustering approach that divides \su s into different clusters. Malady et al.~\cite{malady2008clustering} propose several approaches for grouping \su s into clusters in distributed \crn s to keep bandwidth and power requirements manageable. Their methods are based on different criteria including \su s' positions with respect to a given reference or to \pu~if \pu's position is known. In~\cite{ding2012decentralized}, Ding et al. propose a decentralized clustering-based user selection algorithm that relies on unsupervised learning to group \su s with best detection performance together to lead the sensing process. As discussed previously, the clustering information could be exploited to localize \su s during the cooperative spectrum sensing process. Another way for selecting \su s for spectrum sensing, which has just started to gain some interest in the context of \crn s, is {\em crowdsourcing} as we have explained earlier. Crowdsourcing may, however, give rise to some privacy risks, especially in terms of location privacy as shown by Jin et al.~\cite{jin2016privacy}. The selection process in this case relies on an open call, made by \fc, for users in order to contribute to the sensing at a specific location. This makes it easy for \fc~to associate users with the location of interest.

\end{itemize}

\paragraph{Geolocation database}
\label{sourcesDB}
With this architecture, \su s are not anymore required to perform spectrum sensing to learn about spectrum opportunities. Instead, they only need to query a geolocation spectrum database to get the list of available channels in their vicinity. This brings new privacy challenges that are completely different from the ones that emerge from the cooperation in spectrum sensing. In the following, we investigate the different sources of location information leakage that may arise from this specific \crn~architecture.

    \textbf{Query:}
    This is the most implicit source of location information, as every \su~needs to include its precise location every time it queries \db~for available channels. This information is usually sent in a plaintext form, allowing eavesdroppers to retrieve it. And even if the communication channel between \su s and \db~is authenticated; i.e. it eliminates the risk of an eavesdropper, there is still the risk of having a malicious \db.

    \textbf{List of available channels in the query's response:}
    This information could also be used by an adversary to narrow down the locations where a target \su~might possibly be. Indeed, knowing which channels are available for a certain \su~allows a malicious entity to attribute this \su~to multiple \pu s coverage areas especially if the adversary, \db~for example, is aware of these \pu s' activities and status.

   \textbf{ Maximum transmit power ($\boldsymbol{MTP}$):} The $MTP$ over a specific spectrum band is included in \db's response to \su, and is assigned to it based on its distance from its corresponding \pu. It is usually calculated as follows
\begin{equation}
P = \begin{cases} 0, & d \leq r_0 \\ h(d-r_0) , & d > r_0 \end{cases}
\label{mtp}
\end{equation}
where $d$ is the distance between the querying \su~and its closest \pu, $r_0$ is the protected contour radius of the channel of interest and $h(.)$ is a continuous, monotonically increasing function. As shown in Equation~\eqref{mtp}, $MTP$ is highly correlated to the distance of \su~from \pu. In situations where  \pu s' positions are publicly known, an attacker could exploit $MTP$ values that \su s receive from  \db~to infer \su s' locations.

These vulnerabilities and sources of leakage could become actual threats when exploited solely or combined together, and can occur in the following scenarios:
\begin{itemize}
\item \textbf{Querying $\boldsymbol{\db}$:} When a \su~interacts with \db~to learn about spectrum availability, its location can easily be revealed as it is included in the query. Even if, somehow, a privacy-preserving scheme is implemented to make \db~unable to retrieve \su's location information from its query but at the same time can still provide it with the spectrum availability information at its vicinity, an adversary can still localize \su~by exploiting the information included in \db's response as we discuss next.

\item \textbf{$\boldsymbol{\db}$'s response:} \db's response to a \su's query includes information like the list of available channels, and the maximum transmit power over each of those channels. This information could be used as explained earlier by a malicious \db~or an external adversary to infer the location of a target \su.

\item \textbf{Commitment phase:} Some implementations of the database-based \crn s require that a \su, upon receiving the response from \db, informs \db~about the channel that it chooses to operate on. This will make \su's usage information available at least to \db. Hence, \su s in database-based \crn s will be prone to attacks that could exploit the vulnerabilities arising from spectrum utilization information as we explain in Section~\ref{sourcesMobility}.

		\end{itemize}

\subsection{Location information leakage in spectrum analysis}
\label{sourcesDec}

This is an important step in the cognition cycle as it allows to analyze the information obtained from spectrum sensing to gain knowledge about spectrum holes (e.g. interference estimation, and
duration of availability). Spectrum analysis usually consists of two major components: spectrum characterization, and reconfiguration. In this section, we explain each of these two components and discuss their sources of location information leakage.

\subsubsection{Spectrum characterization}
Available spectrum bands may have different channel characteristics that vary over time. In order to determine the most suitable spectrum band, one needs to characterize these channels. Such a characterization requires the monitoring and observation of the RF environment, as well as the monitoring and awareness of \pu s activities in these channels~\cite{masonta2013spectrum}.

\paragraph{RF environment characterization}
This process estimates some of the following key parameters to characterize the different spectrum bands.

\begin{itemize}
\item {\em Interference:} It is crucial to estimate and model the interference caused by \su s at the primary receiver to derive the permissible power of a \su~and ensure coexistence between \su s and \pu s. Rabbachin et al.~\cite{rabbachin2011cognitive} propose a statistical model for aggregate interference generated by \su s in a limited or finite region by taking into consideration the shape of the region and the position of \pu. The interference signal at \pu~generated by the $i^{th}$ \su~is modeled as~\cite{rabbachin2011cognitive}:
\begin{equation}
\label{interference}
I_i = \sqrt{P_I} R_i^{-b} X_i
\end{equation}
where $P_I$ is the interference power at the near-far region limit; $R_i$ is the distance between the $i^{th}$
\su~and \pu; and $X_i$ is the
per-dimension fading channel path gain of the channel from
the $i^{th}$ \su~to \pu.

\item {\em Path loss:} This is closely related to distance and frequency. Path loss increases as the operating
frequency increases, resulting in a decrease in the transmission
range. Increasing the transmission power may be used to compensate for the increased path loss, and hence for the decrease in transmission
range. But this may increase interference at other \su s and \pu s. According to~\cite{rappaport1996wireless}, the average path loss of a channel could be expressed using a path loss exponent $\alpha$. This exponent measures the rate at which the \rss~decreases with distance, and its value depends on the specific propagation environment~\cite{mao2007path}. It is also considered as a key parameter in the distance estimation based localization algorithms, where distance is estimated from the \rss~\cite{dantu2005robomote}.

\item {\em Channel switching delay:} This is basically the delay introduced by switching from one channel to another. In \crn s, the channel switching could be triggered by several events, such as the detection of \pu s, the return of \pu s to their channels, and/or the degradation of received \qos~in the current channel, as we discuss in Section~\ref{specMobility}.

\item {\em Channel holding time:} It is the expected duration \su s can occupy a licensed channel before getting interrupted.

\item {\em Channel error rate:} This is defined as the rate of data elements incorrectly received from the total number of data elements sent during a time interval. This rate may vary depending on the modulation scheme and the interference level of the channel~\cite{masonta2013spectrum}.

\end{itemize}

\paragraph{\pu~activity modeling}
As spectrum availability depends not only on the RF environment characteristics but also on the activities of \pu s, it is crucial that \pu~activities are taken into account when characterizing the spectrum bands. This is essentially done by accounting for how long and how often \pu s appear on their licensed spectrum bands. Existing approaches adopted for modeling this activity mainly rely on measured data obtained from the numerous spectrum measurement campaigns that have been conducted worldwide to quantify and study the \pu~spectrum utilization and assess the current status of the spectrum~\cite{chen2016survey,saleem2014primary,hoyhtya2016spectrum}. These measurements are also performed to improve the accuracy of spectrum databases.
Many of these works consider only simple but important statistics of the spectrum occupancy, such as the maximum or the minimum and the average of power levels, the spectrum occupancy, and the duty cycle~\cite{chen2016survey}. These statistics are simple and reliable, but provide an incomplete model of the \pu s' activities. Other approaches consider more advanced statistical models, such as probability
function models (e.g. CDF and PDF), Markov chains and linear regressions. These measurement-based modeling methods describe the statistical behaviors of the spectrum occupancy as a whole, but do not give the actual state
of the spectrum occupancy, i.e. whether a channel is busy or available.

Some other significant research models the
\pu~activity as a Poisson process with exponentially distributed
inter-arrivals~\cite{lee2011spectrum,saleem2014primary}. However, such approaches fail to capture the short-term temporal fluctuations or variations exhibited by the \pu~activity, and do not consider correlations and similarities within the monitored data~\cite{saleem2014primary}.

There are also approaches that try to predict future \pu~activities and thus locate future spectrum opportunities by using learning techniques and by exploiting the history of spectrum band usage~\cite{masonta2013spectrum,saleem2014primary}. However, the prediction may go wrong resulting in harmful interference to \pu s.


%

\paragraph{Sources of location information leakage}
\label{sourcesCharact}
As mentioned earlier, spectrum characterization consists of building knowledge about the radio environment and \pu~activities. This knowledge, however, could be exploited (maliciously or un-maliciously) to leak location information of \su s, as discussed next.


\textbf{Interference:} As shown in Equation~\eqref{interference}, the interference is highly correlated to the distance that separates \su~from a \pu. An adversary that has access to the characteristics of the interference caused by \su s can exploit this information to estimate the distance that separates \su~from a \pu.

\textbf{Radio environment map ($\boldsymbol{\rem}$):} This is a widely used method to characterize the spectrum. It is an integrated database that could be deployed in \crn s to store information about the radio  environment's interference, signal properties, geographical features, spectral regulations, locations and activities of radios, policies of \su s and/or service providers, and past
experiences~\cite{zhao2006overhead,zhao2006network}. The main functionality of a \rem~is the construction of dynamic interference map for each
frequency at each location of interest. This could be done in two different ways, either via field measurements or via propagation modeling. In the first approach, a \rem~collects spectrum measurements from nodes with spectrum sensing capabilities. These nodes could be actual \su s or dedicated spectrum sensors~\cite{yilmaz2013radio}. Since it is impractical to have measurements all the time at all possible locations, \rem~fuses
the collected measurements to estimate the
interference level at locations with no measurement data by means of spatial and temporal interpolation~\cite{yilmaz2013radio}.
The field measurement approach is believed to provide the highest location accuracy but not without a price. Its price lies in the need to perform drive tests whenever changes occur in the radio environment to keep the \rem~up to date. The second approach, propagation modeling, relies on mathematical models for radio propagation prediction, which allow easy, fast and inexpensive updating for the \rem. Indeed, whenever there is a change in the radio environment, we only need to rerun the propagation models with the new parameters to update the \rem~\cite{zekavat2011handbook}.

This is different from the spectrum geolocation database in that \rem~generates spectrum map by processing the data collected from multiple sources with its cognitive engine, and therefore can easily adapt to dynamic operating environments whereas \db~stores quasi-static information. \rem~introduces environment awareness that would be harder to acquire by individual \cogr~capabilities via extensive spectrum analysis.
Hence, \rem~can also be seen as the network support turning simple nodes into intelligent ones~\cite{yilmaz2013radio}.

This radio map, when it is in the hands of some malicious entity in the network, could be exploited to localize a querying \su~that sends its measurement to the REM manager in order to learn about spectrum availability. One way to exploit this information is based on fingerprinting localization technique which basically estimates the target position by simply finding the best-matched pattern or fingerprint for the measurement provided by \su~within a certain map~\cite{zekavat2011handbook}. Machine learning techniques could be used to build the radio signal map during the training phase and then to compare the online measured \rss~to the preconstructed map during the localization phase~\cite{zekavat2011handbook}. Obviously the map that could be used for the localization is the REM itself. As the REM could be used in a distributed or a centralized manner, either a malicious \bs~or a malicious \su~could exploit it to localize a target \su.


\subsubsection{Reconfiguration}
After the channel of choice has been characterized, \su's transceiver parameters have to be reconfigured to adapt to channel conditions and satisfy the \qos~requirements and regulatory policies. These parameters include:

\begin{itemize}
\item Transmission power: Controlling this parameter aims to achieve several objectives that include minimizing energy usage, reducing co-channel interference, etc.~\cite{tragos2013spectrum,hoven2005power}.

\item Operating frequency: This parameter represents the capability of \su s to reconfigure their central frequency in response to variations in the RF environment.

\item Channel bandwidth: This refers to the width of the spectrum over which a \su~operates. It is essential for \su s to have variable channel adaptation capabilities to be able to operate in heterogeneous networks.

\item Communication technology: This allows interoperability between different communication technologies such as GSM, LTE, etc.
\end{itemize}

\textbf{Sources of location information leakage:}
Some of the reconfigurable parameters that we have listed could leak some information about \su s' location especially if these parameters are controlled in a shared way.
\begin{itemize}

\item {\em Power control}: This process may present a threat to \su s' location privacy. Most of the existing approaches for power control rely on the {\em signal-to-noise ratio} (\snr) or the {\em signal-to-interference-plus-noise ratio} (\sinr) metric when solving the power control problem~\cite{islam2008joint,ghorbel2015power,hoven2005power,chakchouk2011traffic}. For example, Hoven et al.~\cite{hoven2005power} use local \snr s of primary signals measured by \su s as a metric to design an effective power control rule. Other works use \sinr~as a constraint or a requirement to minimize the total transmission power of the \crn~as in~\cite{islam2008joint} and maximize the spectrum utilization of the \crn~as in~\cite{hoang2006maximizing}. Yang et al.\cite{yang2010optimal} model this problem as a game with \sinr-based utility function. Power control might become threatening to the privacy of \su s as information like \snr~and \sinr~is usually correlated to the distance that separates a \su~from a \pu. This is problematic especially when the power control process is intended to achieve a system-level goal like minimizing the total transmission power~\cite{islam2008joint} or maximizing the overall spectrum utilization~\cite{hoang2006maximizing} of \crn s. In that case, power control will have to be performed jointly between \su s in a centralized~\cite{islam2008joint,qian2007power} or distributed~\cite{islam2008joint,hoang2006maximizing,yang2010optimal,qian2007power} way, thereby exposing local \snr~and \sinr~values, for example, to other \crn~entities or intruders, putting \su s' location information at risk.

\end{itemize}
\subsection{Location information leakage in spectrum sharing}
\label{sourcesSharing}
Multiple \su s may try to access the same spectrum bands at the same time, thus necessitating multiple-access coordination mechanisms that allow multiple \su s to share the same spectrum~\cite{jiang2015effective}. Spectrum sharing consists then of enabling coexistence of multiple \su s while avoiding interference (among \su s themselves as well as between \su s and \pu s) and maintaining some target \qos~levels. Broadly speaking, this functionality is composed of three elements: resource allocation, spectrum access and spectrum trading.

\subsubsection{Resource allocation}
Enabling dynamic spectrum sharing is crucial to the success of \crn s. It allows users to select, use, and share spectrum bands adaptively with the aim of maximizing the overall spectrum utilization efficiency while not causing harmful interference to legacy users~\cite{tragos2013spectrum,nie2007game,hamdaoui2009adaptive,peng2006utilization,khalfi2015dynamic}. In this section, we discuss two resource allocation functions: {\em spectrum selection and assignment} and {\em power control and beamforming}.
%

\paragraph{Spectrum selection and assignment}
\label{selectionassignment}
Once the spectrum holes are analyzed and characterized, the most suitable channel is selected based on \qos~requirements of \su s, as well the characteristics of the channels~\cite{peng2006utilization,ghorbel2014resources,tragos2013spectrum}. Several criteria may be taken into account while assigning spectrum bands to \su s. These include minimizing interference to \pu s, maximizing overall spectrum efficiency, maximizing \su s' throughput, minimizing network delay, and increasing network connectivity, just to name a few~\cite{khalfi2015distributed,tragos2013spectrum}.
Spectrum assignment could be done in a centralized or a distributed way, and there have been many proposed approaches, both centralized and distributed, that address the spectrum assignment and selection problem in \crn s~\cite{tragos2013spectrum,nie2007game,ehsan2012radio,hamdi2015implementation,nie2006adaptive,bkassiny2013survey,alsaleh2011enabling,tan2012channel}. Generally speaking, these approaches are mainly based on one of the four mature concepts: graph theory, game theory, learning and adaptation, and optimization theory. Next, we explore these four concepts and investigate the sources of location information leakage that may arise from using them.

\subparagraph{Graph theory}
Graph theory has been extensively used to address the spectrum assignment problem especially when the structure of the \crn~is assumed to be known a priori~\cite{tragos2013spectrum}. Here the network is modeled as a graph, where the vertices usually represent \su s and the edges model the connection between these \su s. To solve the graph-based spectrum assignment problem, network conflict graphs and graph coloring are widely used~\cite{tragos2013spectrum}.

\begin{itemize}

\item Network conflict graph: This models and captures the interference among \su s caused by concurrent transmissions of nearby \su s communicating on the same or neighboring channels~\cite{tragos2013spectrum}.
    The vertices of the graph represent the communication links among \su s, whereas the edges represent the pairs of links whose concurrent communications interfere with one another when assigned the same or adjacent spectrum bands~\cite{tragos2013spectrum,teotia2015conflict,peng2006utilization}. Conflict graphs are mostly used in centralized topologies, where a central entity (\bs~or \fc) constructs the graph and uses it to assign channels among \su s.

\item Graph coloring: In this approach, the \crn~is mapped to a graph that could be either unidirectional or bidirectional depending on the algorithm's characteristics. The vertices in this graph represent \su s that need to share the spectrum, and the edges model the interference between \su s. \pu s could also be included in the graph with pre-assigned colors. The spectrum assignment problem using graph coloring is equivalent to coloring each vertex (or edge) using different colors from a specific set of colors, each often representing an available spectrum band~\cite{tragos2013spectrum,yang2009historical,peng2006utilization}. The goal is to improve spectrum efficiency by increasing frequency reuse while meeting interference constraints by ensuring that two connected vertices (\su s) cannot be assigned the same color, i.e. the same band.
\end{itemize}

\textbf{Sources of location information leakage:} We identify two main sources of leakage that arise from graph-based approaches during the spectrum selection process: the topology and the connectivity information.
\begin{itemize}
\item {\em Topology:} The topology of the network that could be learned via the graph-based spectrum assignment techniques could be explored to infer \su s' location. In fact, some works have already used this information to localize nodes in wireless sensor networks~\cite{priyantha2003anchor,wymeersch2009cooperative}.

\item {\em Connectivity:} This information basically tells which nodes are located within each other's transmission range (i.e., connected to one another). Many approaches have used this information for positioning purposes~\cite{shang2003localization,shang2004localization,lederer2009connectivity,wang2009connectivity} and some of them can be used to localize target nodes even from connectivity information among the nodes themselves only~\cite{shang2003localization,shang2004localization}.
\end{itemize}

\subparagraph{Game theory}
Game theory has also been extensively used to solve the spectrum assignment problem in \crn s~\cite{wang2010game,nie2007game,nie2006adaptive}. 
A game could be seen as a way of interaction between multiple players competing with each other while trying to adjust their strategies to optimize their utilities~\cite{hossain2009dynamic}.
Game theory is suitable for the spectrum assignment problem in \crn s as the spectrum allocation decision of one \su~has a direct impact on the performance of other neighboring \su s~\cite{tragos2013spectrum}.

Spectrum selection games in \crn s~usually consist of three components: The players which represent \su s and may include \pu s, the action space and the utility function(s). The players have a set of functions representing available frequency bands. The action space is the Cartesian product of the sets of actions of all players. Each player has a utility function that is used to translate the action space into the real world needs, e.g. the frequency bands that meet \su's~requirements~\cite{tragos2013spectrum}. The goal of the game is to maximize each \su's utility function. This takes into consideration the impact of each \su's decisions on the other players. For games with specific characteristics, there is always a steady state solution (i.e., a Nash equilibrium), and any unilateral change of a player leads to a lower utility for that specific player~\cite{tragos2013spectrum,wang2010game}.

\textbf{Sources of location information leakage:}
Games may require that \su s share their channel selection decisions among one another. This information, just like the case of spectrum availability, could be used for \su~localization. In fact, this information reveals the list of channels that a \su~may be interested in using; i.e. the list of available channels in its vicinity. Sharing this list with other \su s may put into risk \su's own privacy, as this information could be used by an adversary to estimate its position especially if this adversary has a global knowledge of the \crn.

\subparagraph{Learning and adaptation}
\crn s employ software-defined radios, which are capable of executing complex computational tasks through a specialized software module called the cognitive engine~\cite{huang2010modeling,bkassiny2013survey}. This engine has learning capabilities that allow \su s to make spectrum selection decisions and perform tasks in a distributed manner by only relying on what \su s learn from the environment~\cite{bahrak2013security,bkassiny2013survey}. This is usually done by means of machine learning techniques, which have recently attracted significant attention in the context of \crn s~\cite{noroozoliaee2013efficient,clancy2007applications,oliaee2013adaptive}. For example, in~\cite{baldo2009neural}, the authors propose a cognitive engine based on artificial neural network (ANN) that learns how environmental measurements and the status of the network affect the \crn~performance on different channels. Based on this, the cognitive engine can dynamically select the best channel, expected to yield the best performance for \su s.
Li et al.~\cite{li2009multi} use a multi-agent Q-learning approach, a model-free type of reinforcement learning, to address the problem of channel selection in multi-user and multi-channel \crn s.
Each \su~considers both the channel and the other \su s as its environment, updates its Q values continuously, and uses the Q-table to select the best channel. 
NoroozOliaee et al.~\cite{noroozoliaee2013efficient,noroozoliaee2011achieving} derive new private objective functions suitable for supporting elastic traffic that can be used by learning algorithms to enable cognitive users to locate and exploit
unused spectrum opportunities in a distributed manner while maximizing their received throughput. 
These same authors also derive learning-based objective functions for the inelastic traffic model with non-cooperative~\cite{hamdaoui2012coordinating,hamdaoui2011aligning} and cooperative~\cite{noroozoliaee2012maximizing,noroozoliaee2011distributed} users.
Yau et al.~\cite{yau2009context} propose a context-aware and intelligent dynamic channel selection scheme that enables \su s to adaptively  select channels for data transmission to enhance QoS.

\textbf{Sources of location information leakage:} The learning process may also lead to some location information leakage. This is mainly due to:
\begin{itemize}
\item {\em Environmental measurements:} In centralized \crn s, the learning agent, usually \fc, needs to collect environmental measurements during the training phase~\cite{baldo2009neural} to be able to select the best channels for secondary transmissions. In the case of distributed \crn s, the learning process involves multiple agents, which often need to exchange measurement information among themselves. As we have shown previously, this information, when shared among the different \crn~entities, may reveal significant information about \su s' location.

\item {\em Activity prediction:} Prediction strategies through machine learning techniques could also be used to predict both \pu~and \su~activities based on past measurements and experience~\cite{akter2008modeling,xing2013spectrum}. This can allow a malicious entity to predict which channels a \su~might be using in the future. Combining this information with the learned activity model of \pu s and their coverage areas, it becomes possible to predict a \su's location, just as explained in Section~\ref{sourcesMobility}.
\end{itemize}

\subparagraph{Optimization theory}
Optimization techniques (e.g., convex optimization, linear programming, non-linear programming, etc.) have also been widely used to solve the spectrum assignment problem in \crn s. For instance, Tan et al.~\cite{tan2012channel} formulate the channel assignment problem as an integer optimization with the aim to maximize throughput, and propose two greedy non-overlapping and overlapping channel assignment algorithms to solve it. Bkassini et al.~\cite{bkassiny2010optimal} model the channel assignment problem as a weighted bipartite graph, where \pu s and \su s constitute the two disjoint sets of vertices in the bipartite graph. The authors use the well-known Hungarian method~\cite{kuhn1955hungarian} to solve this problem in polynomial time. Ding et al.~\cite{ding2010distributed} formulate the joint spectrum and power allocation problem as a convex optimization problem, and propose a distributed algorithm to solve it. 
Ben Ghorbel et al.~\cite{ghorbel2016distributed,ghorbel2014distributed} propose 
two-phase optimization heuristics also for joint allocation of the spectrum and power resources. Their proposed heuristics split the spectrum and power allocation problem into two sub-problems, and solve each of them separately. The spectrum allocation
problem is solved during the first phase using learning, whereas the power allocation is formulated as a
real optimization problem and solved, during the second phase, by traditional optimization solvers.
Salameh et al.~\cite{salameh2011throughput} formulate the joint rate/power control and channel assignment as a mixed-integer program with the aim to maximize the sum-rate achieved by all contending \su s over all available spectrum opportunities. Due to the NP-hardness nature of this problem, they transform it into a binary linear programming problem which they solve in polynomial time. In~\cite{xin2009joint}, the authors formulate the joint QoS-aware admission control, channel assignment, and power allocation as a non-linear NP-hard optimization problem. In~\cite{salameh2008distance} the channel assignment problem is expressed as an Integer Linear Programming (ILP) problem. These approaches rely on heuristics to solve the spectrum assignment due to the complexity of the formulated optimization problems.

\paragraph{Power control and beamforming}
\label{powerall}
Power control and beamforming are effective methods for mitigating co-channel interference and thus boosting the system capacity. The challenge with power control and beamforming in \crn s lies in making sure that \su s' transmissions do not cause the received interference at \pu s to exceed a tolerable limit. In light of this, a number of beamforming and power allocation techniques have been proposed for \crn s with various objectives, such as capacity maximization~\cite{zhang2008joint} and transmit power minimization.

For instance, Le et al.~\cite{le2008resource} propose to formulate the joint rate and power allocation problems for the secondary links as optimization problems with both QoS and interference constraints under low network load conditions. This work relies on two popular fairness criteria, namely, the max-min and the proportional fairness criteria. Kim et al.~\cite{kim2008joint} develop joint admission control and rate/power allocation methods subject to QoS and minimum rate requirements as well as maximum transmit power and fairness constraints for \su s in MIMO ad hoc \crn s.

Zhang et al.~\cite{zhang2008joint} consider beamforming
and power allocation jointly for SIMO-MAC, and formulate it as two optimization problems: sum-rate maximization and $\sinr$ balancing. These problems are solved using a water-filling based algorithm and constraint decoupling techniques. The goal is to obtain the suboptimal power allocation strategy and to maximize the minimal ratio of the achievable $\sinr$s relative to the target $\sinr$s of the users in the system under a sum-power constraint. Zheng et al.~\cite{zheng2009robust} propose beamforming designs for a multi-antenna \crn, with the aim of allowing multiple \su~transmissions concurrently with the \pu~presence, to achieve also $\sinr$ balancing subject to the constraints of the total \su s transmit power and the received interference power at the \pu s. This is achieved by optimizing the beamforming vectors at the \su~transmitter based on imperfect
channel state information (CSI).

\subsubsection{Spectrum access}
Spectrum access of \crn s is responsible for the sharing of the spectrum among \su s by handling medium contention, interference avoidance, multi-user coexistence, etc.~\cite{de2012survey}.


\paragraph{Access paradigms}
There are three spectrum access paradigms in \crn s:

\textbf{Spectrum underlay:}
This paradigm mandates
that \su s can transmit concurrently with \pu s only if doing so generates an amount of interference at the primary receivers that is below some acceptable threshold~\cite{goldsmith2009breaking,kim2008joint}.

\textbf{Spectrum overlay:}
Spectrum overlay paradigm also allows concurrent primary and secondary transmissions. But \su s are assumed to have knowledge about certain primary transmission parameters to avoid interference with the primary transmissions. The enabling premise for overlay systems is that \su s are allowed to use the spectrum for their own transmissions as long as they are willing to use some of their power to relay some of \pu s' transmissions~\cite{srinivasa2007cognitive}.

\textbf{Spectrum interweave:}
This paradigm is based on the opportunistic spectrum access idea, which has been one of the main drivers for cognitive radio access. Different from the two previous paradigms, this paradigm does not allow simultaneous secondary and primary transmissions on the same frequency band. Instead, it allows \su s to access and use the licensed spectrum only when the spectrum is vacant~\cite{goldsmith2009breaking}.

\paragraph{Spectrum access techniques}
Many \mac~protocols have been proposed  to coordinate \su s to access and share the available channels and to avoid (or reduce) collisions among users\cite{zhang2008joint}. Such a coordinated access could be performed in a distributed or a centralized way~\cite{de2012survey}.
These protocols can either be cooperative~\cite{kondareddy2008synchronized,hamdaoui2008mac} in that they require coordination among \su s to enable efficient sharing of spectrum and thus improve spectrum utilization, or contention-based~\cite{ma2005dynamic,jia2008hc} in that no coordination is required among users.
%
In contention-based protocols, cognitive senders and receivers exchange their sensing results through handshaking mechanisms to negotiate which channel they will use for their communications~\cite{de2012survey}. Tan et al.~\cite{tan2012channel} propose an overlapping channel assignment algorithm and design a MAC protocol to resolve the access contention problem when multiple \su s attempt to exploit the same available channel. Salameh et al~\cite{salameh2009mac} propose a contention-based protocol that tries to satisfy QoS constraints by limiting the number of used channels per \su.

In coordination-based protocols, each \su~shares its channel usage information with its neighbors to increase sensing reliability, and to improve overall system performance~\cite{de2012survey}. For instance, Hamdaoui et al.~\cite{hamdaoui2008mac} propose a coordination-based MAC protocol that adaptively and dynamically seeks and exploits opportunities in both licensed and unlicensed spectra and along both the time and the frequency domains. Zhao et al.~\cite{zhao2005distributed} propose a heterogeneous distributed MAC protocol that permits distributed coordination of local clusters in a multi-hop \crn~through a local common channel.

\paragraph{Sources of location information leakage}
 The sharing of information during this coordination process, though needed for enabling efficient multiple access, could expose the location information of \su s to one another.

\textbf{Sensing outcomes:} Contention-based \mac~protocols may require \su s to share their sensing outcomes with one another to negotiate their access to the spectrum. However, as we have shown in Section~\ref{coopSources}, these sensing outcomes can potentially leak \su s' location information.

\textbf{Channel usage information:} Channel usage information, when shared among \su s as in coordination-based MAC protocols, is shown to leak details about their location; this will be discussed later in Section~\ref{sourcesMobility}.

\subsubsection{Spectrum trading}
Spectrum trading could be seen as the economic aspect of spectrum sharing~\cite{maharjan2011economic}. It aims to maximize the revenue of the spectrum owners, i.e. \pu s, while maximizing the satisfaction of \su s~\cite{niyato2008spectrum} that compete for gaining access to the spectrum. Spectrum trading can be done between \pu s and \su s or among \su s only~\cite{maharjan2011economic}. It relies mainly on two concepts: Auction theory and market theory. Next, we highlight these two concepts and investigate their sources of leakage.

\paragraph{Auction}
A typical dynamic spectrum auction has three main phases: 1) {\em Spectrum discovery phase:} \su s obtain spectrum availability information through one of the spectrum opportunity discovery approaches, explained in Section~\ref{specDisc}, and determine the bid price for each available channel based on its quality. 2) {\em Bidding phase:} each \su~submits its bids and its location along with its ID to the auctioneer. 3) {Channel assignment:} once the auctioneer collects all the bids from \su s, it distributes channels among them and charges the winners accordingly~\cite{liu2013location}. This is suitable for situations when the price of the spectrum is undetermined and depends on \su's requirements~\cite{niyato2008spectrum}. Auction-based spectrum sharing for \crn s has been studied intensively in literature (e.g.,~\cite{khaledi2013auction,wang2010spectrum,kasbekar2010spectrum}).

\paragraph{Market theory}

\textbf{Monopoly Market:}
This is the simplest market structure as there is only one seller, i.e. \pu, in the system. Based on \su s' demand, the seller can optimize the trading process to obtain the highest profit~\cite{maharjan2011economic},\cite{tran2015joint,do2014optimal}.

\textbf{Oligopoly Market:}
This is a type of market that lies between full competition and no competition (or monopoly) and is defined as a market with only a small number of firms and with substantial barriers to entry in economics~\cite{hossain2009dynamic}. These firms or primary service providers compete with each other independently to achieve the highest profit by controlling the quantity or the price of the supplied commodity which is the spectrum resource in this case. Unlike the monopoly case, in oligopoly, there are multiple firms that provide the same service, making it necessary for firms to consider each other's strategy~\cite{maharjan2011economic}. The most basic form of oligopoly is duopoly, where only two sellers exist in the market~\cite{tran2015joint,do2014optimal}. 

\textbf{Market-equilibrium:}
In this spectrum trading model, the primary service provider or spectrum seller is assumed to be not aware of other service providers, which could be due to the lack of any centralized controller or information exchange among each other. This makes the spectrum seller naively set the price according to the spectrum demand of \su s. This price reflects the willingness of the spectrum seller to sell its spectrum which is generally determined by the supply function. On the other hand, the willingness of a \su~to buy spectrum is determined by the demand function~\cite{wang2010game}. Market-equilibrium aims at giving a price for which spectrum supply from a primary service provider is equal to spectrum demand from \su s~\cite{hossain2009dynamic}. This price achieves two goals: the spectrum supply of the primary service provider meets all spectrum demand of \su s, and the spectrum market does not have an excess in the supply~\cite{wang2010game}.

\paragraph{Sources of location information leakage} \label{sourcesTrading}
Spectrum trading may also introduce some sources of location information leakage as we discuss next.

\textbf{Location information:}
During the bidding phase of spectrum auction, \su s may need to submit their locations to the auctioneer as suggested in~\cite{liu2013location}. This is clearly an obvious source of location information leakage as it exposes the location information of \su s to the auctioneer and to an external adversary that may be eavesdropping the communications of \su s during the auction process.

\textbf{Bid channels:}
\su s here need to submit their bids for their channels of interest to the auctioneer (or spectrum broker) . An adversary aiming to infer a \su's location can deduce, from the list of channels \su~bids for, that \su~is located somewhere where these channels are available. Simple intersection of the availability areas of these channels can easily locate \su~\cite{liu2013location}.

\textbf{Bid prices:}
For each channel available for auction, a \su~can first evaluate its quality and, depending on the channel's quality, establish a price for it. It then submits its bid for the channel to the broker. These prices are shown to be a potential source of \su s' location information leakage~\cite{liu2013location}.

\subsection{Location information leakage in spectrum mobility}
\label{specMobility}
\su s communicating on a licensed spectrum band may need to vacate their current band at any time, for instance, due to the return of \pu s to their licensed band. When this happens, \su s need to find and switch their ongoing communications to another vacant band to avoid the disruption of their ongoing transmissions. This is known as {\em spectrum mobility} or {\em  spectrum handoff}~\cite{akyildiz2009spectrum}. 
There are several events that could trigger spectrum handoff in \crn s, and next, we list some of them:
\begin{itemize}
\item {\em \pu's return:} Whenever a \pu~returns to its channel, \su~is forced to vacate it and switch to another available one, if any. This initiates the handoff process. Finding a new available channel often requires \su~to perform spectrum sensing, making handoff more challenging~\cite{chengyu2013spectrum}.


\item {\em \su's mobility:} Because spectrum availability is location dependent, moving while having an ongoing communication may trigger spectrum handoff, as current channel may no longer be available in \su's new location~\cite{lee2012spectrum}.


\item  {\em Quality degradation:} Spectrum handoff could be triggered by the degradation of the channel quality. It can be triggered when, for example, the \qos~level received by \su~goes below a certain threshold, forcing it to find and switch to another channel.
\end{itemize}



\subsubsection{Spectrum handoff strategies}
Based on the handoff triggering timing, spectrum handoff techniques could be classified into four categories or strategies: Non-handoff strategy, reactive handoff strategy, proactive handoff strategy, and hybrid handoff strategy~\cite{christian2012spectrum,kumar2015spectrum}. We first explore these different strategies, then we investigate their sources of location information leakage.

\paragraph{Non-handoff strategy} In this strategy, when one of the triggering events for handoff occurs, \su s stop transmitting over the current channel and choose not to switch to another channel. Instead they remain idle until the channel becomes available again~\cite{wang2012modeling}, as introduced in the non-hopping mode of the IEEE 802.22 WRAN standard~\cite{hu2007cognitive}. How good this handoff strategy is depends on the activities and loads of \pu s. It causes very little to no \pu~interference but the waiting latency to resume secondary transmission could be unpredictably very large, as it depends on when \pu~leaves the spectrum. This strategy is best suited for systems with short \pu~transmissions~\cite{christian2012spectrum}.

\paragraph{Pure reactive handoff strategy}
In this strategy, the target channel selection and the handoff are performed reactively after a spectrum handoff triggering event occurs~\cite{wang2008spectrum,kumar2015spectrum}. Here, \su s need to perform spectrum sensing in order to find the target backup channel to which communication is to be transferred. Several reactive handoff strategy-based approaches are proposed in the literature~\cite{willkomm2005reliable,wang2010modeling}. In general, this strategy has less handoff latency than that of the non-handoff strategy, but has larger latency when compared to the proactive spectrum handoff strategy~\cite{kumar2015spectrum,christian2012spectrum} (described next). The handoff performance of this strategy depends on the accuracy and speed of the spectrum sensing process in identifying a vacant target channel.

\paragraph{Pure proactive handoff strategy}
In this approach, the handoff and the target channel selection are performed proactively before a spectrum handoff triggering event takes place~\cite{song2012prospect,nejatian2013proactive}. \su s do so by periodically observing all channels to obtain spectrum usage statistics which allow them to determine the candidate channels for spectrum handoff~\cite{wang2008spectrum}. The selection of the target free channel for future spectrum handoff is usually made based on \pu~traffic characteristics~\cite{kumar2015spectrum}, where \su s can predict \pu~arrivals in the target spectrum band in advance. Hence, the handoff latency is reduced considerably when compared to the reactive spectrum handoff strategy, which requires taking action after the handoff triggering event takes place. However, if the prediction of \pu~traffic is inaccurate or if the target backup channel is obsolete, for instance due to being occupied by other \su s at handoff time, this could lead to poor handoff performance~\cite{christian2012spectrum}. This makes this strategy best suited to networks with well-modeled \pu~traffic characteristics.



\paragraph{Hybrid handoff strategy}
This approach combines proactive spectrum sensing with reactive spectrum handoff as suggested by Christian et al.~\cite{christian2012spectrum}. It performs proactive spectrum sensing to decide on the backup target channel in advance and before the handoff is triggered, and makes a reactive handoff decision after the triggering event takes place. Thus, it reduces the handoff latency when compared to the reactive handoff strategy. This hybrid approach could be seen as a tradeoff between reactive and proactive handoff strategies.

\subsubsection{Sources of location information leakage}
\label{sourcesMobility}
Spectrum mobility can also leak some location information about \su s, as highlighted next:

{\bf Handoff:}
Recall that a \su~utilizing a \pu~channel is forced to vacate the channel (and possibly switch to another) when \pu~returns to and claims its channel. \pu~(and easily other entities) knows, in this situation, that \su~is located within its coverage area. Handoff can thus lead to leakage of location information of \su~performing handoff.

{\bf Spectrum utilization information:}
A \su's spectrum usage history (e.g., sequence of channels \su~has used over some period of time) could easily be used to localize \su~(or to track it if it is moving). Recall that when a \su~is communicating over a \pu~channel, it means that \su~is outside the coverage areas of all ON \pu s associated with that channel, or inside the area of an OFF \pu.
%
%
Now, for instance, by tracking which channels \su~has used over a period of time and by knowing when and which \pu s are OFF/ON during that time period, an adversary can easily narrow down the area where \su~is located at by intersecting the areas associated with \pu s~\cite{gao2013location}.
%
Spectrum utilization history information could then be a significant source of location information leakage.

{\bf Sensing reports:}
Before handoff, a \su~may need to sense the spectrum to identify a new target channel (using one of \pu~detection techniques identified in Section~\ref{pudetection}). If cooperation is further required to select the appropriate channel for handoff, \su s will have to share their sensing reports, which can compromise their location privacy.


Location privacy-preserving protocols should therefore be designed with the objective of hiding information that can leak \su's location during the handoff process and also reducing, as much as possible, the occurrences of handoff events.

\subsection{Summary}
In this section, we identified the sources of location privacy leakage emerging from the different components of \crn s, namely, spectrum discovery, spectrum analysis, spectrum sharing, and spectrum mobility. We highlighted the different functionalities of each of these components, and discussed how some of these functionalities can present some vulnerabilities that could be exploited to localize \su s. In the next section, we will go over a family of renowned privacy enhancing technologies and generic crypto schemes that we believe are the most relevant to \crn s. We will also discuss to which extent these technologies could be applied to design location privacy-preserving protocols that could prevent attacks exploiting the identified vulnerabilities.


\section{Limitations of generic privacy enhancing technologies in CRNs}
\label{limitGeneric}

Location privacy preservation is a mature technology for many wireless systems, such as sensor~\cite{conti2013providing}, vehicular~\cite{wei2010safe,tang2008privacy}, WiFi~\cite{jiang2007preserving},
cellular~\cite{gorlatova2011managing}, and others~\cite{gorlach2005survey}.
Depending on the wireless system and application at hand, location information can be leaked through various means, ranging from wireless signal localization~\cite{jiang2007preserving,conti2013providing} to traffic monitoring and analysis~\cite{xi2006preserving}. For instance, in sensor networks, location information can be inferred by monitoring packet reception times~\cite{xi2006preserving} or
analyzing packet traffic~\cite{jian2007protecting,ozturk2004source} of source nodes.
Countermeasure solutions for these attacks have also been proposed, ranging from introducing randomness to multi-hop path selection~\cite{deng2005countermeasures,ngai2013providing} to making the source nodes move randomly~\cite{xi2006preserving} to confuse the attackers.
Unlike other wireless systems, location privacy preservation that addresses vulnerabilities in \crn s has not, however, received much attention, though several works related to spectrum sensing~\cite{kasiri2015privacy,gao2013location,troja2014leveraging,troja2015efficient,li2015agent,zhang2015optimal,li2012location}, spectrum auction bids~\cite{huang2015general,liu2013location}, subscriber identification~\cite{reddy2014method}, and database-driven \dsa~\cite{gao2013location,zeng2014location,troja2014leveraging,troja2015efficient,li2015agent,zhang2015optimal} have been proposed.

\subsection{Adaptation of existing privacy enhancing technologies}
Direct adaptation of existing Privacy Enhancing Technologies (PETs), such as Searchable Encryption (SE) (e.g.,~\cite{DSSE:EfficientUpdate:CCS2014:Hahn,DSSE:Yavuz:SAC:2015,DSSE:MultiKeyword:Fuzzy:Infocom:2014,DSSE:NDSS2014DavidCash,DSSE:SP2014MuhammadNaveed}) and Oblivious Random Access Memory (ORAM) (e.g.,~\cite{ORAM:Goldreich:1996:SPS} \cite{Stefanov_TowardPracticalORAM_NDSS12,ORAM:RevisitedPinkas:2010}), which enable a client to outsource its data to a database in an encrypted form so it can perform search queries on it, cannot, for example, be used as they are in database-driven \dsa~to enable private spectrum information retrieval. There have also been proposed cryptographic techniques that enable generic (e.g., Fully Homomorphic Encryption (FHE)~\cite{FHE:IntegerRevisited:2015:Eurocrypt,FHE:overInteger:vanDijk:2010,FHE:Smart:2014:FHESIMD}) or specific (e.g., functional encryption~\cite{FunctionaEnc:ShenShiWaters:PredicatePrivacy:TTC:2009,FunctionalEnc:Garg:CandiateIndistObfus:FOCS:2013}) data processing over encrypted data, and these existing PETs cannot be directly adapted either to fit the \crn~context, so that \su s' location privacy is preserved while still querying the spectrum database for availability information in an effective manner.
Architectural differences and performance requirements of \crn s make direct adaptation extremely ineffective. Privacy-preserving search/access techniques, such as SE or ORAM, are specifically designed for a data outsourcing model~\cite{Curtmola:2006:SSE,DSSE:NDSS2014DavidCash,DSSE:SP2014MuhammadNaveed}, in which a client encrypts {\em its own data} with {\em its private key} and then outsources it to the database. However, in database-driven \dsa, a third party owns and manages the spectrum database. Therefore, it is impractical for database owners to generate a searchable encrypted copy of the database for each single user (note that the initialization phase of these PETs are highly costly~\cite{DSSE:Yavuz:SAC:2015,ORAM:RevisitedPinkas:2010}). Existing, fully generic techniques such as FHE~\cite{FHE:IntegerRevisited:2015:Eurocrypt,FHE:overInteger:vanDijk:2010}) are, on the other hand, extremely costly and therefore impractical for \crn s.

That is said, there have been several attempts that aimed to adapt existing PETs to fit the \crn~context. In the case of database-driven \dsa~for example, the proposed techniques that aim to protect the location information of \su s when they are querying databases for spectrum availability information rely on either {\em $k$-anonymity}~\cite{samarati2001protecting,khoshgozaran2009private} or \pir~({\em private information retrieval})~\cite{chor1998private,wang2010generalizing}.
{\em $k$-anonymity} approaches (e.g.,~\cite{zhang2015optimal}) essentially rely on a third party, known as the anonymizer, to ensure that the probability of identifying the location of a querying user remains under $1/k$, where $k$ is the size of the anonymity set to be received by the untrusted database (alternatively, the anonymity set can be constructed distributedly instead of relying on a third party).
{\em $k$-anonymity} approaches are known to suffer from one major problem: they cannot achieve high location privacy without incurring substantial communication/computation overhead (e.g., higher privacy means higher $k$). They often compromise the location privacy at the benefit of lowering the incurred overhead, or vice-versa~\cite{peddinti2011limitations}.
\pir-based approaches~\cite{troja2015efficient,gao2013location,troja2014leveraging}, on the other hand, offer much better privacy than $k$-anonymity approaches, but also incur substantial overhead, thus limiting their practical use for \crn s~\cite{ghinita2008private}. Proposed approaches relying on these technologies will be discussed in more details in later sections.

In what follows from this section, we take a closer look at some of the most known and generic PETs and discuss why they cannot be used off-the-shelf as they are in the context of \crn s to protect \su s from location inference attacks that exploit the vulnerabilities identified in Section~\ref{sec:sources}. These techniques, include {\em homomorphic encryption}, {\em oblivious transfer}, {\em private information retrieval}, {\em data outsourcing-based techniques}, {\em differential privacy}, and {\em secure multiparty computation}.

\subsection{Homomorphic encryption}
Homomorphic encryption is a special form of encryption that allows computations to be performed on ciphertexts. It generates an encrypted result whose decryption matches the result of operations performed on the plaintexts. There are two kinds of homomorphic encryption: full and partial.

\subsubsection{Fully homomorphic encryption}
This is a special type of homomorphic encryption which allows the computation of arbitrary functions on encrypted data without decrypting it. This concept was first introduced by Gentry~\cite{gentry2009fully}  and is based on the properties of ideal lattices. Theoretically speaking, this is a very powerful concept as it permits the construction of a program that performs all kind of operations on the ciphertexts. Since such a program does not need to decrypt its inputs, it can be run by an untrusted party without revealing its inputs and internal state, making it an attractive tool for preserving privacy.

This might seem applicable in the context of \crn~to hide, for example, the observations of \su s (proven to leak information about \su s location as discussed in Section~\ref{coopSources}) during the spectrum sensing phase and share them with \fc~(or other \su s) without worrying about \su 's location privacy. 
The main issue, however, with this type of encryption is that it involves high computation and requires large storage, making it unpractical. Another major issue with this encryption is that the search time resulting from using fully homomorphic encryption is linear in the length of the dataset. This again makes it unpractical, especially for applications with large datasets like spectrum geolocation databases.

\subsubsection{Partially homomorphic encryption}

A partially homomorphic cryptosystem is an encryption scheme that, unlike fully homomorphic encryption, can only perform either multiplication or addition on the ciphertexts, but not both.  Several cryptosystems with homomorphic propoerties were proposed in the literature. Paillier cryptosystem~\cite{paillier1999public} is one of the most famous additive homomorphic schemes. Examples of multiplicative homomorphic cryptosystems include El Gamal~\cite{elgamal1984public} and RSA~\cite{rivest1978method}. Thanks to their homomorphic properties, these schemes could be used in situations that require performing some basic operations on sensitive data while hiding user inputs (like when reporting sensing information).

Partially homomorphic encryption is more practical than the fully homomorphic one; however, for them to provide high security level, they incur large communication and computational overhead. This makes it unpractical to use  especially for large \crn s if not used judiciously.

\subsection{Oblivious transfer}
\begin{figure}[h!]
\vspace{-2pt}
\center
\includegraphics[width=0.24\textwidth]{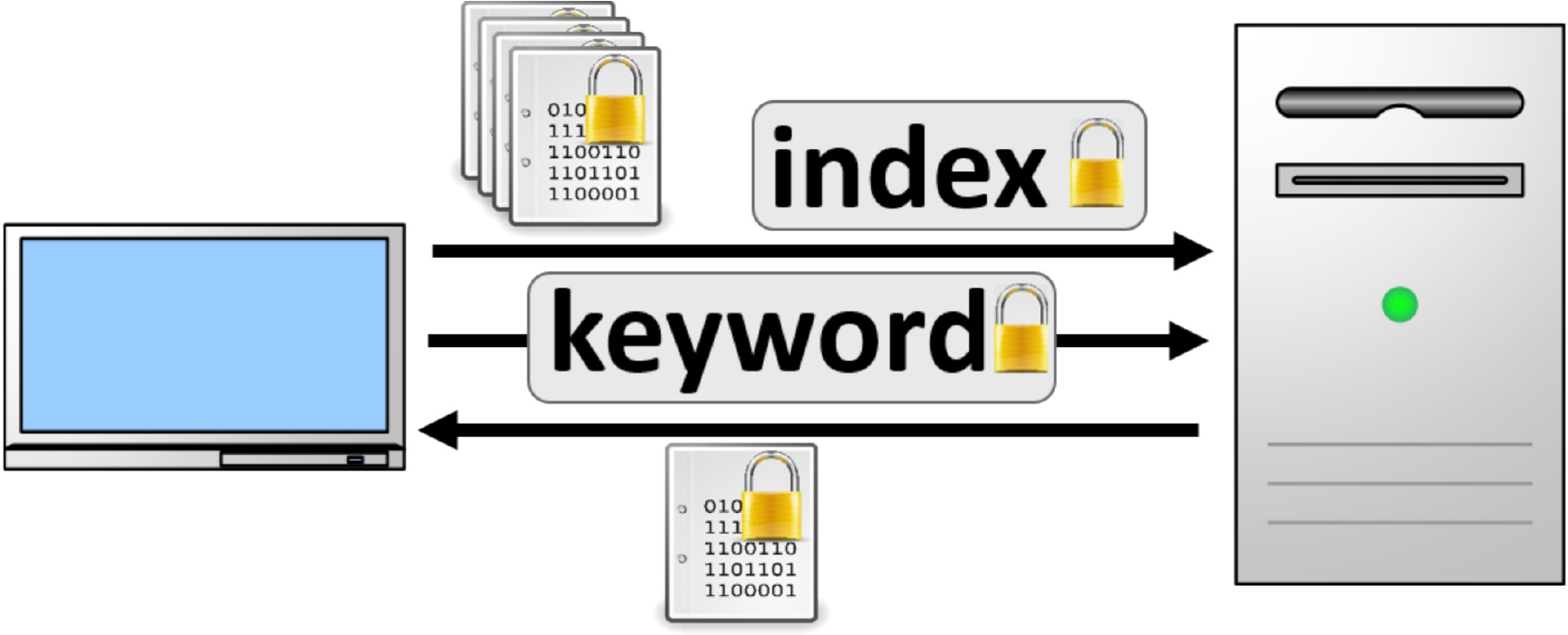}
\caption{\small Oblivious transfer for the case $1$-$out$-$of$-$n$ }
\label{obt}
\end{figure}

Oblivious transfer (\ot) is a privacy enhancing protocol that enables a sender to transfer one of many pieces of data to a receiver, while keeping the sender oblivious as to which piece has been sent and while making sure that the receiver receives only one message. The simplest flavour of this protocol, $1$-$out$-$of$-$2$, was first introduced by Rabin~\cite{rabin2005exchange} and was later generalized to $1$-$out$-$of$-$n$ and $k$-$out$-$of$-$n$ cases. In the $1$-$out$-$of$-$n$ case, as explained in Figure~\ref{obt}, the sender has $n$ messages and the receiver has an index $i$. The receiver wants to learn the $i^{th}$ message without the sender learning $i$. On the other hand, the sender wants that the receiver only learns one message among the $n$ messages. This could be thought of as a suitable approach to use for extracting spectrum availability information from the spectrum \db. This approach, however, incurs very large communication and computational overheads which makes it unpractical in a delay sensitive problem like spectrum availability discovery.

\subsection{Private information retrieval (\pir)}
This concept was first introduced by Chor et al.~\cite{chor1998private}. It allows users to privately retrieve records from a database while preventing the latter from learning which records are being retrieved. This could be thought of as a weaker version of $1$-$out$-$of$-$n$ \ot~which further requires that the receiver does not learn anything about the other entries in the database.

\pir~approaches could be classified into two categories: Information-theoretic \pir~and computational \pir. In information-theoretic setting, the reconstruction of the client's query is impossible no matter how much computation the adversary would perform. A trivial \pir~approach could be to download the entire database. This would offer an information-theoretic privacy, i.e. unbreakable privacy, but on the other hand involves enormous communication overhead. Any information-theoretical \pir~solution has a communication overhead of at least the size of the database as proven by Chor~\cite{chor1998private}. Fortunately, this applies only to the case where the database is stored only on a single server. One way to get around this extensive overhead is by assuming that the database is replicated in several servers that do not communicate with each other. This way, a non-trivial theoretic \pir~solution that has communication overhead smaller than the database size turns out to be feasible. An information-theoretic approach in this model means that an individual database server cannot learn which element was retrieved by the user, no matter how much computation it may perform as long as it does not collude with the other servers~\cite{cachin1999computationally}. Several approaches proposed in the literature considerably reduce the communication overhead of information theoretic \pir~(e.g.~\cite{ambainis1997upper} where the communication cost is $\mathcal{O}(n^{1/2k-1})$ with $k$ is the number of database servers). 

On the other hand, in computational \pir~approaches, the security is based on hard-to-solve well-known cryptographic problems, e.g. discrete logarithm or factorization~\cite{menezes1996handbook}. This makes them secure against computationally bounded adversaries. But an adversary with sufficient computational resources can learn the client's query by breaking the underlying security system. Some computational \pir~approaches are able to provide poly-logarithmic communication complexity~\cite{cachin1999computationally}. Gentry et al.\cite{gentry2005single} propose the most communication efficient \pir~that has a constant communication overhead.

Even though research on \pir~is making progress in terms of reducing the overhead, \pir~approaches still suffer from large overhead that limits their practicality and impedes their off-the-shelf use without adaptation in the context of \crn s.

\subsection{Data outsourcing-based techniques}
These techniques are designed for applications that require secure data outsourcing, where a client's sensitive data is outsourced to a third-party storage provider, e.g. the cloud. Existing access control
solutions focus mainly on preserving confidentiality of stored data from
unauthorized access and the storage provider. Next, we discuss two well known data outsourcing based PETs: {\em searchable symmetric encryption (\sse)} and {\em oblivious random access memory (\oram)}.

\subsubsection{Searchable symmetric encryption (\sse)}
\begin{figure}[h!]
\vspace{-5pt}
\center
\includegraphics[width=0.33\textwidth]{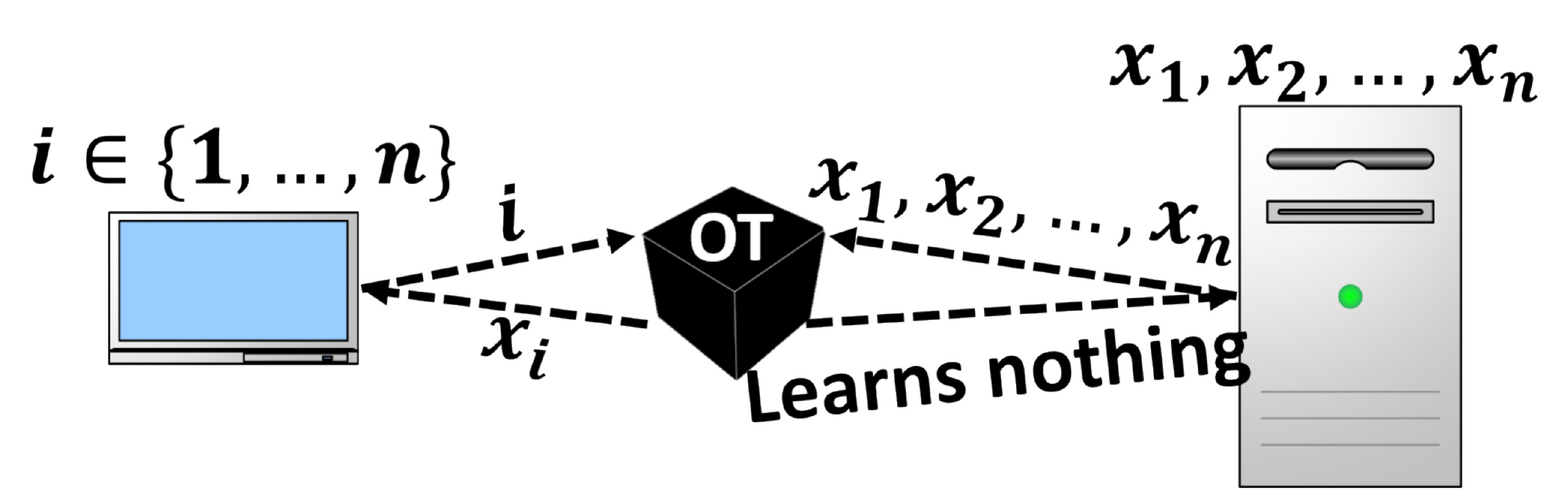}
\caption{\small Searchable symmetric encryption}
\label{sse}
\end{figure}

Searchable symmetric encryption is a PET that is largely deployed to privately outsource one's data to another party while maintaining the ability to selectively search over it~\cite{Curtmola:2006:SSE}. This means that a client needs to outsource its data to a database/server in an encrypted form to be able to later perform private search queries on it as shown in Figure~\ref{sse}.
Despite its efficiency and the high level of privacy that \sse~provides, it cannot be applied to database-based \crn s simply because in \sse, the data has to be outsourced by the client, whereas in database~based \crn s, the data about spectrum availability is generated and provided by the service operator that manages the spectrum database. This means that \su s have no control over this data and, thus, they cannot encrypt it and outsource it to \db~as required by \sse.
\subsubsection{Oblivious random access memory (\oram)}

Encrypting its outsourced data is not sufficient for a user to protect the confidentiality of his/her data content as his/her access pattern to the data remains unprotected which may reveal the user's private information. \oram~is introduced by Goldreich et al.~\cite{ORAM:Goldreich:1996:SPS} to not only  preserve data confidentiality but also to hide a user's access pattern to its outsourced data blocks. Traditionally, \oram~has been designed to arrange the data such that the user never touches the same piece twice, without an intermediate shuffle. This erases the correlation between block locations and obfuscates the memory accesses of data, so that access patterns do not leak information about the stored data. Just like \sse, \oram~can only fit to the problem of data outsourcing which is not suitable to the context of \crn s for the same reasons discussed for \sse.

\subsection{Differential privacy}

This is a recent privacy concept tailored to the statistical disclosure control problem  which is defined as follows: how to release statistical information about a set of people without
compromising the privacy of any individual~\cite{dwork2006calibrating}. Its goal is to assure a good statistical accuracy while preserving individual's privacy. It is a well established definition guaranteeing that queries to a database do not reveal too much information about specific individuals who have contributed to the database as suggested in~\cite{groce2011limits}. The formal definition of this concept could be found in~\cite{dwork2006differential}. The basic idea behind it is that for two almost identical input data sets, the outputs of
the mechanism that provides differential privacy are almost identical.
More precisely, it requires that the probability that a query returns a value $v$ when applied to a database $\mathcal{D}$, compared to the probability to report the same value when applied to an adjacent database $\mathcal{D'}$ ( i.e. $\mathcal{D}$, $\mathcal{D'}$ differ in at most 1 entry) should be within a bound of $\exp^\epsilon$ for some privacy level $\epsilon$. Since differential privacy is a probabilistic concept, any differentially private mechanism is necessarily random.  A typical way to achieve this notion is to add controlled random noise, drawn from a Laplace distribution for instance, to the query output. One benefit of this concept is that a mechanism can be shown to be differentially private independently from any side information that the adversary might have.

However, standard differential privacy techniques usually perform poorly in situations where participants contribute various time-series data that could be aggregated and mined for useful information, due to noise~\cite{rastogi2010differentially}. Examples of time-series data may include users' current locations, weather information or information obtained from other participatory sensing applications like spectrum sensing in \crn s~\cite{rastogi2010differentially}. Moreover, the nature of differential privacy concept makes it poorly suitable for applications that involve a single user, such as spectrum database-based opportunities discovery, where the location of a single user has to be hidden. Thus, it requires that any change in a user's location have negligible effect on the published output of the query, which makes it impossible to communicate any useful information to the service provider~\cite{andres2013geo}. Despite this, some approaches try to adapt this concept to the context of \crn s as we show in Sections~\ref{lpsd} \& \ref{lpoc}.

\subsection{Secure multiparty computation (MPC)}
The concept of secure multiparty computation (\mpc) originates from the works of Yao~\cite{yao1986generate} and Goldreich et al.~\cite{goldreich1987play}. It allows a group of $n$ mutually distrusting parties $P_1, . . . , P_n$, holding private inputs $x_1, . . . , x_n$ to securely compute a joint function $f(x_1, . . . , x_n)=(y_1, . . . , y_n)$ on these inputs~\cite{bogetoft2009secure}. The goal is to make each party $P_i$ learn only $y_i$ but nothing else. This could be achieved through an interactive protocol, executed between these parties, whose execution should be equivalent to having a trusted party that privately receives $x_i$s from $P_i$s, computes $f$ and returns $y_i$s to $P_i$s. This protocol should be able to give the correct result to honest parties even if some parties are dishonest.

In a \crn~context, this could be an attractive tool to provide privacy for any task that involves some computation between several entities. For instance, this could be used in distributed cooperative spectrum sensing during the spectrum discovery phase to allow \su s to collaborate in order to compute statistics over the sensing reports while preserving the privacy of their reports and thus their location. Another potential use of \mpc~could be during the coalition formation process, again in the spectrum discovery phase, to prevent leaking \snr~values that can compromise \su s' location as explained in Section~\ref{coopSources}. \mpc~could also be used in game theoretical approaches during the spectrum sharing phase to prevent the leakage that can arise from the local decisions shared between different \su s during the game. Furthermore, this could be an attractive tool also to protect the bids of \su s during the auction process that is performed to ensure spectrum sharing among \su s. As explained in Section~\ref{sourcesTrading}, the auction process may leak some information about \su s' location which makes it natural to consider leveraging sealed bids or relying on a trusted party for the auction. Ideally, an \mpc~protocol should be equivalent to a trusted third party; hence, \mpc~could play this role and replace an untrusted auctioneer as suggested in~\cite{bogetoft2009secure}.

It is obvious that the potential applications of \mpc~are multifold due to its flexibility to emulate multiple scenarios. However, the bottleneck is its extensive computational and communication overhead, which makes its deployment difficult in practical situations, and more precisely in the context of \crn s, at least for the time being.
\vspace{-5pt}

\subsection{Summary}
In this section, we explored a family of renowned PETs and generic crypto schemes that we believe are the most relevant to \crn s. We highlighted the benefits and limitations of applying these schemes to \crn~off-the-shelf as they are. In the following section, we will present and discuss location privacy preservation approaches proposed for protecting location privacy during the spectrum opportunity discovery process. We will explore the different threat models, location inference attacks, and location privacy preserving techniques that are specific to this spectrum discovery component.


\section{Location privacy preservation for spectrum opportunity discovery component}
\label{lpsd}

In this section, we investigate the different approaches proposed in the literature to deal with the location privacy issue in \crn s during the spectrum opportunity discovery phase. First, we discuss the challenges that face designing \su's location privacy preserving protocols in both cooperative spectrum sensing and geolocation database-based approaches. Then, we list the different threat models that need to be considered in these two approaches. After that, we detail existing and potential attacks that could be performed by malicious entities to localize \su s by exploiting the vulnerabilities that we identified in Section~\ref{specDisc}. Subsequently, we describe existing solutions that are proposed to cope with these attacks and preserve \su s' location privacy. Finally, we explain the performance metrics that are or could be used to assess the performance and reliability of location privacy preserving protocols in \crn s, and present tradeoffs that are considered when designing these protocols.

\subsection{Location privacy in cooperative spectrum sensing}
\label{CPdiscovery}
As discussed in Section~\ref{coopSources}, the cooperation among \su s during the sensing process gives rise to several vulnerabilities that could be exploited to compromise \su s' location privacy.
Thus, location privacy preservation protocols for cooperative sensing need to be designed with several goals in mind:
%
\begin{itemize}
\item {\em Hide sensing information.} As explained in Section~\ref{coopSources}, \su s' sensing reports may leak information about their locations~\cite{bhattacharjee2013vulnerabilities}. Hence, one main goal of these protocols is to hide sensing reports by concealing the observed sensing information from decision makers or any potential external attackers that might eavesdrop \su's communications~\cite{li2012location,mao2015protecting,wang2015privacy,grissa2015location,grissa2016efficient}.
\item {\em Achieve accurate spectrum availability information.} Protocols need to preserve the location privacy of \su s, but without compromising their ability to still provide  accurate spectrum availability information. Achieving this design goal is very challenging,
    due to its conflicting nature: hiding information for the privacy protection purpose may limit the ability to provide accurate spectrum availability information.

   \item {\em Optimize resource usage.} An important limitation that needs to be accounted for when designing privacy preserving protocols is \su s' resource capability. It is then important to design protocols that require minimum computation and storage resources and incur limited communication overheads.
       This, for instance, implies that expensive cryptographic approaches are to be avoided.

\item {\em Hide \snr~values.} Another goal that needs be aimed at is to hide the \snr~values that \su s might need to exchange to form coalitions, for example. As explained in Section~\ref{coopSources}, \snr~may leak significant information about \su s' location, and thus a reliable location privacy preserving scheme needs to conceal these values without hindering the \crn~operations relying on them.

\end{itemize}

\subsubsection{Threat models}
Several threat models are considered in the literature to study and address \su s' location privacy issue in cooperative spectrum sensing:
\begin{itemize}
\item {\em Dolev–Yao threat model.} In this model the adversary, usually an intruder, can overhear, intercept, and synthesize any message that is exchanged between \su s and \fc~or even between \su s themselves during the cooperative spectrum sensing process. The adversary is only limited by the constraints of the cryptographic methods used~\cite{dolev1983security}. This model is considered in~\cite{grissa2015location,grissa2015cuckoo,grissa2016efficient}

\item {\em Semi-honest or honest-but-curious threat model.} This means that the adversary, that could be a \fc~\cite{grissa2015location,grissa2016efficient,li2012location,mao2015protecting}, a \su~\cite{grissa2015location,grissa2016efficient} or an additional entity as in~\cite{grissa2016efficient}, follows the sensing protocol honestly without changing any of its parameters. However, it shows some interest in learning the location information of target \su s by exploiting their sensing reports.

\item {\em Malicious threat model.} Entities in the \crn~may be malicious, meaning that \fc, \su~or any other entity involved in the cooperative spectrum sensing process can change their parameters and lead several attacks to localize a target \su.
\item {\em Non-collusion threat model.} \fc, \su s and any other entities in the \crn~do not collude to infer target \su s' location~\cite{grissa2015location,grissa2016efficient}. This means that these entities do not share what they learned about target \su s' location during the cooperative spectrum sensing process.

\item {\em Collusion threat model.} \fc~or some \su s may collude with other \su s or entities and work together to infer target \su s' location~\cite{li2012location,wang2015privacy} by  exploiting their sensing reports and communication signals.
\end{itemize}
\subsubsection{Location inference attacks}
\label{attackcoop}

Location inference attacks exploit the vulnerabilities and the sources of leakage that we explained in Section~\ref{coopSources} to localize \su s. These attacks could be performed by an internal entity (e.g. another \su~or \fc) or an external attacker that does not belong to the \crn. These attacks can be classified into two categories, based on the information used for localization: Geometric localization and fingerprinting.

\begin{figure}[h!]
\vspace{-2pt}
\center
    \includegraphics[width=0.25\textwidth]{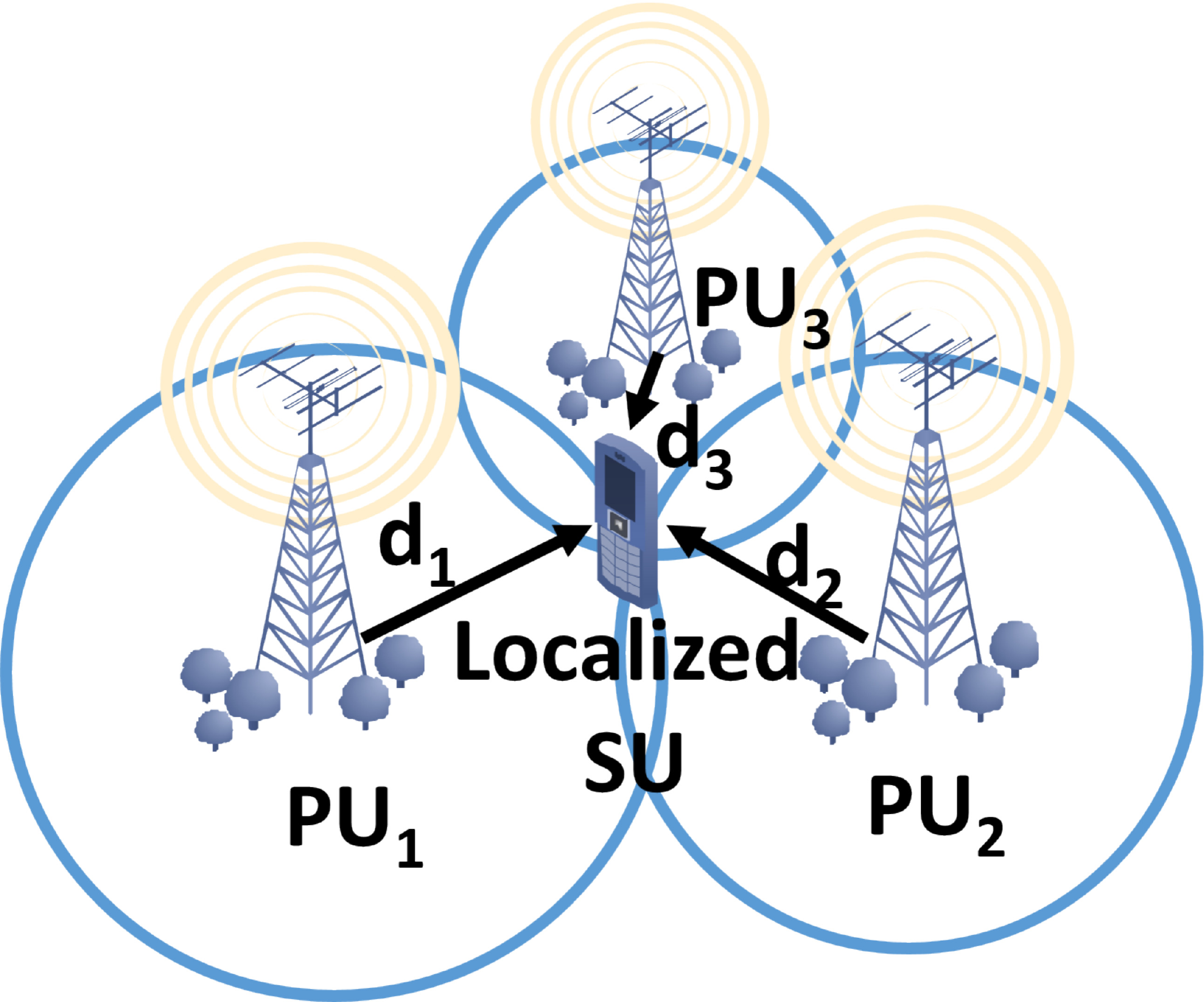}
\caption{{\small Localization of an \su~via Trilateration using the ranges $d_1$, $d_2$ and $d_3$ corresponding to $\pu_1$, $\pu_2$ and $\pu_3$ respectively.} }
\label{trilateration}
\end{figure}

\paragraph{Geometric localization based attacks}
These attacks exploit channel parameter measurements including \rss, \snr, \aoa, \toa~and \tdoa~to localize a  target \su. \rss, \snr~and \toa~could be used to get the range information, as explained in Section~\ref{coopSources}, which is essential for the trilateration localization technique~\cite{zekavat2011handbook,kasiri2015privacy}. Trilateration is a very simple and intuitive approach that computes the position of a target node by finding the intersection of three circles that model the range with respect to at least three anchor nodes as depicted in Figure~\ref{trilateration}.

In the context of \crn, the anchor nodes could be three \pu s whose locations, depending on the situation, could be publicly known. Thus, an attacker that has access to the \rss s that a \su~measures with respect to three channels could exploit this knowledge to localize \su~using trilateration. \snr~could also be used in a similar way, as reported in~\cite{kasiri2015privacy}, for ad hoc \crn s. The attack can occur during the process of forming coalitions and choosing coalition heads as these operations require exchanging \snr~information between \su s. Another attack scenario could involve multiple attackers or colluding nodes that belong to the \crn~and that have a direct communication with the target node.

\begin{figure}[h!]
\center
    \includegraphics[width=0.18\textwidth]{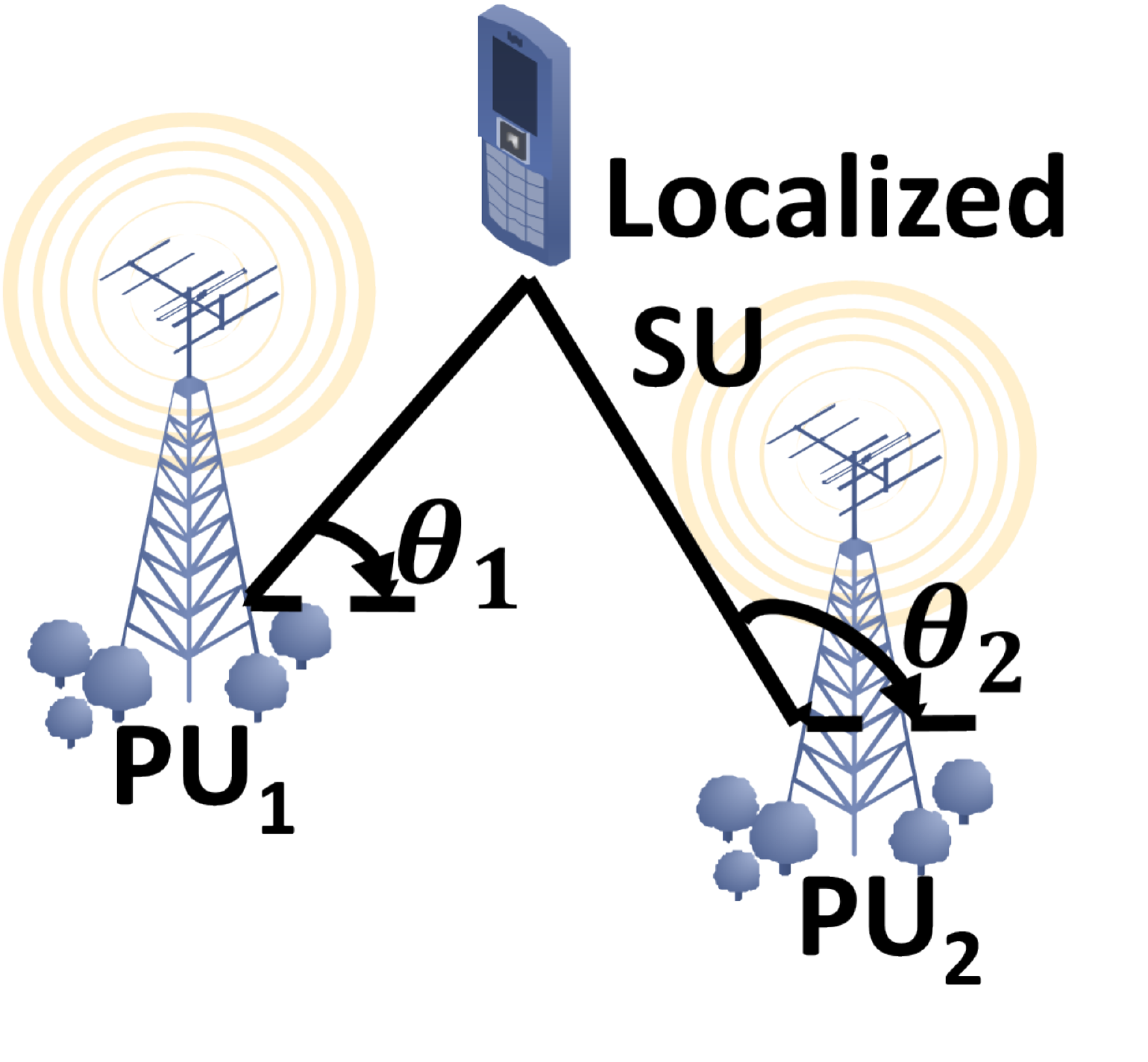}
\caption{\small{ Localization of an \su~via Triangulation using the angles of arrivals, \aoa s, $\theta_1$ and $\theta_2$ of the \su 's signal measured respectively at $\pu_1$ and $\pu_2$}}
\label{triangulation}
\end{figure}

Triangulation is also another technique that exploits channel parameter measurements for localization purposes. It uses angles instead of distances and requires at least two reference nodes to localize the target node~\cite{boukerche2008algorithms}. The two reference nodes measure the \aoa~of the signal coming from the target node. The position of the target node is the intersection of the two lines along the angles from each reference node as in Figure~\ref{triangulation}. As this attack requires a direct communication between the victim and the attackers, this implies that the attackers, which are also the reference nodes in this case, belong to the \crn, e.g. two colluding malicious \su s.

Geometric localization attacks may be performed in \crn s that deploy crowdsourcing (explained in Section~\ref{coop}) for spectrum sensing. For instance, Jin et al.~\cite{jin2016privacy} propose an attack scenario that targets the location privacy of participants in the crowdsourcing process. They consider a special setting where these participants compete to perform spectrum sensing tasks at specific locations via a reverse combinatorial auction operation~\cite{nisan2007algorithmic}. During this auction, participants send their bids, corresponding to their claimed cost of performing the sensing tasks. This cost, as modeled by the authors, involves the round trip distance that a participant needs to travel to perform the sensing tasks and return back to its current location, called base location, which is the target of the proposed attack. This attack exploits the geometric relationship between users bids and the distance they travel to perform the sensing.

\begin{figure}[h!]
\vspace{-2pt}
\center
    \includegraphics[width=0.35\textwidth]{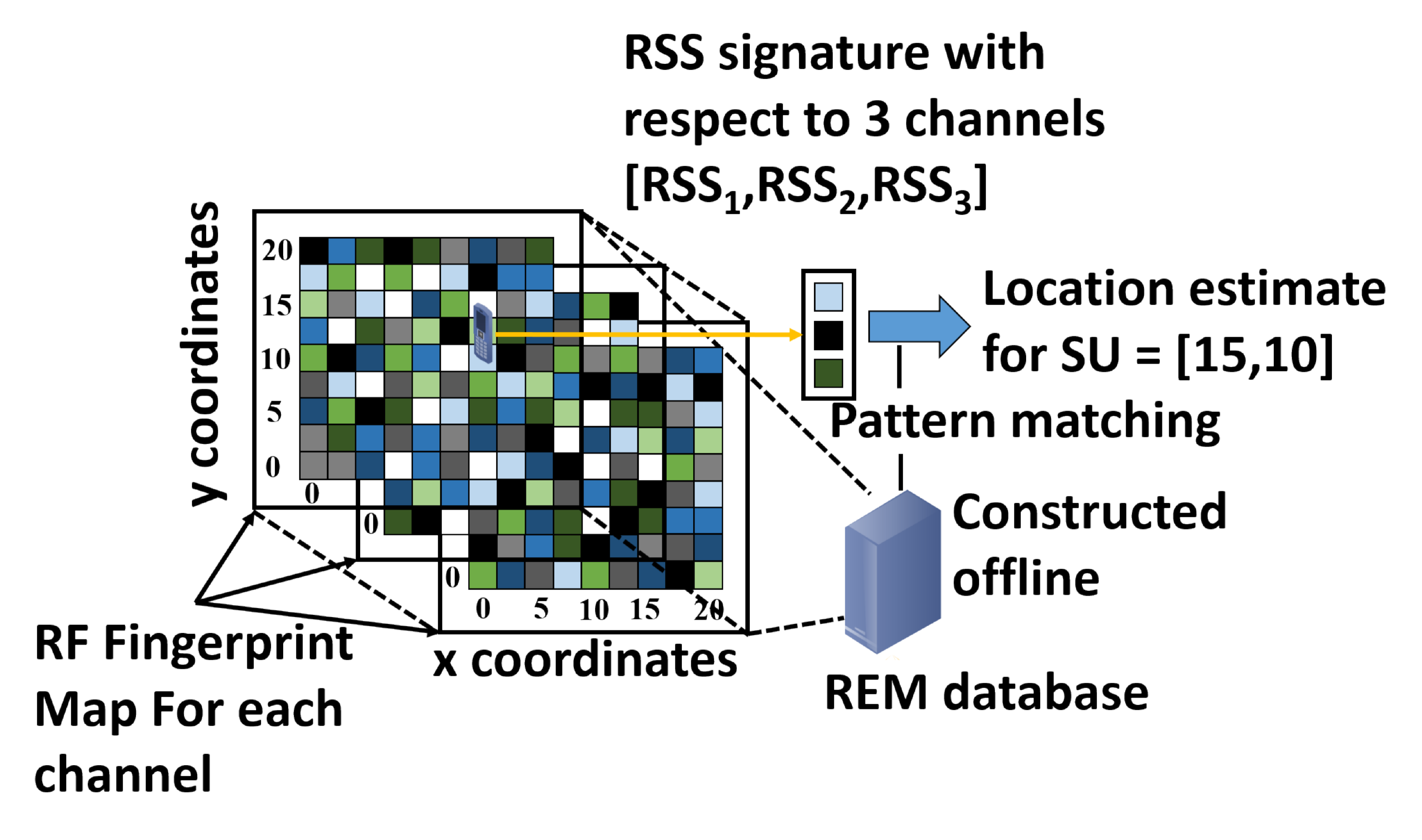}
\caption{{\small Localization of an \su~via Fingerprinting using its \rss~signature $[\rss_1,\rss_2,\rss_3]$ with respect to $3$ channels and the \rem~database.}}
\label{fingerprinting}
\vspace{-5pt}
\end{figure}

\paragraph{Fingerprinting based attacks}
\label{fcattacks}
These attacks are more suitable in situations where the geometric relationships between \su s' positions and measurements cannot be established. It estimates the victim's location by finding the best matched fingerprint for the corresponding measurement within a pre-built RF map. It consists mainly of two phases: An off-line or training phase and an on-line or test phase. In the off-line phase, the RF map is generated. This map could be the \rem~(discussed in Section~\ref{sourcesCharact}) if the attacker is \fc~or a \su~that has access to it, or it could be a map that an external attacker has built by itself. Figure~\ref{fingerprinting} shows a simplified example of how this kind of localization works.

Li et al.~\cite{li2012location} consider two attacks that rely on this principle to localize a \su~based on its \rss~measurements that it shares with \fc~in a centralized \crn. They assume that an attacker constructs a signal propagation model by collecting all the sensing reports transmitted within the network~\cite{li2012location}. The attacker uses machine learning techniques, for example k-means classifier as in~\cite{li2012location}, to partition the \rss~data into multiple sets corresponding to various locations. The first attack, called {\em single report location privacy (SRLP)} Attack, involves an external attacker that eavesdrops \su s' communications or an internal attacker that could be an untrusted \fc~or a compromised \su. Under this attack, the attacker exploits individual \rss~measurements of \su s to localize them by computing the distance between each sensing report and the centroids of each cluster in the signal propagation model that is built beforehand by the attacker. The second attack that they propose is called {\em differential location privacy ({\em DLP})} attack which estimates the sensing report of a \su~during the aggregation process performed by \fc. In this attack, the attacker compares the changes of the aggregation results after a \su~joins or leaves the \crn~and then it infers its location by finding to which cluster the estimated report belongs to, just like in the {\em SRLP} attack.

It is worth mentioning, however, that even though fingerprinting could be attractive for leading location inference attacks, it is not necessarily practical unless the attacker is very powerful with lots of resources. This is due to the fact that the construction of accurate radio maps and fingerprints requires considerable off-line effort and may give rise to several challenges. These include, but are not limited to, the huge number of measurements that need to be taken and also the need to regularly update the radio map due to the inherent time varying nature of wireless channels and networks~\cite{zekavat2011handbook}.

\subsubsection{Location privacy preserving approaches}
As explained in Section~\ref{coop}, \su s in cooperative spectrum sensing \crn s need, first, to share their observations either with \fc~(in centralized \crn s) or with other \su s (in distributed \crn s). These local observations are then combined to make a cooperative spectrum availability decision. These observations could be statistics computed over the signal or just local binary decisions made by each \su~individually. Both cases present some privacy risks to \su s as discussed in Section~\ref{sec:sourcesof}. Thus, research efforts should focus on hiding \su s' observations from the other entities in the network. Most of the existent works that we discuss in this Section consider the location inference attack from the sensing reports that \su s share. We summarize these works in Table~\ref{solCoop} and we discuss them in more details in the following.

{\large
\begin{table*}
\centering
\small
\caption{\small Location privacy preserving schemes in cooperative spectrum sensing}
\label{solCoop}
\resizebox{\textwidth}{!}{%
\renewcommand{\arraystretch}{1.25}{
\begin{tabular}{@{}lp{4cm}lp{5.5cm}p{5.5cm}@{}}
\toprule[1.5pt]
Countermeasures & Attacks Considered                           & Techniques                                                                                                                      & Pros                                                                                                                 & Cons                                                                            \\ \midrule
Li et al.~\cite{li2012location} &  - Location inference from sensing reports (e.g. \rss) &   \begin{tabular}[c]{@{}p{4cm}@{}}- Privacy preserving aggregation with encryption \\ - Dummy report injection \end{tabular} & \begin{tabular}[c]{@{}p{5.5cm}@{}}- Relatively efficient against differential privacy attacks \end{tabular}&  \begin{tabular}[c]{@{}p{5.5cm}@{}}- Very high computational and communication overhead \\ - No fault tolerance \\ - Has a little negative effect on the sensing performance\end{tabular} \\ \addlinespace[5pt] \hline \addlinespace[5pt]

Grissa et al.~\cite{grissa2015location}   &- Location inference from sensing reports (e.g. \rss) & \begin{tabular}[c]{@{}p{4cm}@{}}- Private comparisons using Yao's millionaires protocol\\ - Order preserving encryption\end{tabular} & \begin{tabular}[c]{@{}p{5.5cm}@{}}- Low communication overhead\\ - High location privacy\end{tabular}                       & - Relatively high computational overhead                                        \\ \addlinespace[5pt] \hline \addlinespace[5pt]

Mao et al.~\cite{mao2015protecting}      & - Location inference from sensing reports (e.g. \rss)                                           &  - {\em El Gamal} cryptosystem & - Considers both semi-honest and malicious adversaries &  \begin{tabular}[c]{@{}p{5.5cm}@{}} - High communication overhead \\- Prone to {\em DLP} attack\end{tabular} \\ \addlinespace[5pt] \hline \addlinespace[5pt]

Wang et al.~\cite{wang2015privacy}       & \begin{tabular}[c]{@{}p{4cm}@{}} - Location inference from sensing reports (e.g. \rss) \\- Collusion between service providers \\- Collusion between service providers and \su s \end{tabular}  & \begin{tabular}[c]{@{}p{4cm}@{}} - Cloaking of sensing reports  - Dimension reduction of sensing data through non-invertible projection \end{tabular}& \begin{tabular}[c]{@{}p{5.5cm}@{}} - Considers multiple malicious service providers \\ - Considers collusion between some entities \\ - Provides differential privacy \end{tabular} &                                                                                \begin{tabular}[c]{@{}p{5.5cm}@{}} - Privacy level decreases with the decrease of service providers \\- Privacy level decreases with the increase of \su s\\- Some information distortion during the cloaking process \end{tabular}\\ \addlinespace[5pt] \hline \addlinespace[5pt]

Kasiri et al.~\cite{kasiri2015privacy}       & - Location inference from \snr~during coalition formation &                                                                                                                                - Anonymization of \snr s &                                                                                                                     - Takes into account \su s' mobility &                                                                                \begin{tabular}[c]{@{}p{5.5cm}@{}}- Privacy level decreases as the number of sensed channels increases \\- Providing high location privacy degrades sensing performance \end{tabular}\\ \addlinespace[5pt] \hline \addlinespace[5pt]

Grissa et al.~\cite{grissa2016efficient}   &- Location inference from sensing reports (e.g. \rss) & \begin{tabular}[c]{@{}p{4cm}@{}}- Additional entity in the network\\ - Order preserving encryption\end{tabular}                      & \begin{tabular}[c]{@{}p{5.5cm}@{}}- Very low communication \& computational overhead\\ - High location privacy\end{tabular} & - Additional entity that needs to be managed by a third party for non-collusion \\ \addlinespace[5pt] \hline \addlinespace[5pt]

Jin et al.~\cite{jin2016privacy}    & \begin{tabular}[c]{@{}p{4cm}@{}}- Location inference from sensing cost during reverse auction \\- Location inference from auction result \\ - Location inference from changes in auction participation \end{tabular} & \begin{tabular}[c]{@{}p{4cm}@{}}- Exponential mechanism for differential privacy \end{tabular}&  - Offers differential location privacy  &  - The lower the social cost the higher the location information leakage \\ \bottomrule[1.5pt]
\end{tabular}}}
\vspace{-7pt}
\end{table*}
}

Li et al.~\cite{li2012location} introduce an approach that uses secret sharing and the privacy preserving aggregation process proposed in~\cite{shi2011privacy} to conceal the content of the sensing reports. This scheme uses also dummy report injections to replace the report of a leaving \su~in order to cope with the differential location privacy attack (explained in Section~\ref{fcattacks}) and prevent a malicious \fc~from estimating the sensing report of the leaving \su. Moreover, this scheme can bear collusion attacks involving \fc~and some compromised \su s. Despite its merits, it has several limitations: $(i)$ \fc~needs to collect all the sensing reports in order to be able to decode the aggregated result. Obviously, this could not be fault tolerant, since some reports may be missing due, for example, to the unreliable nature of wireless channels. $(ii)$ It cannot handle network dynamism if multiple \su s join or leave the network simultaneously, as it can only deal with the event of one \su~leaving or joining the network at a time. $(iii)$ The pairwise secret sharing requirement, that this scheme has, incurs extra communication overhead and delay. $(iv)$ The
underlying encryption scheme requires solving the discrete logarithm problem~\cite{menezes1996handbook} for the decryption, which is extremely costly and is only possible for very small plaintext space.

Grissa et al.~\cite{grissa2015location,grissa2017preserving} propose a location privacy preserving protocol that aims to hide \su's sensing reports (specifically \rss) from \fc~and the sensing threshold used for the decision from \su s. This prevents \fc~from trying to localize \su s using their sensing reports and, at the same time, prevents malicious \su s from using the sensing threshold to manipulate their measurements and impact \fc's decision. This scheme relies on {\em order preserving encryption}~\cite{boldyreva2009order} to make \su s encrypt their sensing reports and allow \fc~to learn only the relative order of these reports. Using this order and following a binary search-like technique, \fc~executes at most $\log \: n$ private comparisons between \su s' \rss s and \fc's sensing threshold using {\em yao’s millionaire} protocol~\cite{yao1982protocols}. The order learned by \fc~aims to make the number of private comparisons logarithmic in the number of \su s. This is shown to provide high location privacy to \su s while enabling an efficient sensing performance. However, even though this approach has a low communication overhead and a logarithmic computational overhead as a function of the number of \su s, the computation incurred is still relatively high. This is due to the use of the expensive {\em yao’s millionaire} protocol~\cite{yao1982protocols} that, itself, relies on expensive homomorphic encryption.

Some approaches consider an intermediate node or entity to help addressing the location privacy issue, e.g.~\cite{mao2015protecting,grissa2016efficient}. Mao et al.~\cite{mao2015protecting} provide an approach that requires \su s to encrypt their \rss~values using a derivative of {\em El Gamal}~\cite{elgamal1984public} encryption scheme. In their approach, one of \su s is picked to play the role of a helper to \fc. First, the {\em Helper} and \fc~collaborate to construct a public/secret key pair and each of them keeps a part of the secret key for itself. Then, \fc~and {\em Helper} share the public key with \su s. Subsequently, \su s send their \rss s encrypted using this public key to the {\em Helper} that permutes them, decrypts them with the secret part that it has, and then sends them to \fc~which decrypts them using its part of the key. Once decrypted, \fc~aggregates the \rss~values to make a final decision. The authors consider a semi-honest threat model for \fc~and Helper and a restricted malicious model where only \su s are malicious. However, even though this approach guarantees that individual sensing reports cannot be revealed neither to \fc~nor to the Helper, it incurs high communication overhead. In order to provide high enough security level, the keys of El Gamal cryptosystem, and hence the size of the ciphertexts, need to be very large. This makes the communication cost very high, especially when the number of \su s is large. Moreover, as \fc~can learn aggregated sensing reports of \su s, this scheme is still prone to the {\em DLP} attack explained in Section~\ref{fcattacks}.

Grissa et al.\cite{grissa2016efficient,grissa2017preserving} propose another approach that relies also on {\em order preserving encryption (\ope)} and on deploying an additional node, referred to as {\em gateway (\gw)}. \gw~is deployed to perform private comparisons between \su s' sensing reports and the decision criteria or threshold of \fc. This is done by making each \su~encrypt its \rss, using \ope~and a unique secret key shared with \fc, and send it to \gw. \fc~also sends $n$ encryptions of its sensing threshold, using \ope~and the $n$ keys established with \su s, and sends them to \gw. On top of the \ope~encryption, each entity communicating with \gw~encrypts its data with a key uniquely established with \gw~to secure the communication. \gw~removes the second layer encryption and compares each \ope~encrypted \rss~to its corresponding \ope~encrypted sensing threshold (the one that \fc~has constructed with the same secret key). The main advantage of this approach is its high efficiency in terms of communication and computational complexity due to its reliance on symmetric encryption only. The high efficiency benefits of this technique comes, however, at the cost of needing an additional architectural entity, \gw, that has to be managed by a third party to avoid collusion with \su s or \fc~and to provide the claimed privacy guarantees.

Other approaches consider a different \crn~scenario that consists of multiple service providers (\spr s) that may exchange sensing data among themselves as in~\cite{wang2015privacy}. Wang et al.~\cite{wang2015privacy} propose a framework that aims to preserve \su s' privacy in collaborative spectrum sensing from malicious \spr s. It assumes that the only trustworthy \spr~for a \su~is the one serving it. The remaining \spr s and \su s may collude to infer private information about a target \su, including its location. To preserve \su s' privacy, this framework hides individual sensing data of \su s by making each \spr~transform sensing reports of corresponding \su s into cloaks. To find the optimal cloaking strategy, each \spr~projects its original sensing data to a single-dimensional space, with minimal data distortion~\cite{wang2015privacy}, using a privacy-preserving non-invertible projection and shares statistical information of the projected data with one \spr~picked as a leader. The leader uses this information to decide about the optimal cloaking strategies and shares it with the other \spr s. The authors rely on dynamic programming to obtain the optimal cloaking strategy that minimizes information distortion and that is obtained through collaboration between \spr s. This scheme considers collusion between different malicious entities and provides {\em differential privacy} to \su s. However, its privacy level decreases with the decrease of the number of \spr s and the increase of the number of \su s. It also introduces some distortion to the sensing information which may impact the sensing accuracy.

Some works try also to address the location privacy issue in distributed cooperative sensing. For example, Kasiri et al.~\cite{kasiri2015privacy} address this issue in multi-channel cognitive radio {\em MANET}s. They propose a scheme that relies on the notion of anonymization to prevent location information leakage from \snr~values that are exchanged between \su s for coalition formation purposes. Anonymization is achieved by means of random manipulation and distortion of the exchanged \snr s, which can leak information about the location of \su s as shown in Section~\ref{coopSources}. Each \su~creates an anonymization area with respect to each sensed channel. However, a major limitation of this scheme is that the more channels sensed by a \su~the more likely it is to be located as the adversary can intersect the anonymization areas to narrow down \su's location. Another limitation is that it cannot achieve high location privacy without degrading the sensing performance of the \crn. Indeed, the authors present a tradeoff between privacy and performance as both cannot be maximized together.

Some works try also to preserve the location privacy of users that participate in the crowdsourcing process, which is used to recruit distributed mobile users to sense a given channel around specific locations. For instance, Jin et al.~\cite{jin2016privacy} formulate participants selection process as a reverse auction problem where participants compete to perform spectrum sensing tasks in return for rewards. Each participant's true cost for performing the sensing tasks is closely related to its current location as explained in Section~\ref{attackcoop}. The authors rely on the exponential mechanism to protect the location information and prevent the attack that they have identified (explained in Section~\ref{attackcoop}). Users are selected iteratively for each sensing sub-task following the exponential mechanism to guarantee differential privacy for their bids, and consequently differential location privacy. While protecting location privacy, this approach aims to minimize the social cost that represents the sum of the real costs of users completing all the sensing tasks. However, minimizing this cost deteriorates the location privacy level, which is the main limitation of this approach.

\subsubsection{Performance metrics and tradeoffs}
\paragraph{Performance metrics}
\label{coopPerf}

\noindent \textbf{Computational complexity:} This is an important metric as \su s are usually resource constrained. Thus, it is paramount to consider this when designing a location privacy preserving scheme for \crn s. This metric usually accounts for the overhead resulting from the various operations required by the scheme (e.g., cryptographic operations) and incurred by all different entities involved in the privacy preserving protocol, and could be measured separately for each entity or as a whole for the entire system. Computational complexity has a direct impact on the delay that a \su~may experience before getting the decision about the spectrum availability.
Computational complexity is considered in most of the research works that address the location privacy issue in cooperative spectrum sensing in \crn , e.g.~\cite{li2012location,mao2015protecting,kasiri2015privacy,grissa2015location,grissa2016efficient}.

\noindent \textbf{Communication overhead:}
Communication overhead is another important metric that needs to be considered. Location privacy preserving schemes must not overwhelm the network by incurring high communication overhead that may lead to the degradation of the overall system performance, especially provided that bandwidth and/or energy resources are often limited. Encryption, which most proposed solutions rely on to ensure privacy, tends to incur, depending on the size of ciphertexts, heavy communication overheads. Another factor that also tends to contribute to this overhead is the number of \su s involved in the cooperative sensing task.

\noindent \textbf{Spectrum availability accuracy:}
It is important to protect \su s' location privacy, but while making sure that doing so does not interfere with the cooperative sensing task. Therefore, another important metric is the ability of these privacy preserving schemes to perform the sensing task accurately. This is quantified, for example in~\cite{kasiri2015privacy}, using the detection probability to capture the impact of the privacy preserving scheme on detecting \pu s presence.


\noindent \textbf{Location privacy level:}
As the ultimate goal of any location privacy preserving protocol is to preserve the location privacy of \su s, it is then paramount to have a metric that can be used to assess and quantify the privacy level. There are several metrics that could be used for capturing this:

\begin{itemize}
\item {\em Anonymity level}: This measures the level of anonymity provided by the cloaking algorithm and usually refers to the size of the area to which a \su~generalizes its location to achieve anonymity. One way to quantify this is by computing a relative measure normalized by the anonymity level required by a \su. Kasiri et al.~\cite{kasiri2015privacy} rely on a similar approach and define the location privacy level of a specific \su~as the ratio between the anonymized area with respect to all \pu s and the maximum anonymized area of that \su. The privacy level for the whole network is obtained by computing the average of the location privacy levels over all \su s.

\item {\em Entropy}: This shows how uniform the probability of locating a \su~at a specific position is and it is used to measure the uncertainty level that an adversary has~\cite{shokri2011quantifyingsymp}. Li et al.~\cite{li2012location} have used this concept to quantify the location privacy level of their schemes. The area covered by the \crn~is divided into sub-regions, forming a set $\mathcal{G} = \{g_1,g_2,\cdots,g_m\}$. The uncertainty of the adversary, and thus the location privacy level of a \su~$i$ involved in the cooperative spectrum sensing, is then defined as:
\vspace{-6pt}

\begin{equation}
\label{uncertainty}
\mathcal{A}(i) = - \sum_{b=1}^{m} p_{i|b} \log(p_{i|b})
\end{equation}
where $p_{i|b}$ is the probability that \su~$i$ is located in sub-region $g_b$. The location privacy level for the overall system is then given by
$\mathcal{A} = \sum_{i=1}^{n} \mathcal{A}(i)$,
where $n$ is the number of \su s. If an attacker can uniquely infer that \su~$i$ is located at sub-region $g_b$, then $p_{i|b} = 1$, i.e. $\mathcal{A}(i) = 0$. On the other hand, if the attacker is unable to tell which sub-region \su~is located in, which means \su~could be located at any region with equal probability $p_{i|b} = 1/m$, then the privacy level for \su~$i$ would be $\mathcal{A}(i) = \log m$, which is the maximum privacy level it can get when participating in the cooperative sensing.

\item {\em $\epsilon$-differential privacy}:
This concept is based on the differential privacy concept (discussed in Section~\ref{limitGeneric}). A mechanism $\mathcal{M}$ is said to provide $\epsilon$-differential privacy for a \su~$i$ if for any two sets of sensing reports, $R = [r_1,\cdots,r_i,\cdots,r_\nbr]$ and $R' = [r_1,\cdots,r'_i,\cdots,r_\nbr]$, that differ only on $i$'s sensing report, we have:

\begin{equation}
 \vert\ln \frac{Pr[\mathcal{M}(R)= \mathcal{O}]}{Pr[\mathcal{M}(R')= \mathcal{O}]} \vert \leq \epsilon
\end{equation}

for all $\mathcal{O} \in Range(\mathcal{M})$ with $Range(\mathcal{M})$ is the set of all possible outputs of $\mathcal{M}$~\cite{wang2015privacy} .
The privacy level is controlled by the parameter $\epsilon$ with higher privacy is ensured by lower $\epsilon$ values. Very low values of $\epsilon$~ensure that $Pr[\mathcal{M}(R)= \mathcal{O}]$ and $Pr[\mathcal{M}(R')= \mathcal{O}]$ are roughly the same, meaning that the output $\mathcal{O}$ is not sensitive to the changes of any single \su's sensing reports.

\end{itemize}

Location privacy could also be quantified using the concepts of {\em inaccuracy} and {\em incorrectness} introduced by Shokri et al.~\cite{shokri2011quantifyingsymp}. These concepts could be redefined to fit the context of location privacy in \crn s as done in~\cite{bahrak2014protecting}. First, let $\Theta$~denote the observed sensory information that could be used to localize a \su, and $x$ and $x_c$ represent the location estimated by the attacker and the actual \su's location, respectively. Let also $p(x|\Theta)$ be the probability distribution of all possible values of the target \su's location given the observed information. Essentially, this probability models the adversary's extracted information from its observations.
\begin{itemize}

\item {\em Inaccuracy}: This is the discrepancy between the posterior distributions $p(x|\Theta)$ and $\hat{p}(x|\Theta)$ which basically quantifies the difference between \su's real location distribution and the adversary's estimated location distribution.

\item {\em Incorrectness}: This is the distance (or expected distance) between the true \su's location and that inferred by the attacker. This metric is shown in~\cite{shokri2011quantifyingsymp} to be the most appropriate for quantifying location privacy. The expected distance, which is the adversary's expected estimation error, can be written as
$\sum_x \hat{p}(x|\Theta) \Vert x - x_c\Vert$,
%
where $\Vert\cdot\Vert$ is a distance, e.g. euclidean, between $x$ and $x_c$.
\end{itemize}




\paragraph{Performance tradeoffs}
Several performance tradeoffs could be made when designing location privacy preserving schemes for cooperative spectrum sensing:

\textbf{Scheme overhead vs. hardware cost:}
Scheme overhead in terms of communication, computation, and/or energy could be reduced at the cost of additional architectural components. For example, Grissa et al.~\cite{grissa2016efficient} introduce and rely on an extra network entity to reduce both communication and computational overheads while also improving privacy.
This reduction in overhead is achieved by means of this new entity, introduced to carry out the private comparisons between \su s and \fc~without disclosing \rss~values.
Without such an entity, these comparisons would have been very expensive, resulting in an excessive scheme overhead.
%

\textbf{Privacy level vs. scheme overhead:}
Achieving higher location privacy at the cost of deploying more expensive cryptosystems with higher communication and/or computation overhead is another tradeoff researchers often make. For example, the works in~\cite{li2012location,mao2015protecting,grissa2015location} make such tradoffs in order to improve the location privacy of their schemes.
%

\textbf{Privacy level vs. sensing accuracy:}
Higher location privacy can also be obtained at the cost of willing to degrade the sensing performance of the \crn. For example, such a tradeoff is made in the approach proposed by Kasiri et al.~\cite{kasiri2015privacy}, where the anonymization area, capturing the privacy level, is increased but at the cost of decreasing the average detection probability, representing the \crn~sensing performance.


\subsection{Location privacy in database-based spectrum discovery}
\label{DBdiscovery}

Here, the location privacy issue is completely different from that of the cooperative sensing-based \crn s. In fact, as explained in Section~\ref{db}, each \su~is now required to send its exact location to \db
~in order to learn about spectrum opportunities in its vicinity. This makes preserving the location privacy of \su s more challenging, since an adversary does not need to perform any extra computation to estimate the position, and the location information here could be easily extracted from the query itself.
Thus, location information preserving schemes for database-based \crn s need to be designed with two conflicting goals: $i)$ hiding or not including \su's location information in the query to be sent to \db,
and $ii)$ in response to a \su's query, \db~needs to inform \su~about spectrum availability in \su's vicinity.
The second goal above somehow entails that \db~needs to know where \su~is located at, and thus,
meeting these two conflicting requirements is very challenging. As we will see later, this cannot be achieved without making some performance tradeoffs.

\subsubsection{Threat models}
Several threat models are considered in the literature to study and address \su s' location privacy issue in database-driven \crn s:

\begin{itemize}
\item {\em Dolev–Yao threat model}: The adversary, usually an intruder, can overhear, intercept, and synthesize any message exchanged between \su s and \db. More specifically the adversary can learn the location of an \su~from the query that the latter sends to \db~to learn spectrum opportunities. The adversary here is only limited by the constraints of the used cryptographic schemes~\cite{dolev1983security}. This model has been considered in several works~\cite{gao2013location,zhang2015privacy}.
\item {\em Semi-honest} or {\em honest-but-curious threat model}: The adversary, usually \db, follows the sensing protocol honestly without changing any of its parameters, but shows some interest in learning the location of target \su s~\cite{zhang2015optimal,li2015agent,troja2015efficient,gao2013location}. This means that it responds to \su s queries with correct spectrum availability information, but at the same time tries to learn their whereabouts.
\item {\em Malicious-entity threat model}: \db, or an intermediate \bs, may be malicious, i.e. they can change protocol parameters to localize a target \su~that is querying \db. In some situations, the malicious entity could even be a sophisticated adversary that has considerable resources and has access to information from \db~\cite{zhang2015achieving}.

\end{itemize}

\subsubsection{Location inference attacks}
%
The most straightforward and basic attack is based on \su's query content. A \su~needs to include its exact location in its query to \db. This makes it vulnerable to an intruder, that can learn its location by eavesdropping its queries, or even to \db~that has access to these queries.
Typically, \db's response to a \su's query contains spectrum availability information; e.g., the list of available channels in \su's vicinity and the maximum allowed transmit powers in each of these available channels. An adversary that has access to this information could localize a target \su~by overlapping the availability areas of the different channels available at \su's location as explained in Section~\ref{sourcesDB}. This kind of attack assumes that the adversary has knowledge about the RF environment covered by \db~as well as the activity and coverage of \pu s. The adversary can also exploit the fact that the allowable secondary transmit powers are highly correlated to the relative distance between a \su~and a \pu~as discussed in Section~\ref{sourcesDB}. This has been exploited by Zhang et al.~\cite{zhang2015optimal} to identify a unified attack framework to localize both \su s and \pu s based on the $MTP$ function introduced in~\cite{bahrak2014protecting}. The $MTP$ calculated by \db~is divided into several levels based on the distance between \su~and \pu. Specifically, when this distance is less than a certain protection radius, \su~is not permitted to transmit on \pu's channel. Beyond the protection radius, \su~can transmit at an increased power level as its distance from \pu~increases until it reaches the maximum allowed transmit power as regulated by FCC.

\subsubsection{Location privacy preserving approaches}

We summarize the approaches that are proposed in the literature to cope with the location privacy issue in database-based spectrum discovery in Table~\ref{solDB} and we discuss them in more details in the following. Generally speaking, most existing techniques attempt to protect \su s' location privacy by adopting one of two techniques/concepts: {\em k-anonymity}~\cite{sweeney2002k} or {\em \pir} ({\em private information retrieval})~\cite{chor1998private}.

As discussed in Section~\ref{limitGeneric}, {\em $k$-anonymity}-based approaches try to ensure that the probability of identifying the location of a querying \su~remains under $1/k$, where $k$ is the size of the anonymity set to be received by the untrusted \db. {\em k-anonymity}-based approaches are known to suffer from one major problem: they cannot achieve high location privacy without incurring substantial communication/computation overhead. Furthermore, it has been shown in a recent study led by Sprint and Technicolor~\cite{zang2011anonymization} that anonymization based techniques are not efficient in providing location privacy guarantees, and may even leak some location information.

For instance, Zhang et al~\cite{zhang2015optimal} rely on the {\em k-anonymity} concept to provide a location privacy preserving mechanism to protect the location privacy of both \pu s and \su s. The proposed scheme requires that each \su~queries \db~by sending a square cloak region that includes its actual location instead of just sending this location. \su~keeps querying \db~using the same cloak region to avoid further location information leakage. This scheme requires a tradeoff between high location privacy and spectrum utility, which means that achieving a high location privacy level results in a decrease in spectrum utility. This limits the applicability of this kind of approaches as they impact the main goal of \crn s which is optimizing spectrum utilization efficiency. As discussed earlier, a good approach should provide location privacy to \su s but without hindering the functioning of \crn s.

{\em $k$-anonymity} is also used by Li et al.~\cite{li2015agent} to protect \su s' location privacy during the commitment phase in which \su s have to register the channels that they are planning to use as explained in Section~\ref{sourcesDB}. In this approach, \su s first send their channel requests to the \bs~that they are associated with, using pseudonyms that are randomly generated by a certification authority. \bs, then, queries \db~on behalf of the querying \su s using their pseudonyms. After that, \db~performs hash matching of \su s' pseudonyms with a hash matrix provided by the certification authority to verify \su s' pseudonyms. Subsequently, \db~assigns a set of channels to \bs~based on the latter's location. \bs~then allocates the channels to its \su s using a coloring model to prevent interference between them. Finally, \bs~registers the used channel of each \su~in \db~by including dummy information to provide {\em k-anonymity} to the utilization information. This is done by registering more channels than the number of \su s' requests to confuse attackers and prevent them from using the utilization information to localize \su s. Using \bs~to register the used channels helps cutting off the relation between the registered channels and \su s' identities, which makes it harder for \db~to associate this information to corresponding \su s and, hence, localize them. Thus, the proposed scheme can decrease the probability of localizing \su s. However, it requires that \bs~is trustworthy or it would not be able to protect \su s' location. This assumption is not usually realistic as it is hard to guarantee trustworthiness in practice. It suffers from the fact that the probability of localizing \su s increases as the number of switching events increases or as the number of \bs s decreases.

\begin{table*}
\centering
\caption{\small Location privacy preserving schemes in database-driven spectrum opportunities discovery}
\label{solDB}
\resizebox{\textwidth}{!}{%
\renewcommand{\arraystretch}{1.25}{
\begin{tabular}{@{}lp{4cm}lp{5.5cm}p{5.5cm}@{}}
\toprule[1.5pt]
Countermeasures & Attacks Considered                                                                                                       & Techniques                                                                                       & Pros                                                                                           & Cons                                     \\ \midrule

Zhang et al.~\cite{zhang2015optimal}    &                                                                                                                         \begin{tabular}[c]{@{}p{4cm}@{}}- Location inference from maximum transmission power\\ - Location inference from channel switch\end{tabular} &                                                                                                \begin{tabular}[c]{@{}p{4cm}@{}} - Cloaking the query of \su~within a square region based on {\em k-anonymity}\end{tabular} &                                                                                               - Provides location privacy for both \su s and \pu s &             - High location privacy degrades spectrum utility                             \\ \addlinespace[5pt] \hline \addlinespace[5pt]

Li et al.~\cite{li2015agent}   & - Location inference from spectrum utilization information  & \begin{tabular}[c]{@{}p{4cm}@{}}- Intermediate base stations to forward \su s' queries to \db\\- Intermediate base stations for spectrum allocation \\ - {\em k-anonymity} for registering used channels\end{tabular} & \begin{tabular}[c]{@{}p{5.5cm}@{}} - Adversaries cannot link usage information to \su s \\ - Decreases \su s' geolocation probability \end{tabular} &  \begin{tabular}[c]{@{}p{5.5cm}@{}} - The probability of geolocating \su s increases with the number of available channels. \\ - The probability of geolocating \su s increases with the number of switching events \end{tabular} \\ \addlinespace[5pt] \hline \addlinespace[5pt]

Gao et al.~\cite{gao2013location}      & \begin{tabular}[c]{@{}p{4cm}@{}}- Location inference from query\\ - Location inference from spectrum utilization information\end{tabular} & \begin{tabular}[c]{@{}p{4cm}@{}}- Query blinding via \pir\\ - Spectrum mobility reduction\end{tabular} & \begin{tabular}[c]{@{}p{5.5cm}@{}}- Low communication overhead\\ - Reduces the localization probability of \su s \end{tabular} & \begin{tabular}[c]{@{}p{5.5cm}@{}} - High computational overhead                    \end{tabular} \\ \addlinespace[5pt] \hline \addlinespace[5pt]

Grissa et al.~\cite{grissa2015cuckoo}    &                                                                                                                         - Location inference from query &                                                                                                 \begin{tabular}[c]{@{}p{4cm}@{}}- Sending portion of \db~to \su~using cuckoo filter \end{tabular}&                                                                                                \begin{tabular}[c]{@{}p{5.5cm}@{}}- Very low computational overhead\\ - Provides ideal location privacy\end{tabular} &                     - Large communication overhead if \db~is huge                     \\ \addlinespace[5pt] \hline \addlinespace[5pt]

Troja et al.~\cite{troja2014leveraging}    &  - Location inference from query  &   \begin{tabular}[c]{@{}p{4cm}@{}} - Collaboration between \su s \\ - {\em private information retrieval} \end{tabular} &                                                                                               \begin{tabular}[c]{@{}p{5.5cm}@{}} - Minimal number of \pir~queries via collaboration between \su s \\ - Takes into account \su's mobility \end{tabular} &     \begin{tabular}[c]{@{}p{5.5cm}@{}}- Large communication overhead                                    \\ - Relatively high computational overhead \end{tabular} \\ \addlinespace[5pt] \hline \addlinespace[5pt]
Troja et al.~\cite{troja2015efficient}    &                                                                                                                         - Location inference from query & \begin{tabular}[c]{@{}p{4cm}@{}}- Hilbert space filling curve indexing of \db \\ - {\em private information retrieval} \end{tabular}&  \begin{tabular}[c]{@{}p{5.5cm}@{}} - Takes into account \su's mobility \\ - Minimal number of \pir~queries via trajectory prediction \end{tabular} & - Relatively high computational overhead                                          \\ \addlinespace[5pt] \hline \addlinespace[5pt]

Zhang et al.~\cite{zhang2015achieving}    &   - Location inference from query &                                                                                                 \begin{tabular}[c]{@{}p{4cm}@{}}- Random obfuscation using Laplacian noise \end{tabular}&  - Provides differential location privacy for both \su s and \pu s &               - Increasing the location privacy level decreases the utility of both \pu s and \su s \\  \bottomrule[1.5pt]

\end{tabular}}}
\vspace{-8pt}
\end{table*}
\pir-based approaches~\cite{gao2013location,troja2014leveraging,troja2015efficient}, on the other hand, offer much
better privacy than {\em $k$-anonymity}-based approaches, but incur substantial computation and communication overhead, thus limiting their practical use for \crn s~\cite{ghinita2008private}, unless used judiciously as discussed in Section~\ref{limitGeneric}. For instance, Gao et al.~\cite{gao2013location} propose a \pir-based location information preserving scheme by adopting the \pir~protocol of Trostle et al.~\cite{trostle2010efficient}. Instead of sending its location, \su~hides its coordinates within other locations and transforms this information in such a way that \su~is the only one that can revert it. Upon receiving the blinded query, \db~multiplies it with the spectrum availability information matrix and sends the outcome back to \su. \su~will be able to only retrieve the availability information in its location using the secure parameters that it used to transform the original query. \su~is the only one who knows the blinding factors and the transformation used to transform the original query. Hence, only \su~can recover the spectrum availability information from the result sent by \db. However, this approach suffers from large computational overhead which is due to the use of the \pir~protocol, known to be expensive to execute as we highlighted earlier.

Grissa et al.~\cite{grissa2015cuckoo} propose an approach that offers an unconditional privacy to \su s within the \db's coverage area. This approach uses set membership data structure, more precisely {\em cuckoo filter}~\cite{fan2014cuckoo}, to send a compressed version of \db~to~\su. In this scheme, \su~only sends its characteristics, but not its location, to \db, which it uses to adapt the content of the {\em cuckoo filter}. After receiving the filter, \su~constructs a query that includes its location and a combination of other parameters (e.g.  band frequency, transmission power level, etc) and queries the filter to check whether it contains the constructed query. If it is the case, \su~can deduce that the channel is available and can use it by following the parameters specified in the query. Otherwise, \su~concludes that the specified combination does not exist in \db~and keeps querying the filter with different combinations until it finds one or reaches the filter's capacity. Obviously, the main advantage of this scheme is that it provides optimal location privacy to \su s as opposed to the other approaches. However, it incurs a relatively large communication overhead especially when the size of \db~is huge. The authors try to address this issue by proposing to sacrifice one of \su's coordinates to considerably reduce the size of the filter while providing reasonable privacy. This is not needed when the size of \db~is not large.

Troja et al.~\cite{troja2014leveraging} propose another \pir-based approach to protect the location privacy of mobile \su s. The \pir~mechanism used in this work allows a \su~to learn spectrum availability in multiple-cell block containing its current cell. As they move, \su s gradually develop a trajectory-specific spectrum knowledge cache, via a series of \pir~queries. \su s within communication range of each other form groups and interact in a peer-to-peer (P2P) manner to privately exchange their anonymized cached channel availability information. This reduces considerably the number of \pir~queries as less \su s need to query \db~since they could learn opportunities from \su s within their group. However, this still incurs large communication cost and relatively high computational overhead, especially when the group size is relatively large.

Troja et al.~\cite{troja2015efficient} propose another \pir-based privacy-preserving protocol that relies on the Hilbert space filling curve which is a continuous fractal
that maps space from {\em 2-D} to {\em 1-D}~\cite{kamel1993packing}. \db~is indexed based on this curve to address \su s' mobility which allows neighboring cells to be stored in consecutive locations in \db. \db~is split into multiple disjoint segments which enables \su~to retrieve channel availability information for a large number of consecutive cells surrounding \su's location with a single \pir~query. \su s use trajectory information, known a priori or generated on the fly via a prediction mechanism, to minimize the number of future \pir~queries as a \su~can obtain availability information for current and future positions in just one query. Despite its merit in providing location privacy to mobile \su s with efficient communication overhead, this approach incurs relatively large computational overhead. The main advantages of this scheme are that it considers mobile \su s and exploits trajectory information to reduce the number of \pir~queries to \db~in order to reduce overhead. However, it still suffers from one of the well known limitations of \pir-based approaches, i.e. the high computational overhead, despite its nice effort in reducing the number of required queries.

\vspace{-2pt}
Other approaches try to adapt the {\em differential privacy} concept, explained in Section~\ref{limitGeneric}, and apply it in the context of database-driven \crn s. For instance, Zhang et al.~\cite{zhang2015achieving} propose an approach to protect bilateral location privacy of both \pu s and  \su s. \su s obfuscate their location using a two dimensional {\em Laplacian} distribution noise satisfying the {\em $\epsilon$-geo-indistinguishability} mechanism, derived from {\em differential privacy}, introduced in~\cite{andres2013geo}. The obfuscation depends on the privacy preserving level that is decided by both \su s and \pu s by solving an optimization problem that maximizes their bilateral utility. \su~sends its obfuscated location along with the privacy level which represents the maximum distance that separates the sent location from the actual location. Based on these parameters, \db~decides about the transmit power and radius or distance from \pu~that~\su~cannot exceed. The main advantage of this approach is that it provides differential location privacy for both \pu s and \su s while allowing them to adjust their privacy level to maximize their utility. However, as this approach aims to maximize both the utility and privacy level, which are always conflicting, increasing the privacy level of both \pu s and \su s often results in decreasing their utility, and striking a balance is challenging.

\subsubsection{Performance metrics and tradeoffs}
\paragraph{Performance metrics}

\textbf{Computational complexity:}
Making sure that these schemes do not require heavy computation at both ends, \su~and \db, is crucial to the design of such schemes. This is important merely because these \su~devices, again, are usually resource constrained (in both energy and CPU), and the applications running on them may not tolerate delays. In addition, it is highly desirable not to overwhelm \db~by involving it in heavy computations, which can lead to congestion.
Several works (e.g.,\cite{troja2015efficient,gao2013location,grissa2015cuckoo,troja2014leveraging}) use this as a metric for assessing the effectiveness of their proposed approaches. For example, Troja et al.~\cite{troja2015efficient} captures the computation overhead by measuring the average cumulative response time that their proposed scheme leads to. This time includes the query generation time at \su, the processing time at \db, the network transfer time, and the resulting extraction time at \su.

\textbf{Communication overhead:}
Another crucial performance metric is to assess how much network data the proposed scheme generates. This assesses whether adding a privacy preserving scheme would inundate the network and degrade its performance. Indeed, a large communication overhead may introduce a considerable delay that may leave the spectrum availability outdated and cause interference to \pu s if \su s decide to use channels based on this outdated information.

\textbf{Location privacy level:}
In addition to the privacy concepts already discussed in Section~\ref{coopPerf}, the following can be used to assess the privacy level of any given scheme.
\begin{itemize}
\item {\em Localization probability:} This is basically the probability that a \su~is geolocated successfully by an attacker under a given scheme. It may be influenced by different parameters, e.g. the number of channel switching events, the number of \bs s in the network, etc. Some approaches like~\cite{li2015agent} have considered this metric to evaluate their approach's privacy level.
\item {\em Size of possible location set:} This measures the granularity of the location that an attacker can infer about a \su. A privacy preserving scheme fails completely to protect the location of a \su~if the size of this set is equal to $1$, which means that the attacker has succeeded to determine the exact cell in which \su~is located~\cite{gao2013location}.

\end{itemize}

\paragraph{Performance tradeoffs}
\textbf{Location privacy vs. spectrum utilization:} This tradeoff consists on sacrificing some utility to provide high location privacy guarantees. This means that seeking a higher privacy level will necessary reduce the utility in question. For instance, Zhang et al.~\cite{zhang2015optimal} make a tradeoff between the location privacy of both \su s and \pu s, and spectrum utilization. \su s and \pu s can adjust their privacy levels to maximize their utilities. In this case, increasing the location privacy level would decrease the spectrum utilization and vice versa.

\textbf{False positive rate vs ideal privacy:} Some approaches, like~\cite{grissa2015cuckoo}, use set membership data structures to construct a compact representation of \db~and make \su s query it for spectrum availability. However, this kind of data structures, despite its efficiency in compacting large sets of data, could introduce some false positives when it is queried. This means that the result of query may reveal that a channel is available while in reality it is not. Some data structures, like the {\em cuckoo filter} used in~\cite{grissa2015cuckoo}, give the possibility to control this rate. Minimizing this rate will, however, increase the communication overhead. So the tradeoff here is to allow some false positives in the filter to guarantee ideal privacy to \su s.

\subsection{Summary}
In this section, we discussed the location privacy issues in the spectrum opportunity discovery component for both cooperative spectrum sensing-based and database-driven spectrum discovery. We detailed the different threat models and attacks that target the location information of \su s. We then presented the different approaches that are proposed in the literature to deal with these issues. Finally, we explained the different performance metrics that are or could be used to assess the efficiency and the privacy level of location privacy preserving protocols in \crn s. In the following section, we will follow the same structure and reasoning to discuss the location privacy issues in the remaining \crn~components.

\section{Location privacy preservation in other \crn~components}
\label{lpoc}

In this Section, we investigate \su s' location privacy issue in the remaining \crn~components of the cognition cycle. Unlike the spectrum opportunity discovery component, much less attention has been given by the research community to the location privacy issue in these components. The design goals of privacy preserving schemes for each of these components are then to address the sources of location information leakages discussed in Section~\ref{sourcesDec} (spectrum analysis), Section~\ref{sourcesSharing} (spectrum sharing), and Section~\ref{specMobility} (spectrum mobility).

\ccomment{
\subsection{Design challenges of location privacy preserving protocols in the remaining components}
There are some challenges that may impede designing privacy preserving protocols to protect the location information of \su s during spectrum decision, spectrum sharing and spectrum mobility phases. Here we list some of these challenges.

\begin{itemize}
\item {Hide bids information during spectrum trading:} As this is shown to be a significant source of location information leakage, a privacy preserving protocol should be able to conceal this information while preserving the trading process for spectrum sharing.

\item {Leverage accurate decision} As discussed in Section~\ref{sourcesDec}, several sources of location information leakage could be exploited, e.g. interference, \rem, topology, connectivity information, etc. A location privacy preserving spectrum decision protocol needs to determine the best spectrum bands to be used among \su s but at the same time needs to address the aforementioned vulnerabilities.

\item {Hide sensitive information during spectrum mobility} As highlighted previously, several vulnerabilities may arise from the decision of a \su~to switch to another channel. A location privacy preserving protocol has to allow \su s perform the spectrum handoff, if needed, while minimizing the location information that could be leaked from this process.
\end{itemize}
}

\subsection{Threat models}
The same threat models that we have discussed previously in the spectrum opportunity discovery phase apply to the remaining components of the cognition cycle. Thus, we skip these threat models here and we refer the reader to Sections~\ref{CPdiscovery} (cooperative spectrum sensing) and~\ref{DBdiscovery} (database-based spectrum opportunity discovery) for more details.

\subsection{Location inference attacks}
\label{attacksOther}
\begin{figure}[th!]
\centering
\includegraphics[width=0.4\textwidth]{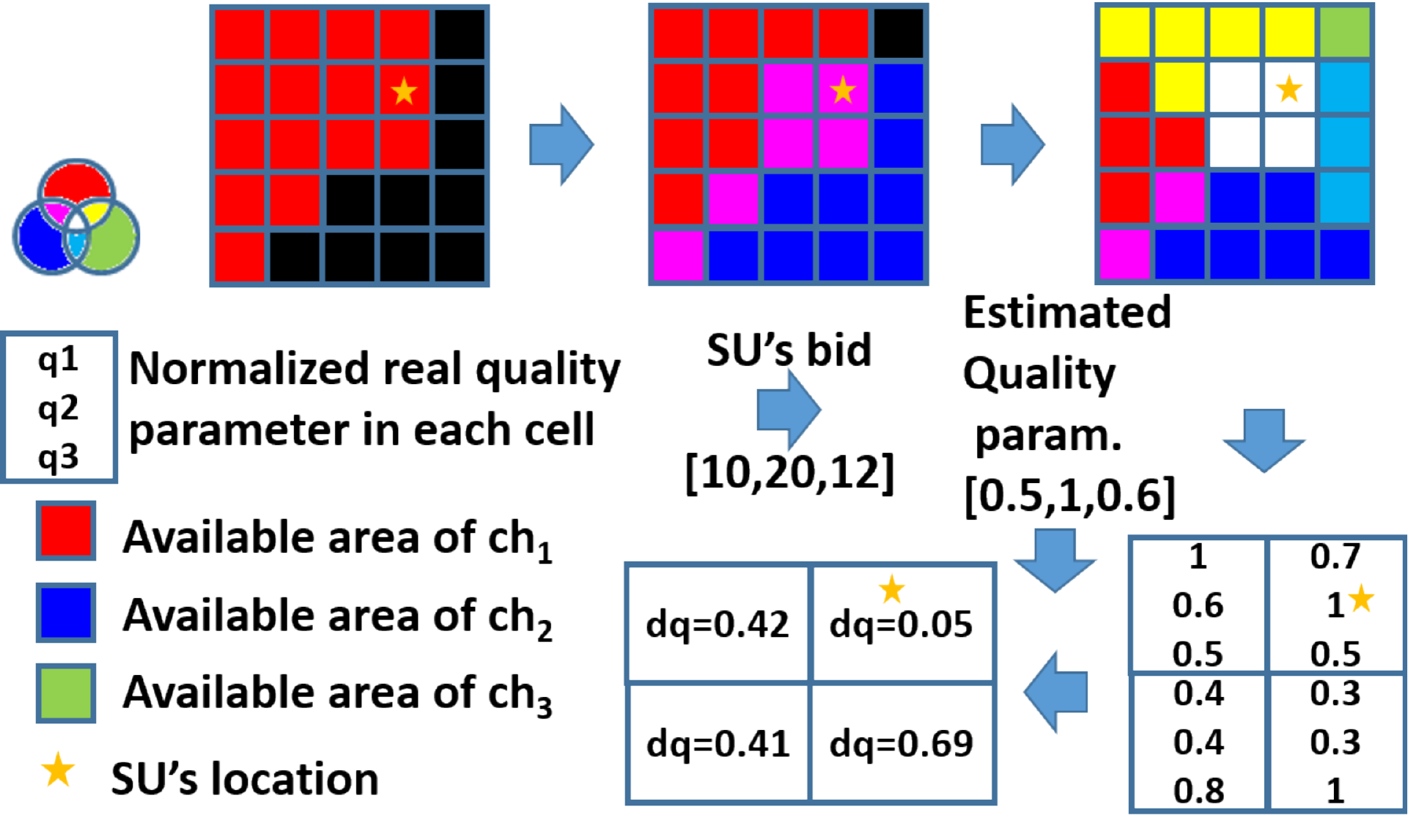}
\caption{{\small An example of the attacks identified in~\cite{liu2013location} which first estimate the position of an \su~to be in the intersection of the available areas of channels $1$, $2$ and $3$. Then, the attacker further narrows down the estimated area by picking the cell having the smallest distance between the exact channels' qualities and those estimated from bid prices.}}
\label{bpmbcm}
\end{figure}

Some of these attacks may target \su's location during the dynamic spectrum auction process. For instance, Liu et al.~\cite{liu2013location} identify an attack that exploits two sources of leakage, highlighted in Section~\ref{sourcesTrading}: bid channels and bid prices. The first attack uses bid channels (i.e. channels that are bid for by a \su). As explained earlier, a \su~bids only for channels that are available for it, i.e. \su~belongs to the complement area of each corresponding \pu's coverage. Hence, a malicious auctioneer can use the \su's available set of channels, obtained from the \su's bids, to decrease its possible location range by intersecting the complements of the corresponding \pu's coverage areas as shown in Figure~\ref{bpmbcm}. 
The second attack exploits the bid prices, which depend on the quality and characteristics of the spectrum known to be highly correlated to \su's location. It could be used after the first attack to further narrow down the possible location area of the target \su. A higher bid price means that the \su~perceives a high spectrum quality, and hence, the auctioneer can estimate the channel quality perceived by a \su~from the \su s' bid price information. Since an attacker can easily have (or can reasonably be assumed to have) access to the statistics of channels' qualities in each cell, it can then compute the distance between these exact channels' qualities and those estimated from bid prices. The cell with the minimum distance corresponds then to \su's location with high probability, as depicted in Figure~\ref{bpmbcm}.

\begin{figure}[h!]
\centering
\includegraphics[width=0.4\textwidth]{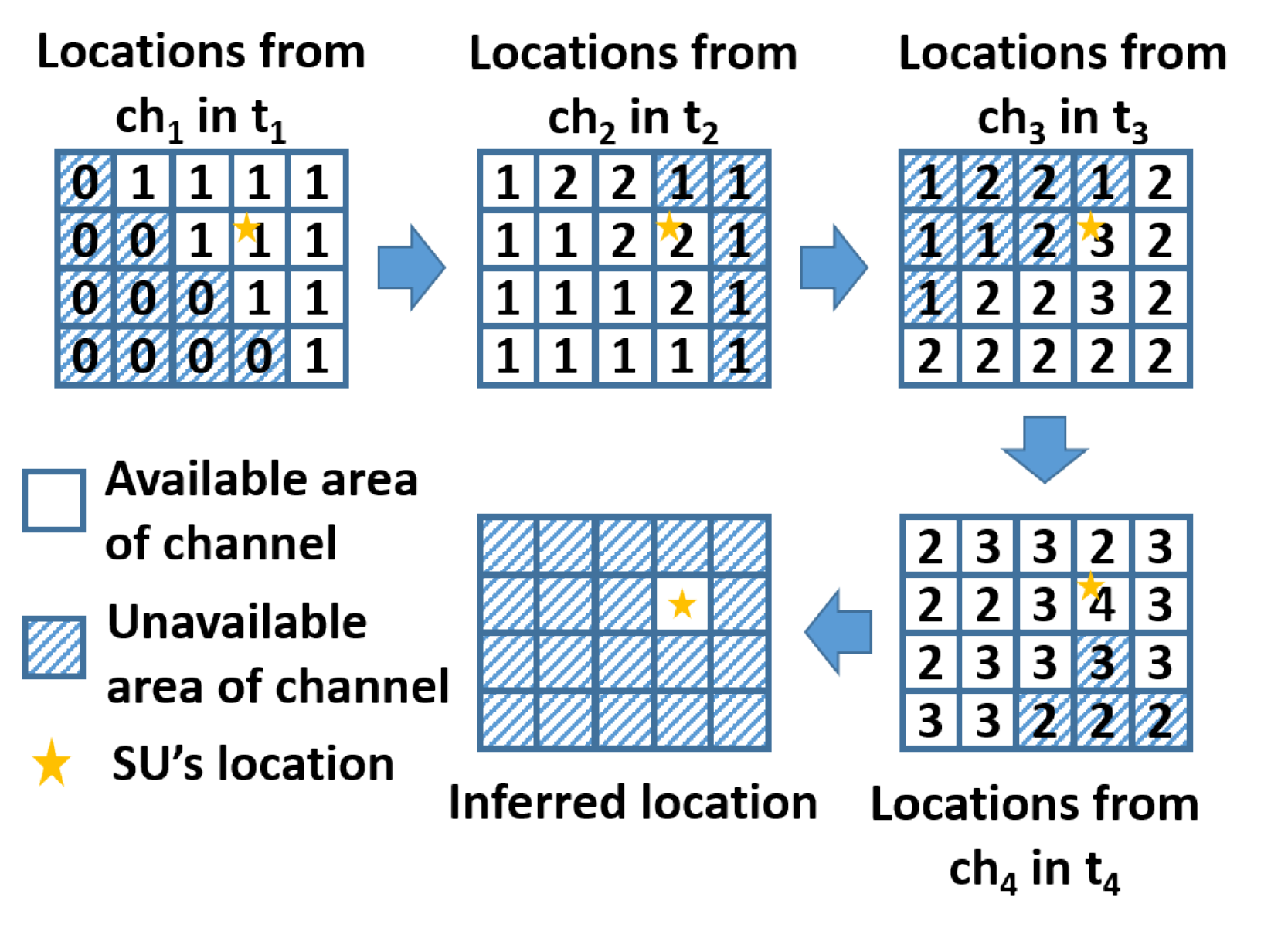}
\caption{{\small An example of the attack identified in~\cite{gao2013location} which uses the complement of the coverage area of each transmitting \pu~to gradually localize an \su~by incrementing a score for each cell situated outside the coverage area of each \pu. The inferred location willl be the cell with the highest score.}}
\label{pucc}
\end{figure}

Other attacks may exploit the spectrum utilization information to localize \su s as explained in Section~\ref{specMobility}. Gao et al.~\cite{gao2013location}, for example, identify an attack that infers \su s' location in database-driven \crn s by exploiting the channels' utilization information.
The first component of the proposed attack arises from the fact that a \su~cannot access a \pu~channel if the \pu~is present, and hence, if a \su~is active in the presence of a \pu, then the \su~must be outside the \pu's coverage area. This gives the attacker a clue that the \su~is located at the complement of the \pu's coverage area.
If the \crn~covered area is modeled as a grid, as shown in Figure~\ref{pucc}, the adversary keeps incrementing a score, initially initialized to $0$, for each cell that belongs to an available area of a specific channel. The location of the target \su~will be the cell with the maximum score, which represents the area where all available areas of the channels overlap as illustrated in Figure~\ref{pucc}. 
The second component of the proposed attack relies on the fact/event that a \su~plans to switch from some channel $\ch_{k_1}$ to another channel $\ch_{k_2}$ when $\pu_{k_1}$ returns to its channel. In this situation there are two possible scenarios: First, when $\pu_{k_2}$ is also present and is using its channel $\ch_{k_2}$. In this case, since \su~cannot interfere with $\pu_{k_2}$, the attacker can learn that the target \su~is situated in the $\pu_{k_1}$ coverage area  and the complement of $\pu_{k_2}$ coverage area. Second, when $\pu_{k_2}$ is absent. In this case, the adversary can learn that \su~must be within the coverage area of $\pu_{k_1}$, as it must have switched to $\ch_{k_2}$ after $\pu_{k_1}$'s return. This same attack is also used by Zhang et al.~\cite{zhang2015optimal} as a second component of their attack framework.

Physical-layer information based attacks are also possible during the spectrum sharing process.
In fact, an adversary can directly extract position-related parameters like \rss, \aoa, \toa, etc, from \su s' signals and exploit them to locate \su s, as explained in Section~\ref{attackcoop}. As an example, this kind of attacks is considered by Zhang et al.~\cite{zhang2015privacy}.

\subsection{Location privacy preserving approaches}
\label{lpparc}
Few works have addressed the location privacy issue in spectrum sharing and mobility but none, to the best of our knowledge, have addressed this problem during spectrum analysis phase. These works are summarized in Table~\ref{solSharing}.

\begin{table*}
\centering
\caption{\small Location privacy preserving schemes in spectrum sharing and spectrum mobility}
\label{solSharing}
\resizebox{\textwidth}{!}{%
\renewcommand{\arraystretch}{1.25}{
\begin{tabular}{@{}lp{4cm}p{4cm}p{5.5cm}p{5.5cm}@{}}
\toprule[1.5pt]
Countermeasures                                & Attack Considered           & Techniques                & Pros                    & Cons                                                                             \\ \midrule

Liu et al.~\cite{liu2013location}   & - Location inference from Bid channels and prices & \begin{tabular}[c]{@{}p{4cm}@{}} - Prefix membership matching \\ - HMAC \end{tabular} &  \begin{tabular}[c]{@{}p{5.5cm}@{}}- Efficient in thwarting attacks that use bid prices  \\ - Defends to some extent against attacks that exploit bid channels \end{tabular} &   \begin{tabular}[c]{@{}p{5.5cm}@{}} - Requires a {\em trusted third party} \\ - Requires a tradeoff between location privacy and auction performance \end{tabular} \\ \addlinespace[5pt] \hline \addlinespace[5pt]

Zhang et al.~\cite{zhang2015privacy}    & - \rss-based PHY-layer attack & - Random power perturbation & - Mitigates a PHY-layer attack which is usually hard to thwart &  - High location privacy level incurs significant degradation of network throughput \\ \addlinespace[5pt] \hline \addlinespace[5pt]

Gao et al.~\cite{gao2013location}      & - Location inference from spectrum utilization information &  - Spectrum mobility reduction & \begin{tabular}[c]{@{}p{5.5cm}@{}}- Low communication overhead\\ - Reduces the localization probability of \su s \end{tabular} & \begin{tabular}[c]{@{}p{5.5cm}@{}} - High computational overhead \\ - The localization probability of \su s increases with the increase of channel switches.                   \end{tabular} \\
 \bottomrule[1.5pt]
\end{tabular}}}
\vspace{-5pt}
\end{table*}

\subsubsection{Spectrum sharing}

Some approaches try to prevent the location information leakage by hiding sensitive information exchanged during spectrum auction, e.g. location, bid channels, and bid prices, as discussed in Section~\ref{sourcesSharing}. Liu et al.~\cite{liu2013location} propose an approach that aims to preserve the location privacy of the \su s that participate in spectrum auction. This approach consists of two main components: The first component enables \su s to submit their encrypted locations and bid prices, while allowing the auctioneer to construct the conflict graph (explained in Section~\ref{selectionassignment}) and determine the maximum bid price. This is done using {\em HMAC}~\cite{krawczyk1997hmac} and the prefix membership verification scheme proposed in~\cite{chen2010safeq}. The second component enables the auctioneer to launch the auction using a greedy spectrum allocation algorithm to allocate the spectrum among \su s and a charging algorithm to securely determine the winning bids with the help of a trusted third party. Despite its merit in reducing the effectiveness of some of the attacks presented in Section~\ref{attacksOther}, and increasing the location privacy of \su s by hiding the bid prices and channels, this scheme suffers from some limitations. First, it relies on a trusted third party which is not always realistic. Second, it cannot achieve high location privacy without degrading the auction's performance.

Other approaches try also to prevent physical-layer based attacks during spectrum sharing, where attackers can capture the target \su s' transmitted signal when they try to access the spectrum and use it to extract position related measurements like \rss, \toa, \aoa, etc, as explained in Section~\ref{coopSources}. For instance, Zhang et al.~\cite{zhang2015privacy} try to prevent attackers from measuring \rss~and using it to localize \su s following some of the  approaches presented in Section~\ref{attackcoop}. The authors propose to rely on a random power perturbation approach where \su s perturb their power transmission level to obfuscate their \rss~values measured at the adversary side. This perturbation consists of reducing the transmission power to prevent an attacker from correctly estimating \su s' positions. They also provide a design of a socially-aware spectrum sharing algorithm that can operate well together with the power perturbation based privacy protection approach. The main advantage of this scheme is that it tries to address a physical-layer attack that is usually hard to prevent. However, the main shortcoming of this approach comes from the fact that the higher the privacy level, the more significant the degradation of network throughput. This means that using their scheme to preserve the location privacy of \su s would degrade system performance.

\subsubsection{Spectrum mobility}
\label{solMobility}
Spectrum mobility necessarily involves the usage of different spectrum bands over time and as \su s move. However, as explained in Section~\ref{sourcesMobility}, spectrum utilization information can become a serious source of location information leakage especially when the number of used channels increases.
Gao et al.~\cite{gao2013location} propose a technique to prevent this in database-driven \crn s by relying on two observations: The first is that higher location information leakage takes place during the channel switching process; i.e., when \su~switches from one channel to another. This means that if there is a way to make a \su~only switch to a channel that it has already used previously, then this would not give extra information that could be exploited by the adversary. The second is that \su s that choose the most stable channels are less likely to switch channels. Based on these two observations, each \su~constructs a list that stores its used channels and a prediction list that contains the prediction of the duration of channels availability. \su~chooses a channel from the first list, containing the usage history, if it is available. Otherwise, \su~uses the second list containing the predicted availability duration of each channel to make sure that it picks the one with the best estimated duration, i.e. the most stable. Despite its merit in reducing the localization probability of \su s, this approach does not completely thwart the attack based on \su's spectrum mobility. It just reduces the action space of the adversary which is still able to approximate \su's location when it tunes to other channels. Hence, as the number of channel switching events increases, the localization probability increases. In addition, it suffers from a relatively high computational overhead.


\subsection{Performance metrics and tradeoffs}
\subsubsection{Performance metrics}

\textbf{Computational complexity:}
This is again an essential metric that needs to be used to evaluate any proposed scheme. It has already been discussed in previous sections.

\textbf{Communication overhead:}
This is also an essential metric due to bandwidth constraints in \crn s, and has also been discussed in previous sections.

\textbf{Privacy level:}
The approaches used here are very similar to the approaches stressed in the previous sections. For instance, Liu et al.~\cite{liu2013location} rely on the previously discussed concepts of {\em uncertainty} and {\em incorrectness} (see Section~\ref{coopPerf}) to assess the privacy level of their proposed scheme.
Another metric could be the {\em number of used channels} as it is important to minimize the frequency of \su s' switching events to avoid attacks relying on the channel utilization as explained in Section~\ref{attacksOther}. So, the number of used channels could be seen as a suitable metric to evaluate how a privacy-preserving scheme performs in preventing such attacks as done in~\cite{gao2013location}.

\subsubsection{Performance tradeoffs}
As in the spectrum discovery phase, designing location privacy preserving protocols for spectrum analysis, sharing and mobility may require some tradeoffs between providing location privacy and maintaining some utility. For example, Zhang et al.~\cite{zhang2015privacy} consider making tradeoffs between achieving high location privacy and maintaining high network throughput. Indeed, increasing the location privacy level using their approach, as explained in Section~\ref{lpparc}, is equivalent to increasing the perturbation level on the transmission power of \su s to prevent the adversary from accurately localizing them. However, as the perturbation level increases, and so does the privacy level, the network throughput decreases, hindering thus the \crn~performance.

\subsection{Summary}
In this section, we discussed the location privacy issues in the spectrum analysis, spectrum sharing and spectrum mobility components. We detailed the different threat models, location inference attacks, and location privacy preserving approaches that are proposed in the literature to protect the location privacy in \crn s with a focus on the aforementioned components. Finally, we explained the different performance metrics that could be used to assess the efficiency and the privacy level of location privacy preserving protocols in these components. In the following section, we will discuss some of the open research problems and challenges with respect to the location privacy in \crn s.


\section{Open research problems}
\label{openproblems}
There are still open research problems that could be further investigated when it comes to location privacy in \crn s. The following is a list of some of these challenges.

{\bf Location privacy in spectrum analysis:}
Location privacy issues arising during the spectrum analysis process have received little attention by the research community in spite of, as discussed in Section~\ref{sourcesDec}, the several vulnerabilities and sources of location information leakage this process has. Much work still needs to be done when it comes to investigating inference attack models that can exploit these sources of leakage, as well as developing countermeasure solution protocols that tackle those inference attacks.
For instance, an attack framework could combine information like topology, connectivity, interference and \rem~to localize \su s, since this information could be accessible during the spectrum analysis process as highlighted in Section~\ref{sourcesDec}. To the best of our knowledge, none of the existing works have exploited these vulnerabilities, nor did they try to defend them.

{\bf Location privacy in spectrum sharing and mobility:}
Not many approaches in the literature have addressed the location privacy issue in these components of the cognition cycle despite the amount of information that could be leaked during spectrum sharing and mobility as stressed in Sections~\ref{sourcesSharing}~\&~\ref{specMobility}. This is still an open issue that requires further efforts from the research community.

{\bf Location privacy in distributed cooperative sensing:}
The research efforts on providing location privacy to \su s in cooperative spectrum sensing have focused on centralized approaches but little has been done to address this issue for distributed cooperative sensing. Little work has been done in this regard (e.g.~\cite{kasiri2015privacy}); this research area is still not mature enough and requires further investigation.

{\bf Location privacy with malicious adversaries:}
Most of the existing location privacy preserving protocols in \crn s consider attack scenarios that assume no collusion between the different network entities; for example, in the context of cooperative spectrum sensing, it is almost always assumed that there is no collusion between \fc~and some \su s. However, it is not unrealistic to assume that different entities can collude with one another to infer location information, especially that collusion often leads to better inference. Techniques that address colluding attackers still need to be developed and investigated, as not much has been done in this regard.


{\bf Location privacy for crowdsourced spectrum sensing:}
Crowdsourcing is an emerging tool that is gaining lots of interest in the context of \crn s. It enables the discovery of spectrum opportunities in regions with insufficient presence of \su s. In such cases, one can rely on other users (not necessary \su s) to assess which and whether other channels are available, mainly through an open call kind of process. To participate, these other users can be encouraged through various types of incentives (e.g., monetary, credit, etc.). In the context of \crn s, crowdsourcing suffers from location privacy risks that may expose the whereabouts of participating mobile users. Dealing with this issue is still an open problem and only a few works in the literature have dealt with it~\cite{jin2016privacy}.

{\bf Location privacy of ${\boldsymbol \pu}$s:}
This is another direction that is worth investigating, as the location of \pu s could be of paramount importance, especially in the case of military incumbent systems that have stringent requirements in terms of security and privacy. Also, \crn~solutions that rely on the cooperation of \pu s may fail or poorly perform if \pu s are concerned about their location privacy.
Addressing the location privacy of
\pu s is still in its infancy, and more still needs to be done~\cite{bahrak2014protecting,zhang2015achieving,zhang2015optimal,clarkcan2016can}.

{\bf Location privacy in emerging ${\bm \cogr}$-based technologies:} Emerging \cogr-based technologies~\cite{wang2011emerging} may bring additional location privacy challenges on top of the ones that we have discussed in this paper. For instance, in cognitive radio-based cellular networks~\cite{elsawy2013stochastic,thilina2015cellular,guizani2015large}, multiple base stations may localize or track \su s as they move across different cells. The relatively small size of the cells in this kind of networks could make it easier to localize \su s. In \crn-enabled smart grids~\cite{khalfi2014optimal,khan2016cognitive,bicen2012spectrum}, smart meters act as \su s and opportunistically search for the available spectrum to transmit their data. The location privacy concern here is quite different as it does not involve tracking a user but can lead to identifying his own personal address if a smart meter is localized. The location information when augmented with power consumption data sent by the smart meters can further reveal the presence or absence of home owners and could lead to burglary for example. Another emerging \cogr-based technology is cognitive radio sensor networks ($CRSN$)~\cite{akan2009cognitive,bukhari2016survey} where the sensor nodes are required to sense the environment and also the spectrum. Depending on the spectrum availability, sensor
nodes, acting as \su s, transmit their readings in an opportunistic manner to their next hop cognitive radio sensor nodes, and ultimately, to the sink. As the sensor nodes exchange their sensing results of both the spectrum and the environment with other nodes, this presents considerable threats to the location privacy of these nodes and makes $CRSN$ inherit the location privacy issues of both $WSN$s and \crn s. All of these technologies share similar privacy threats but also have their unique vulnerabilities as well. Thus, there cannot be a one-fits-all solution to address the location privacy in these technologies, and further research efforts need to be made to investigate and address issues that are specific to each of these technologies.

{\bf Location privacy in multi-database-driven ${\bm\crn}$s:} As \fcc~has already approved several companies to administrate, operate and manage spectrum databases, leveraging the existence of these multiple databases (which are inherent to spectrum database-driven dynamic spectrum sharing) opens up a new class of very promising, spectrum access techniques that can guarantee the protection of users' location privacy information yet without incurring significant overhead. This area has not been explored yet, and research efforts need to be made to investigate the potential of such an approach.

\section{Conclusion}
\label{con}
In this survey, first, we have investigated \su s' location privacy issues in \crn s by exploring each functional component and identifying its inherent vulnerabilities. Then, we have discussed when and why generic and well known privacy enhancing approaches cannot be applied off-the shelf to provide location privacy for \su s. After that we have explored existing attacks and approaches for providing location privacy solutions in the different \crn~components. Finally, we have highlighted some related open research problems that require future investigation and attention.

\section*{Acknowledgment}
This work was supported in part by the US National Science Foundation under NSF award CNS-1162296.
The authors would like to thank the editor and the reviewers for their valuable feedback that has improved this survey paper greatly.
\small{
\bibliographystyle{IEEEtran}
\bibliography{IEEEabrv,references,references-CRNs,refs-security-privacy-sensing-14-15,references_Attila,refs-bechir-privacy-wireless-systems}
}

\vspace{-30pt}
\begin{IEEEbiography}[{\includegraphics[width=1in,height=1.25in,clip,keepaspectratio]{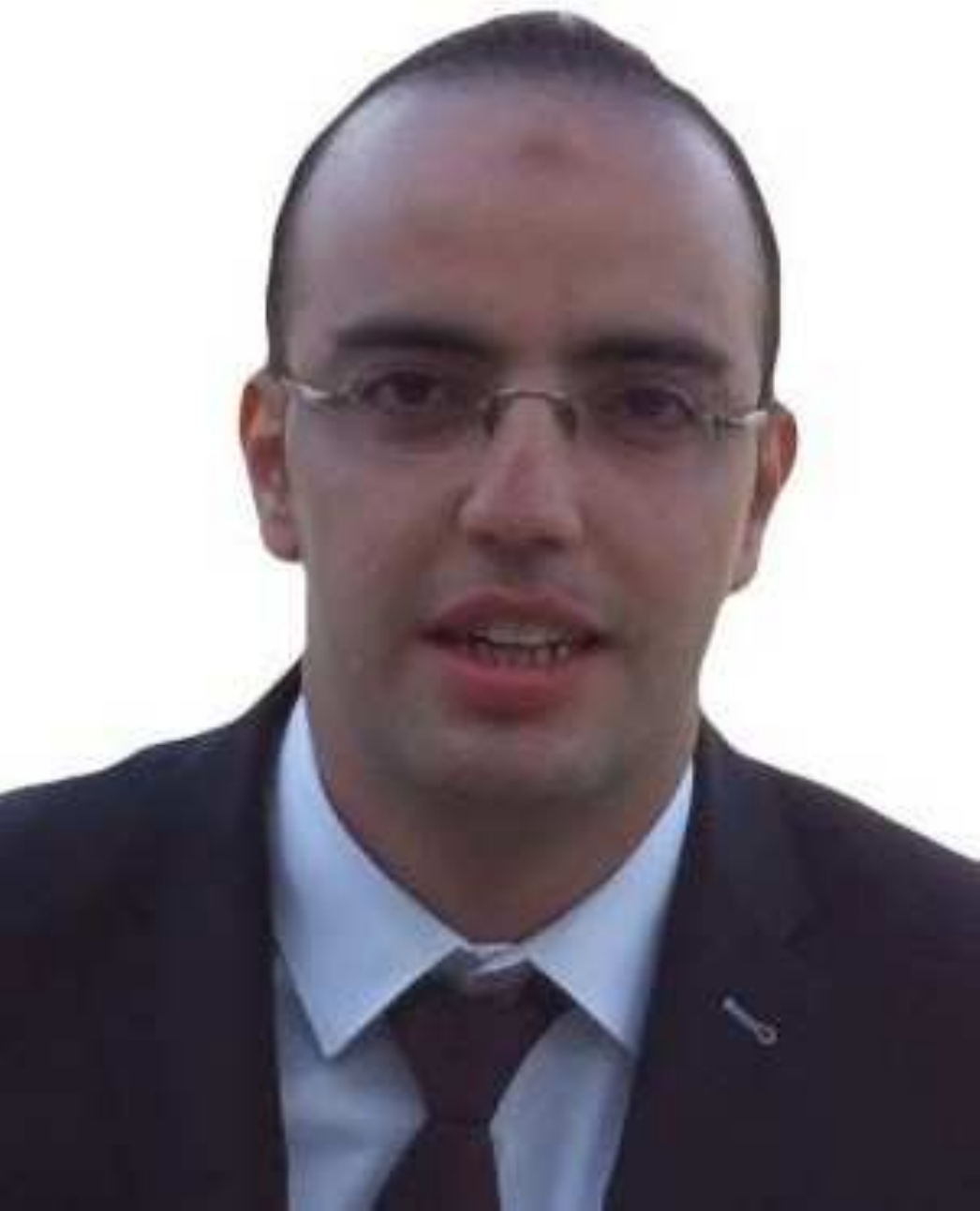}}]{Mohamed Grissa}(S'14) received the Diploma of Engineering (with highest distinction)
in telecommunication engineering from Ecole
Superieure des Communications de Tunis, Tunis,
Tunisia, in 2011, and the
M.S. degree in electrical and computer engineering
(ECE) from Oregon State University, Corvallis, OR, USA, in 2015. He is currently working
toward the Ph.D. degree at the School of Electrical
Engineering and Computer Science (EECS), Oregon
State University, Corvallis, OR, USA.

Before pursuing the Ph.D. degree, he worked as a Value Added Services Engineer at Orange France Telecom Group from 2012 to 2013. His research interests include privacy and security in wireless networks, cognitive radio networks, IoT and eHealth systems.
\end{IEEEbiography}

\begin{IEEEbiography}[{\includegraphics[width=1in,height=1.25in,clip,keepaspectratio]{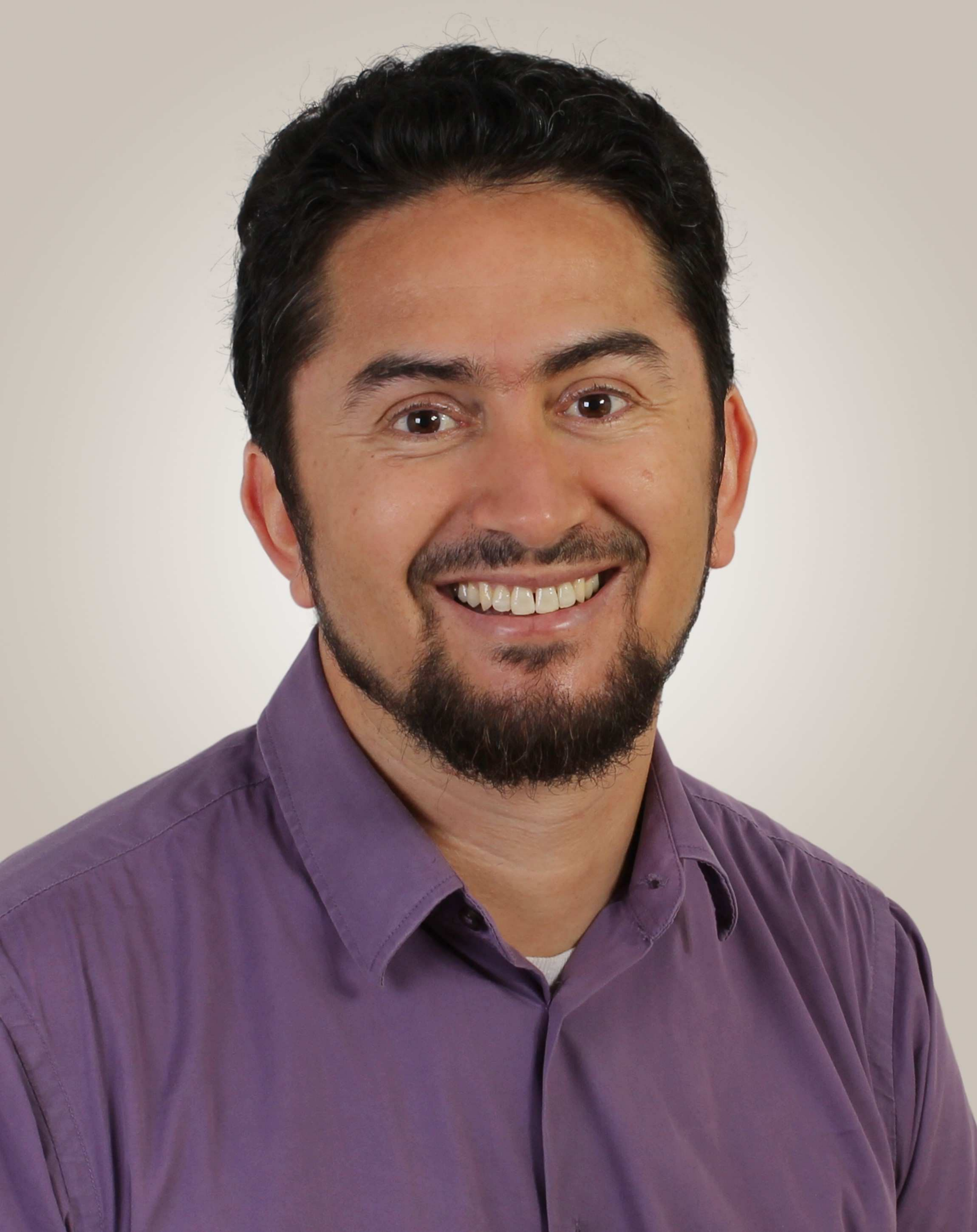}}]{Bechir Hamdaoui} (S'02\textendash M'05\textendash SM'12) is presently an Associate Professor in the School of EECS at Oregon State University. He received the Diploma of Graduate Engineer (1997) from the National School of Engineers at Tunis, Tunisia. He also received M.S. degrees in both ECE (2002) and CS (2004), and the Ph.D. degree in ECE (2005) all from the University of Wisconsin-Madison. His research interest spans various areas in the fields of computer networking, wireless communications, and mobile computing, with a current focus on distributed optimization, parallel computing, cognitive networks, cloud computing, and Internet of Things. He has won several awards, including the 2016 EECS Outstanding Research Award and the 2009 NSF CAREER Award. He is presently an Associate Editor for IEEE Transactions on Wireless Communications (2013-present). He also served as an Associate Editor for IEEE Transactions on Vehicular Technology (2009-2014), Wireless Communications and Mobile Computing Journal (2009-2016), and for Journal of Computer Systems, Networks, and Communications (2007-2009). He is currently serving as the chair for the 2017 IEEE INFOCOM Demo/Posters program. He has also served as the chair for the 2011 ACM MOBICOM's SRC program, and as the program chair/co-chair of several IEEE symposia and workshops, including GC 16, ICC 2014, IWCMC 2009-2017, CTS 2012, and PERCOM 2009. He also served on technical program committees of many IEEE/ACM conferences, including INFOCOM, ICC, and GLOBECOM. He has been selected as a Distinguished Lecturer for the IEEE Communication Society for 2016 and 2017. He is a Senior Member of IEEE, IEEE Computer Society, IEEE Communications Society, and IEEE Vehicular Technology Society.
\end{IEEEbiography}

\begin{IEEEbiography}[{\includegraphics[width=1in,height=1.25in,clip,keepaspectratio]{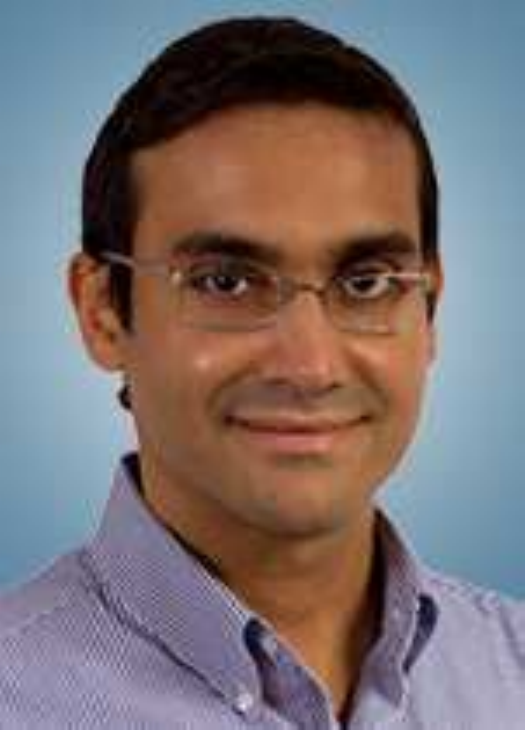}}]{Attila A. Yavuz} (S'05\textendash M'10) received a BS degree in Computer Engineering from Yildiz Technical University (2004) and a MS degree in Computer Science from Bogazici University (2006), both in Istanbul, Turkey. He received his PhD degree in Computer Science from North Carolina State University in August 2011. Between December 2011 and July 2014, he was a member of the security and privacy research group at the Robert Bosch Research and Technology Center North America. Since August 2014, he has been an Assistant Professor in the School of Electrical
Engineering and Computer Science, Oregon State University, Corvallis, USA. He is also an adjunct faculty at the University of Pittsburgh's School of Information Sciences since January 2013.

Attila A. Yavuz is interested in design, analysis and application of cryptographic tools and protocols to enhance the security of computer networks and systems. His current research focuses on the following topics: Privacy enhancing technologies (e.g., dynamic symmetric and public key based searchable encryption), security in cloud computing, authentication and integrity mechanisms for resource-constrained devices and large-distributed systems, efficient cryptographic protocols for wireless sensor networks.
\end{IEEEbiography}
%
%

\end{document}